\PassOptionsToPackage{pdfpagelabels=false}{hyperref} 
\documentclass[fleqn,usenatbib]{mnras}

\usepackage{hyperref}

\usepackage{epsfig}
\usepackage{hyperref}
\usepackage[T1]{fontenc}
\usepackage{natbib}
\usepackage{newtxtext,newtxmath}
\usepackage{graphicx}	
\usepackage{amsmath}	
\usepackage{amssymb}	
\usepackage{lipsum}
\usepackage{float}

\newcommand{\kms}{km~s$^{-1}$ }
\newcommand{\kmsp}{km~s$^{ -1}$}

\newcommand{\heii}{\ion{He}{ii}}
\newcommand{\hei}{\ion{He}{i}}

\newcommand{\mgii}{\ion{Mg}{ii}}
\newcommand{\mgi}{\ion{Mg}{i}}

\newcommand{\hi}{\rm \textsc{H}$^{0}$}
\newcommand{\lya}{Ly$\alpha$}

\newcommand{\fescm}{$f_\text{esc}\left(\ion{Mg}{ii}\right)$}
\newcommand{\fescl}{$f_\text{esc}\left(2803\right)$}
\newcommand{\fescc}{$f_\text{esc}$$\left({\rm LyC}\right)$}
\newcommand{\nmg}{$N_{{\rm Mg}^{+}}$}
\usepackage[nameinlink,capitalise]{cleveref}
\usepackage{ulem}

\newcommand{\nhi}{$N_\mathrm{H^0}$}

\begin{document}

\title[Optically-Thin \mgii\ Emission Maps the Escape of Ionizing Photons]{Optically-Thin Spatially-Resolved \mgii\ Emission Maps the Escape of Ionizing Photons\footnote{The data presented herein were obtained at the W. M. Keck Observatory, which is operated as a scientific partnership among the California Institute of Technology, the University of California and the National Aeronautics and Space Administration. The Observatory was made possible by the generous financial support of the W. M. Keck Foundation.}}
\author[Chisholm et al.]{J. Chisholm$^{1}$\thanks{Hubble Fellow}\thanks{Email: jochisho@ucsc.edu}, J. X. Prochaska$^{1,2}$, D. Schaerer$^{3}$, S. Gazagnes$^{4}$, and A. Henry$^{5}$ \\
$^{1}$  University of California--Santa Cruz, 1156 High Street, Santa Cruz, CA, 95064, USA \\
$^2$  Kavli Institute for the Physics and Mathematics of the Universe, 5-1-5 Kashiwanoha, Kashiwa, 277-8583, Japan\\
$^{3}$  Observatoire de Gen\`eve, Universit\'e de Gen\`eve, 51 Ch. des Maillettes, 1290 Versoix, Switzerland \\
$^{4}$  Kapteyn Astronomical Institute, University of Groningen, P.O Box 800, 9700 AV Groningen, The Netherlands \\
$^5$ Space Telescope Science Institute, 3700 San Martin Drive, Baltimore, MD, 21218, USA}

\date{Accepted 2020 August 12. Received 2020 August 11; in original form 2020 June 25}
\pubyear{2019}
\label{firstpage}
\pagerange{\pageref{firstpage}--\pageref{lastpage}}
\maketitle
\begin{abstract}
{Early star-forming galaxies produced copious ionizing photons. A fraction of these photons escaped gas within galaxies to reionize the entire Universe. This escape fraction is crucial for determining how the Universe became reionized, but the neutral intergalactic medium precludes direct measurement of the escape fraction at high-redshifts. Indirect estimates of the escape fraction must describe how the Universe was reionized. Here, we present new Keck Cosmic Web Imager spatially-resolved spectroscopy of the resonant \mgii~2800~\AA\ doublet from a redshift 0.36 galaxy,   {J1503+3644}, with a previously observed escape fraction of 6\%. The \mgii\ emission has a similar spatial extent as the stellar continuum, and each of the \mgii\ doublet lines are well-fit by single Gaussians. The \mgii\ is optically thin.  {The intrinsic flux ratio of the red and blue \mgii\ emission line doublet, $R=F_{2796}/F_{2803}$}, is set by atomic physics to be two, but Mg$^+$ gas along the line of sight decreases $R$ proportional to the \mgii\ optical depth. Combined with the metallicity, $R$ estimates the neutral gas column density. The observed $R$ ranges across the galaxy from  {0.8-2.7}, implying a factor of 2 spatial variation of the relative escape fraction. All of the ionizing photons that escape  { J1503+3644} pass through regions of high $R$.  We combine the \mgii\ emission and dust attenuation to accurately estimate the absolute escape fractions for  {ten} local Lyman Continuum emitting galaxies and suggest that \mgii\ can predict escape fraction within the Epoch of Reionization.}
\end{abstract}

\section{Introduction}
At redshifts between 6--10 the first galaxies emitted a sufficient number of photons with $\lambda < 912$ (or Lyman Continuum, LyC, photons) to reionize all of the neutral hydrogen between galaxies in the Universe \citep{becker, fan2006, banados}. This "Epoch of Reionization" (EoR) marked the first time that galaxies exerted their influence over the entire Universe. However, observations have not established \textit{how} galaxies reionized the universe because the sources of ionizing photons from within these first galaxies are, at the moment, observationally unconstrained.

The two major sources of ionizing photons are massive stars and accretion onto black holes (Active Galactic Nuclei or AGN). The total emissivity ($j_\text{ion}$[photon s$^{-1}$ Mpc$^{-3}$]) of either AGN or star-forming galaxies can be observationally estimated as the product of their FUV luminosity function ($\rho_\text{UV}$[erg s$^{-1}$  Hz$^{-1}$ Mpc$^{-3}$]), the intrinsic production of ionizing photons per FUV luminosity for each source ($\xi_\text{ion}$[photon erg$^{-1}$ Hz]), and the fraction of ionizing photons that escape a galaxy (the escape fraction; \fescc). Numerically, this is
\begin{equation}
    j_\text{ion} = \text{\fescc}\xi_\text{ion} \rho_\text{UV} .
    \label{eq:emis}
\end{equation}
If \fescc, $\xi_\text{ion}$, and $\rho_{\rm UV}$ can all be measured then observations can determine $j_\text{ion}$ of both star-forming galaxies and AGN within the EoR. A direct comparison to the cosmological matter density, with an assumption about the clumpiness of the early Universe, determines whether a given source produces sufficient ionizing photons to reionize the Universe. High-redshift AGN are an appealing source of ionizing photons because their high-ionization and high-luminosity means that \fescc~$\approx100$\%. However, current observations find too few AGN (low $\rho_\text{UV}$) to reionize the Universe \citep{hopkins08, willott, ricci, onoue, matsuoka18, shen20}, but there may be numerous unobserved, low-luminosity AGN \citep{Giallongo15, Grazian16, matsuoka19}.

Meanwhile, vigorously star-forming galaxies must also have large reservoirs of cold neutral gas. Neutral gas efficiently absorbs ionizing photons, reducing the number of ionizing photons that escape a typical star-forming galaxy. As such \fescc\ is often observed to be less than 5\% 
in the local Universe \citep{grimes09, vanzella10, leitherer16, naidu18}. However, models using typical estimates of $\xi_{\rm ion}$ and $\rho_\text{UV}$ indicate that \fescc\ must be $>$5-20\% for star-forming galaxies to reionize the universe \citep{ouchi09, robertson13, robertson15, finkelstein19, naidu19}. These escape values are only found in the most extreme local galaxies \citep{leitet11, borthakur,izotov16a, izotov16b, izotov18a, izotov18b}.

Recently, there has been tremendous success in directly measuring \fescc\ from extremely compact, high-ionization, star-forming galaxies \citep{leitet11, borthakur, izotov16b, izotov16a, vanzella16, shapley16, izotov18a, izotov18b, steidel18, fletcher19, rivera19, wang19}, demonstrating that star-forming galaxies \textit{can} emit ionizing photons. The physical parameters of the recently discovered local emitters of ionizing photons ( {low stellar mass, high specific star formation rate SFR/M$_\ast$}, compact, low-metallicity, and low dust attenuation) are similar to the expected properties of the first galaxies \citep{schaerer10}.  However, it is challenging to be certain whether low-redshift galaxies are actually analogs to EoR galaxies. Direct observations of \fescc\ must determine the sources of reionization.

Direct observations of \fescc\ above $z\sim4$ are statistically unlikely because the neutral intergalactic medium (IGM) surrounding these galaxies efficiently absorbs LyC photons \citep{worseck14}. This high IGM opacity means that we will not directly observe the LyC of star-forming galaxies during the EoR. A major goal for the upcoming \textit{James Webb Space Telescope} (JWST) is to determine the sources of cosmic reionization, but to accomplish this requires indirect methods to determine \fescc. 

The two major sinks of ionizing photons are photoelectric absorption by neutral hydrogen and dust; both can remove similar amounts of ionizing photons  \citep{chisholm18}. Ideal indirect \fescc\ methods must be: unaffected by the high-redshift neutral IGM, bright enough to be observed at high-redshift, and trace  \hi\ column density variations between 10$^{16-17.2}$~cm$^{-2}$ where \hi\ becomes optically thin to the ionizing continuum. There are currently a number of prospects of indirect indicators of ionizing photon escape \citep{heckman2011, alexandroff, verhamme, henry, chisholm18, henry18, steidel18, Mckinney, berg19}. However, each has their own drawbacks: \lya\ is bright but the neutral IGM can absorb a large portion of the \lya\ profile at high-redshifts \citep{verhamme17}; the Lyman Series absorption lines are  {within the Ly$\alpha$ forest at high-redshift and are challenging to disentangle from foreground IGM absorption}; and metal-absorption lines require deep observations to detect the stellar continuum  \citep{steidel18, chisholm18, jaskot19}.

The ionization state of the strong \mgii~2800~\AA\ doublet overlaps with \hi\ ( {the \mgii\ ionization potential is 15~eV}), such that \mgii\ emission  traces neutral gas and may be an ideal indirect indicator of \fescc\ \citep{henry18}. Depending on the metallicity of the intervening gas, \hi\ gas column densities $N_{\rm H^0} < 10^{17}$~cm$^{-2}$ lead to optically thin \mgii\ absorption profiles (see \autoref{theory}). Thus, the absence of \mgii\ absorption suggests a neutral gas column density low enough to transmit ionizing photons. Further, \mgii\ emission of low-metallicity galaxies can be 10-60\% of the observed H$\beta$ flux \citep{guseva13}, indicating that \mgii\ emission may be sufficiently bright to observe with upcoming facilities in the distant universe. Thus, \mgii\ emission may help determine the sources of cosmic reionization. 

Here, we present new spatially-resolved spectroscopic observations of the \mgii~2796, 2803~\AA\ doublet from a previously-confirmed LyC emitter, J1503+3644 \citep{izotov16b}. We use these observations to explore the neutral gas properties within this source of ionizing photons and test whether \mgii\ emission traces the escape of ionizing photons. The outline of the paper is as follows: \autoref{data} describes the data reduction and analysis. We then explore the spatially integrated (\autoref{int}) and spatially resolved (\autoref{resolved}) \mgii\ emission line properties. {The physical implications of the spatial distribution and kinematics of the \mgii\ emission and neutral gas is discussed in \autoref{hi_dist}.} \autoref{theory} explores the relationship between the \mgii\ emission and the neutral gas column densities that we use  {in \autoref{highz} to indirectly infer the \fescm\ and \fescc}. \autoref{future} describes the future prospects to detect \mgii\ at high-redshift to determine the sources of cosmic reionization. 

\section{DATA}
\label{data}

\begin{table}
\caption{Properties of J1503 from \citet{izotov16b}. The first row gives the redshift, the second row gives the logarithm of the stellar mass, the third row gives the O$^{++}$ temperature, the forth row gives the electron density, the fifth row gives the nebular metallicity determined from the [\ion{O}{iii}]~4363~\AA\ line  {(using the T$_e$ method)}, the sixth and seventh row give the H$\beta$ and [\ion{O}{iii}]~5007~\AA\ restframe equivalent width, and the eighth row gives the [\ion{O}{iii}]~5007~\AA\ to [\ion{O}{ii}]~3727~\AA\ flux ratio.}
\begin{tabular}{lc}
Property & Value \\
\hline
z & 0.3557 \\
log(M$_\ast$/M$_\odot$) & 8.22 \\ 
T(\ion{O}{iii}) & 14850~K\\
n$_{\rm e}$ & 280~cm$^{-3}$ \\
12+log(O/H) & 7.95 \\
H$\beta$ Equivalent width & 297~\AA \\
$[$\ion{O}{iii}$]$~5007 Equivalent width & 1403\AA \\
$F$[\ion{O}{iii}]~5007 / $F$[\ion{O}{ii}]~3727 & 4.9 \\
\end{tabular}
\label{tab:props}
\end{table}

\subsection{Observations and Data Reduction}
\label{reduction}
We selected SDSS~J150342.83+364450.75 \citep[hereafter J1503;][]{izotov16b} because the galaxy has the largest \mgii~2800~\AA\ emission flux from public Sloan Digital Sky Survey (SDSS) observations \citep{sdss} in the \citet{izotov16b} LyC emitting sample. At $z = 0.3557$, the \mgii\ 2800\AA\ doublet is observed at wavelengths of 3789.90 and 3799.63\AA, respectively, within the wavelength range of the blue sensitive Integral Field Spectrograph Keck Cosmic Web Imager (KCWI) on the Keck~\textsc{ii} telescope \citep[][]{kcwi}. KCWI is an image slicer that is highly optimized to measure faint diffuse emission, and ideally suited to map the extended \mgii\ structure in J1503. 

KCWI is highly configurable and contains an array of beam-slicers and gratings. The different configurations trade off between field of view, wavelength coverage, spatial sampling, and spectral resolution. J1503 is a compact star-forming galaxy that is unresolved by the SDSS and NUV imaging \citep{izotov16b}, thus a large field of view is not as important as spatial sampling. Similarly, we are predominately interested in spectrally resolving the \mgii\ lines. Thus, we used the small image-slicer, the BM grating with a central wavelength of 4000~\AA, and 1x1 binning to afford a field of view of 8.4\arcsec~$\times$~20.4\arcsec, a spatial sampling of 0.35\arcsec\ perpendicular to the slice direction (seeing limited along the slice direction), a full restframe wavelength coverage of 2700--3300~\AA, and a spectral resolution of R~$=8000$ (37~\kmsp).

On the night of January 31$^{\rm st}$ 2019, we obtained a total of 65 on-source minutes (3900~s) using an AABB dither pattern with a dither separation of 1\arcsec and an average airmass of 1.07. The individual exposures were processed through the standard KCWI KDERP pipeline Version 1.1.0\footnote{\url{https://github.com/Keck-DataReductionPipelines/KcwiDRP}} \citep{kcwi}. The eight major steps of this reduction include: (1) bias, over-scan, and cosmic ray removal, (2) dark and scattered light subtraction; (3) geometric transformation and wavelength calibration using a ThAr arc lamp in each image slice with a mean RMS scatter of the calibration emission of 0.042~\AA\ about the reference values, below the expected 0.2\AA\ for the BM grating; (4) flat fielding each slice by creating a master-flat using 6 dome flats; (5) standard sky subtraction, (6) collapsing the individual slices into spatial intensity and variance cubes in air wavelengths; (7) a differential atmospheric refraction correction based upon the observed airmass of J1503; and (8) flux calibrating the J1503 observations using the standard star BD+26~2606. Below, we provide additional specifics for a few of these steps.

\begin{figure}
\includegraphics[width = 0.5\textwidth]{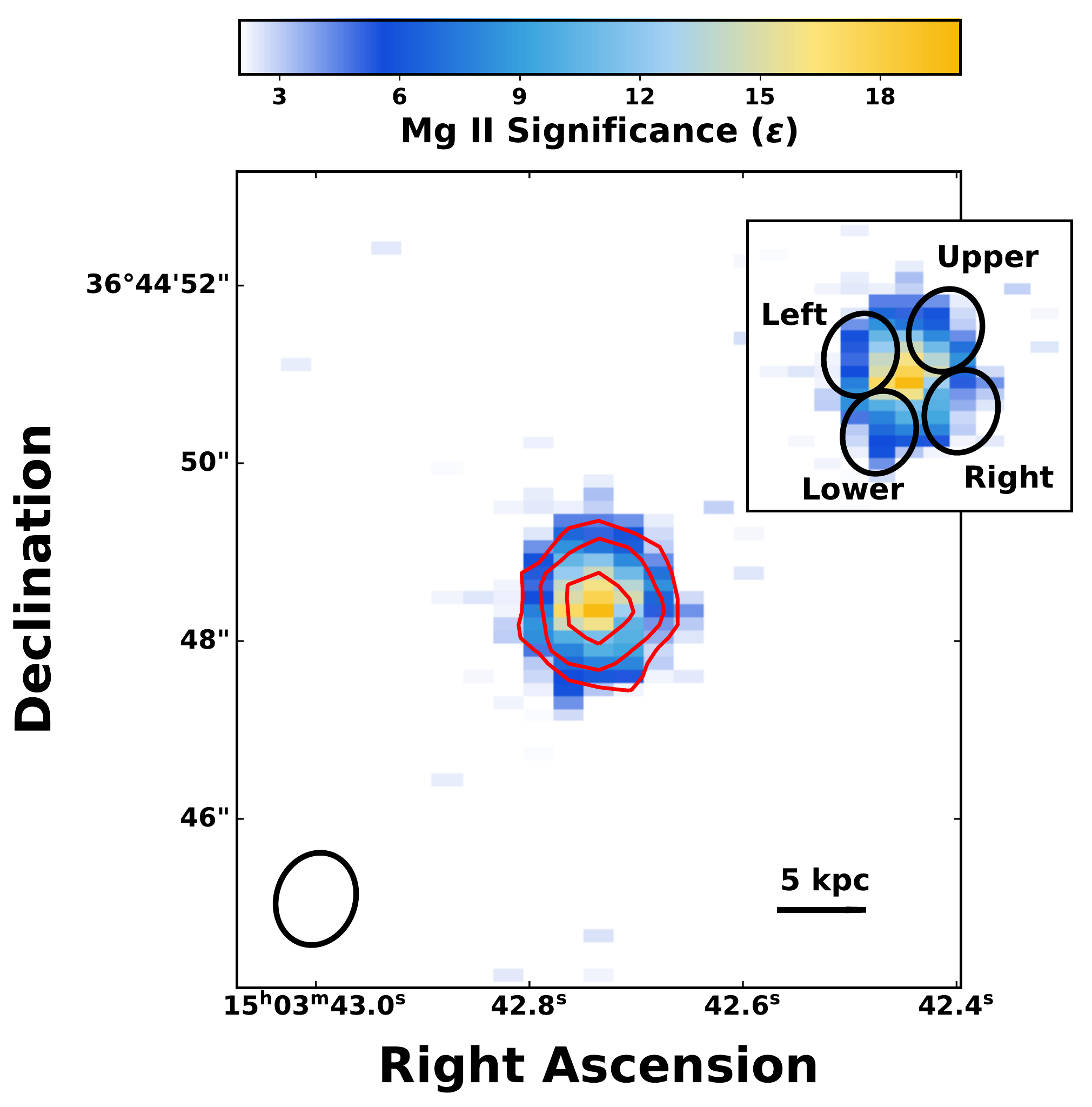}
\caption{ {Spatial map of the integrated, continuum-subtracted \ion{Mg}{ii} emission (the 2796 plus the 2803~\AA\ transitions) from J1503, in units of the \mgii\ statistical significance: $\epsilon = 1.5\times10^{-18}$~erg~s$^{-1}$~cm$^{-2}$ per spaxel. Red contours are the 2, 5, and 10 $\epsilon$ contours of the stellar continuum with $\epsilon_{\rm s} = 1.4\times10^{-19}$erg~s$^{-1}$~cm$^{-2}$~\AA$^{-1}$ per spaxel. The measured 1.04\arcsec$\times$~0.92\arcsec\ seeing, from the standard star observation, is given by the black circle in the lower left. A scale bar is given in the lower right}  { and this figure is a 9\arcsec$\times$9\arcsec cutout of the full { 8.4\arcsec$\times$20.4\arcsec} KCWI field of view. The inset shows the location of the four spatial distinct apertures used in \autoref{tab:spatially_distinct}.}}
\label{fig:sdss_int}
\end{figure}

We used the standard sky subtraction which uses a B-spline to generate a 2-dimensional sky model. Emission from astronomical objects within the field of view were excluded using a 1$\sigma$ clipping. Since J1503 is small on the sky (\autoref{fig:sdss_int}), the 8.4\arcsec~$\times$~20.4\arcsec field of view contains a sufficient amount of emission-free regions to accurately model the sky emission. We tested whether providing a mask around the galaxy before creating the sky model improved the sky-subtraction, but we found negligible differences. After the sky-subtraction, we checked multiple regions within the 2-D data cube for sky over-subtraction and did not find any such indication.

The four individual reduced exposures were shifted into the reference frame of the first exposure and the three subsequent exposures were combined with an inverse variance weighting.  {With the small slicer and 1x1 binning, the spatial-pixels (spaxels) in the resultant data cubes have sizes that are 0.35\arcsec$\times$0.147\arcsec\ in the Right Ascension and Declination directions. This leads to the rectangular pixels shown in the spatial plots.}

Immediately after the J1503 science observations, we observed the standard star BD+26~2606. This blue A5V star is one of the KWCI standard stars and is sufficiently blue to flux calibrate the observed wavelengths. We ran the KCWI pipeline in interactive mode, setting the standard star fitting regions by hand to minimize the residuals, while paying special attention to mask out the stellar Balmer absorption features and sky emission lines. The final inverse sensitivity curves were inspected to ensure continuity and proper fits to the stellar continuum. Finally, the total spatially-integrated, flux-calibrated spectrum of J1503 was compared to the SDSS spectrum. We found a factor of 1.4 difference in the total flux, but the overall spectral shape of the KCWI  observations nicely matched the SDSS spectrum. Also, the relative wavelength calibration was excellent between the two spectra, such that the \mgii\ emission peaks were consistent within the spectral resolution. The spatially integrated KCWI spectrum was multiplied by a factor of 1.4 to align with the SDSS to complete the flux calibration.  {Only the relative fluxing of the KCWI observations matter because we only use observables within the KCWI observations which have the same flux calibration. }

In the spectral dimension, we corrected the observed frame spectra for foreground Milky Way reddening using the \citet{cardelli} attenuation curve and the reddening value of E(B-V) = 0.013 from \citet{green15}. We then shifted the spectrum into the restframe of the galaxy, $\lambda_{\rm rest} = \lambda_{\rm obs}/(1+z)$, using $z = 0.3557$ as measured from the SDSS spectrum \citep{izotov16b}.

We subtracted the continuum from the spatial cube using the Common Astronomy Software Applications \citep[\textsc{casa};][]{casa} routine \textsc{imcontsub}, using a first order polynomial. We tested different orders and found that orders higher than 1 produced similar results. For the \mgii\ emission region, we fit the continuum in spectral regions that, by eye, avoided two [\ion{Fe}{iv}] emission lines at 2829.4 and 2835.7~\AA, respectively, and \mgi~2852~\AA. The continuum was fit at $\lambda_{\rm rest} = 2766-2791$~\AA\ and $\lambda_{\rm rest} = 2808-2825$~\AA\ (see blue regions in \autoref{fig:sdss_full}). Similarly, we fit a first-order polynomial between restframe wavelengths of 3171--3182~\AA\ and 3194--3200~\AA\ for the \ion{He}{i}~3188~\AA\ line to avoid possible contributions from the weakly detected \ion{He}{ii}~3202~\AA\ line. This produced wavelength-calibrated, flux-calibrated, continuum-subtracted cubes from which we analyzed the data. 

To quantify the observed spatial resolution, we fit the size of the standard star, BD+26~2606, in the KCWI cubes using the \textsc{casa} two-dimensional fitting tool at the same wavelength as the \mgii\ emission ($\lambda_{\rm obs} \approx 3790$~\AA). We measured the size of the standard star to be $1.04\pm0.01$\arcsec$~\times~0.92\pm0.01$\arcsec\ at 177$^\circ$. This is consistent with the 1-1.2\arcsec\ seeing reported by the DIMM seeing at the telescope that remained relatively stable during the observations. A single slice width in this configuration is 0.35\arcsec, such that there are at least 2.6 spaxels per seeing PSF. We find a similar standard star size for the spectral region near the \ion{He}{i}~3188~\AA\ line. We use this as our spatial resolution and denote it as a black circle in the bottom left of our spatial plots (e.g., \autoref{fig:sdss_int}). 

We also compare the KCWI observed-frame optical spectrum to the HST/COS G160M \citep{cos} far-ultraviolet spectrum of J1503 \citep[HST project ID: 13744, PI: Thuan;][]{izotov16b, verhamme17}. We downloaded the data from MAST and extracted the spectrum following the methods outlined in \citet{worseck16} that carefully consider the pulse-heights of the extraction region to optimize the extraction of faint objects. The G160M data have nominal spectral resolution of 20~\kmsp, similar to the KCWI observations.

\subsection{Line Profile Fitting}
\label{fits}
We calculated the inferred \mgii\ emission line properties in two ways. The first method used the \textsc{casa} \textsc{immoments} routine to create continuum-subtracted \mgii~2800~\AA\ and \hei~3188~\AA\ integrated intensity maps. For the \mgii\ integrated intensity map, we integrated over both of the \mgii\ emission lines to boost the signal-to-noise ratio (\autoref{fig:sdss_int}). We also obtained a continuum image of each spaxel using the \textsc{imsubcont}  \textsc{casa} routine, which compares the spatial extent of the stellar continuum and the emission lines (red contours in \autoref{fig:sdss_int}). We calculated the error ($\epsilon$) of the integrated intensity maps by taking the standard deviation of off-source regions within the integrated intensity maps. The integrated \mgii\ map has $\epsilon = 1.5\times10^{-18}$~erg~s$^{-1}$~cm$^{-2}$ per spaxel and the continuum map has $1.4\times10^{-19}$erg~s$^{-1}$~cm$^{-2}$~\AA$^{-1}$ per spaxel (or $\sim23$~mag~arcsec$^{-2}$ in the U-band). 

Secondly, we fit the emission lines in each spaxel using single Gaussian profiles. As discussed in \autoref{theory}, the non-trivial observation that the \mgii\ emission lines are well-fit by single Gaussians indicates the lack of resonant absorption and scattering. We fit for the velocity offset ($v$) from zero-velocity as established by the SDSS redshift, total velocity width ($\sigma_\text{tot}$),  continuum flux (F$_{\rm cont}$), and total integrated flux ($F$) of each Gaussian profile as 
\begin{equation}
    G(\lambda) = \frac{F}{\sqrt{2\pi} \sigma_{\rm tot}}e^{-\frac{v^2}{2 \sigma_\text{tot}^2}} + F_{\rm cont}.
\end{equation}
The intrinsic velocity width, $\sigma_\text{int}$, of each emission line is determined from $\sigma_\text{tot}$ by subtracting the instrumental broadening of 37~\kms\ in quadrature (i.e., $\sigma_\text{tot}^2 = \sigma_{\rm int}^2 + \sigma_{\rm inst}^2$). We take the \mgii\ restframe wavelengths in air, $\lambda_0=$~2795.528 and 2802.704~\AA, from the NIST database \citep{nist} to establish the restframe wavelength of each \mgii\ emission line. We fit for $v$, $\sigma$, $F_{\rm cont}$, and $F$ using the Levenberg-Marquardt linear-least squares fitting routine \textsc{mpfit} \citep{mpfit}. The properties ($v$, $\sigma$, $F$) of the 2796~\AA\ and 2803~\AA\ \mgii\ emission lines were fit independently, but simultaneously, to allow for the properties to vary distinctly from transition-to-transition (see \autoref{theory}). We also fit the \hei~2945~\AA\ and \hei~3188~\AA\ profiles with single Gaussians. This spaxel-by-spaxel fitting enables the determination of the emission properties at each spatial location. Whenever we plot the \mgii\ properties, we only include spaxels that have a signal-to-noise ratio (S/N) greater than 2 for the \mgii\ emission line ratio ($F_{2796}/F_{2803}$).

\section{Spatially integrated \ion{Mg}{ii} emission}
\label{int}

\begin{figure*}
\includegraphics[width = \textwidth]{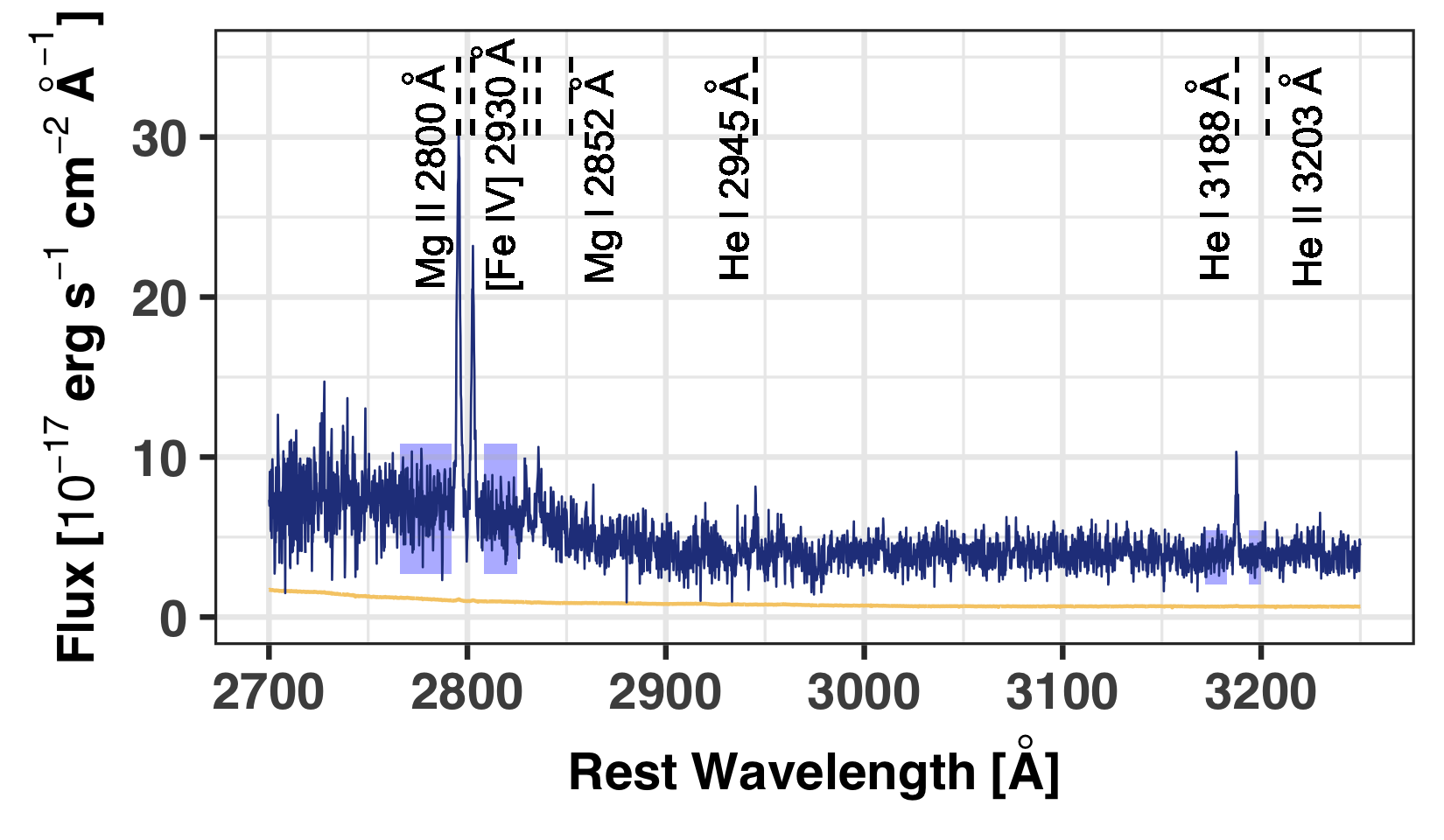}
\caption{The spatially-integrated restframe spectrum extracted from a 3.0\arcsec\ diameter region centered on the bright emission peak in \autoref{fig:sdss_int}. The  {gold} line shows the extracted 1$\sigma$ error on the flux. Emission lines are labelled above the spectrum. Strong \mgii~2796, 2803~\AA\ and \hei~3188~\AA, along with weak [\ion{Fe}{iv}]~2829, 2836~\AA\ and \hei~2945~\AA,  emission lines are detected. We do not detect \mgi\ absorption at any significance. The shaded blue regions are the regions used to subtract the continuum from the spectrum.}
\label{fig:sdss_full}
\end{figure*}

\autoref{fig:sdss_full} shows the integrated spectrum extracted from an aperture with a 3\arcsec diameter centered on the brightest emission peak from J1503. Even though integrating removes spatial information, summing the spaxels produces a high S/N spectrum of the galaxy (the \mgii~2796~\AA\ line is detected at the 21$\sigma$ significance) that is comparable to previous SDSS and HST observations. \autoref{fig:sdss_full} shows a strong \mgii~2800~\AA\ doublet along with moderately strong \hei~3188~\AA\ recombination emission. Among the weaker lines, the high S/N spectrum also shows a weak [\ion{Fe}{iv}]~2830~\AA\ doublet, \hei~2945~\AA, and \heii~3202~\AA\ lines. Notably absent from the spectrum is the \ion{Mg}{i}~2852~\AA\ absorption line. This contrasts with galaxies with strong \mgii\ \textit{absorption}, which often also have strong \ion{Mg}{i} absorption \citep{tremonti07, weiner, martin12, rubin13, finley}. \ion{Mg}{i} has an ionization potential of 7.6~eV; the non-detection indicates that there is negligible gas in lower ionization states than \mgii\ in J1503 (see \autoref{theory}).

The overall spatially-integrated \mgii\ emission flux is comparable to the SDSS flux. Using the same internal extinction correction as \citet{izotov16b} ($E(B-V)$ = 0.09), we find a total observed \mgii~2796~\AA\ flux of $8.5\pm0.4\times10^{-16}$~erg~s$^{-1}$~cm$^{-2}$ while \citet{izotov16b} measured $9\pm1\times10^{-16}$~erg~s$^{-1}$~cm$^{-2}$. To put the strength of these \mgii\ lines into perspective: the integrated extinction-corrected \mgii~2796+2803~\AA\ emission is 40, 53, and  {8}$\%$ of the SDSS [\ion{O}{ii}]~3727~\AA, H$\beta$, and [\ion{O}{iii}]~5007~\AA\ fluxes, respectively. 

\begin{figure}
\includegraphics[width = 0.5\textwidth]{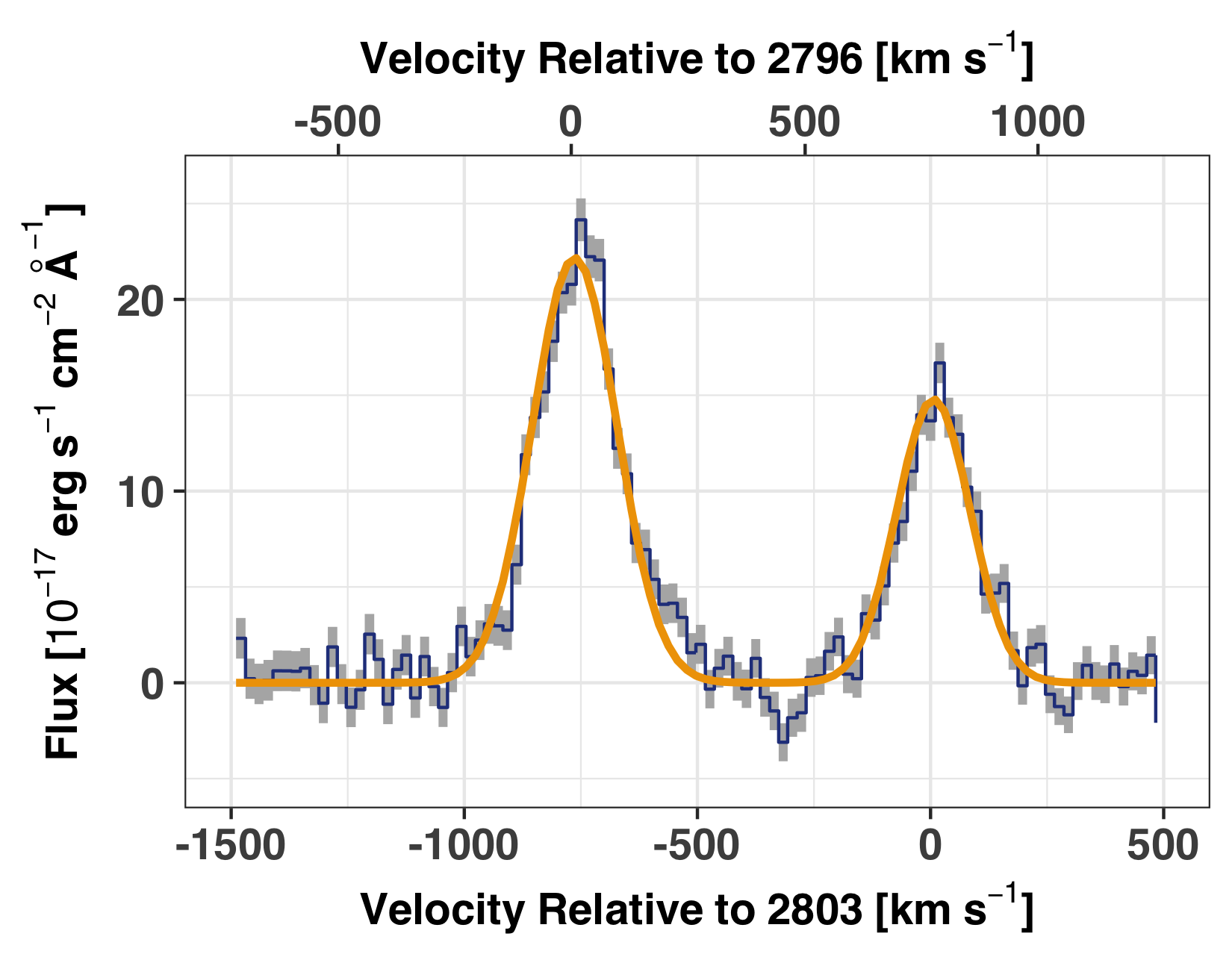}
\caption{Zoom-in on the spatially integrated, continuum-subtracted \mgii~2796, 2803~\AA\ doublet (blue) with a gray ribbon showing the 1$\sigma$ error on the flux. The lower x-axis shows the velocity relative to the \mgii~2803~\AA\ line (the right emission line) and the upper x-axis shows the velocity relative to the \mgii~2796~\AA\ line (the left emission line). The gold line shows a single Gaussian fit to each \mgii\ emission line.  {Both \mgii\ emission lines are within 2$\sigma$ of zero velocity, relative to the optical emission lines from the SDSS spectra (which has a 5~\kms\ redshift error)}, and well fit by a single Gaussian. }
\label{fig:mg2_fit}
\end{figure}

\autoref{fig:mg2_fit} shows the single Gaussian fit to the continuum subtracted \mgii\ doublet from the spatially-integrated profile. The single Gaussian fits match the observed \mgii\ profile. We do not observe any absorption signatures nor do we find strong line profile asymmetries typical of radiatively scattered resonant emission lines \citep[][]{verhamme, prochaska2011, verhamme15, scarlata, orlitova, koki}. This is especially apparent in the weaker \mgii~2803~\AA. The \mgii~2796~\AA\ profile has a slight emission excess at the highest positive velocities ($\sim+200$~\kmsp).  {However, the slight emission excess is only statistically discrepant from the fits at the 2$\sigma$ significance level for two pixels (20~\kms), less than the spectral resolution. On a whole, the fit residuals of 89.6\% (86 of 96) and 99.0\% (95 of 96) of the pixels in the \mgii\ region are less than 2$\sigma$  and 3$\sigma$ discrepant, respectively, in agreement with the expectation of a Gaussian distribution. Therefore, all observed deviations from the fit are fully consistent within the noise of the observations. }

The fitted Gaussian parameters, listed in \autoref{tab:lines}, further describe the relative shapes of the line profiles. The stronger \mgii~2796~\AA\ line is broader than the \mgii~2803~\AA\ line at 2.5$\sigma$ significance, but both are centered at zero velocity. These \mgii\ emission line characteristics are not necessarily true for all observed \mgii\ profiles. \citet{henry18} observed 10 Green Pea galaxies (9 currently  without available LyC observations), and all of those \mgii\ profiles were offset to the red from zero velocity. Some of those galaxies also have asymmetric profiles and most are not well-fit by a single Gaussian. Unlike the Gaussian \mgii\ profiles of J1503, the \mgii\ profiles from \citet{henry18} are heavily modified by resonant scattering.

\begin{table}
\caption{The fitted properties of the emission lines from the 3.0\arcsec\ in diameter spatially-integrated region  {from J1503+3644}. The second column is the velocity offset of the emission lines from the SDSS redshift ($v$), the third column is the spectral-resolution-corrected intrinsic velocity width ($\sigma_\text{int}$), and the last column is the total integrated flux of the emission line (flux-corrected to match the SDSS flux values). The fluxes have been corrected for Milky Way attenuation. No internal attenuation correction has been made.}
\begin{tabular}{lccc}
Line & $v$ &  $\sigma_\text{int}$ & Integrated flux \\
& [\kmsp] & [\kmsp] & [10$^{-16}$~erg~s$^{-1}$~cm$^{-2}$]\\
\hline
\ion{Mg}{ii}~2796\AA\ & $4\pm4$ & $91 \pm 2$ & $6.4\pm0.3$ \\
\ion{Mg}{ii}~2803\AA\ & $7\pm4$ & $78 \pm 3$ & $3.7\pm0.3$ \\
\ion{He}{i}~3188\AA\ & $-10\pm5$ & $57\pm5$ & $1.0\pm0.2$ \\
\hline
\ion{Mg}{ii} Doublet Flux Ratio & $1.7\pm0.1$ \\ 
\end{tabular}
\label{tab:lines}
\end{table}

Finally, in \autoref{theory}, we use the \mgii\ doublet flux ratio to constrain the optical depth of the \mgii\ gas. We define the doublet flux ratio as
\begin{equation}
    R = \frac{F_{2796}}{F_{2803}},
    \label{eq:rat}
\end{equation}
where we measure $R = 1.7\pm0.1$ in the spatially integrated spectrum.

\begin{figure}
\includegraphics[width = 0.5\textwidth]{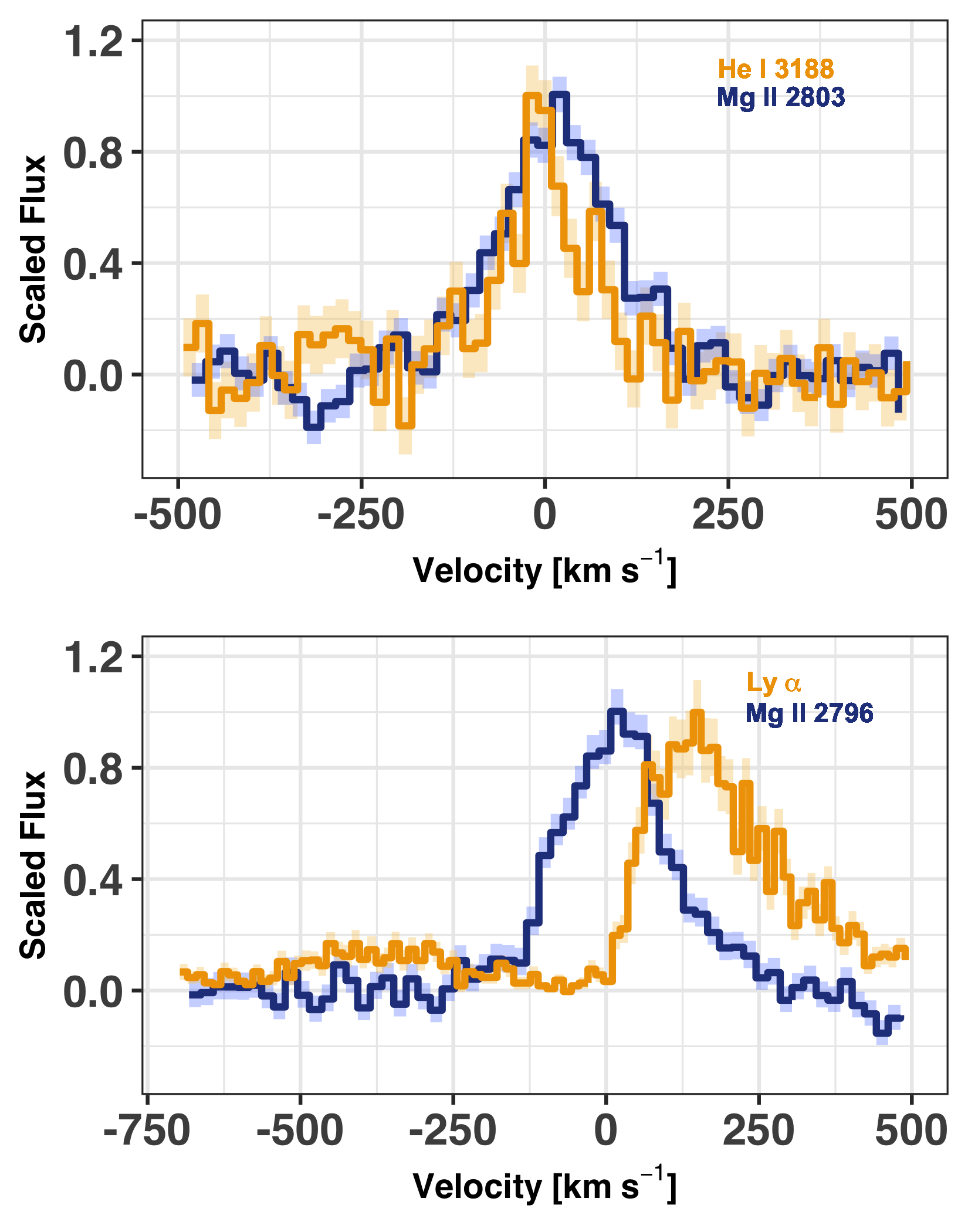}
\caption{Comparison of the \mgii\ emission lines (blue) to the nebular \hei~3188~\AA\ (upper panel) and \lya\ from HST/COS observations (lower panel). Each profile is normalized by the peak flux of the emission line. The 1$\sigma$ errors are shown by a lighter ribbon of the same color. The red portion of the \mgii\ emission is slightly broader than the \hei\ emission.  The \mgii\ does not resemble the \lya\ emission.}
\label{fig:em_comp}
\end{figure}

We compare the spatially-integrated \mgii~ {2803~\AA\ } to the nebular emission line \hei~3188~\AA\ in the top panel of \autoref{fig:em_comp}.  {We compare to the 2803~\AA\ transition because it has a lower $f$-value than the 2796~\AA\ transition, and is, thus, a better comparison to the \hei\ line.} With an intrinsic line width of $\sigma_\text{int} = 57\pm5$~\kmsp, the \hei\ is spectrally resolved, but narrower than the \mgii~2803~\AA\ line at 2.6$\sigma$ significance. The two line profiles are fairly consistent over the blue portion of their profiles, while they diverge slightly in the red portion. Conversely, the moderate-resolution HST/COS G160M \lya\ profile in the bottom panel of \autoref{fig:em_comp} has a very different profile shape from  {the \mgii~2796~\AA\ emission line}. Through the metallicity, there is a factor of $\sim$10$^5$ larger neutral hydrogen column density than Mg$^+$ column density, such that Ly$\alpha$ is more strongly impacted by resonant scattering \citep{neufeld, dijkstra, verhamme, gronke, koki, Michel-Dansac}. Radiative transfer effects are seen in the shape of the \lya\ profile from J1503: the main emission peak is redshifted by $+140$~\kms\ from line center, while a weak blue peak is $-290$~\kms\ from line center \citep{verhamme17}. Meanwhile, the \mgii\ emission does not show these radiative transfer effects: it is centered at zero-velocity and  {well-fit by a single Gaussian profile}.    

 In summary, the spatially-integrated \mgii\ emission-line profiles are well-fit by single Gaussians that are centered at zero velocity (\autoref{fig:mg2_fit}). The fairly symmetric \mgii\ line profiles resemble slightly broader versions of the nebular \hei\ profiles and are markedly different from the double-peaked \lya\ profiles (\autoref{fig:em_comp}). Importantly, the flux ratio of the \mgii\ doublet lines is $R = 1.7\pm0.1$. This ratio will be discussed fully in \autoref{theory}. 

\section{Spatially resolved Mg~II emission}
\label{resolved} 
Here we explore the \mgii\ properties on a spaxel-by-spaxel basis. The main goal of this analysis is to determine the spatial extent of the \mgii\ emission (\autoref{spatial}) and the variation of the \mgii\ properties (\autoref{spat_prop}), with special attention paid to $R$, the doublet flux ratio (\autoref{spat_r}). $R$ encodes information on the \mgii\ optical depth (\autoref{theory}) and the fraction of \mgii\ (and LyC) emission that escapes J1503 (see \autoref{esc_1503}). 

\subsection{The Spatial Extent of Mg~\textsc{ii} Emission}
\label{spatial}
\begin{figure}
\includegraphics[width = 0.5\textwidth]{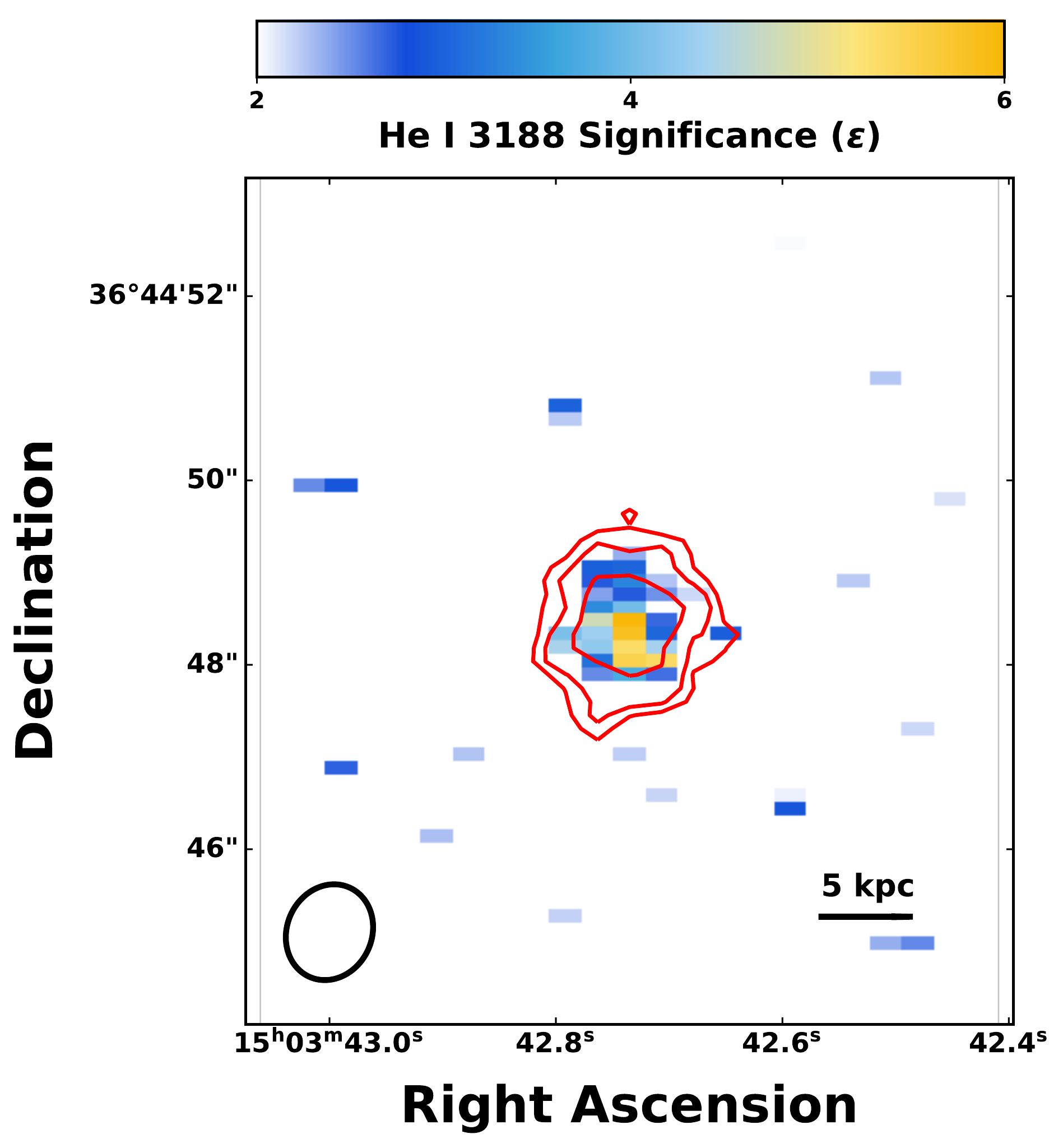}
\caption{Spatial map of the integrated continuum-subtracted \ion{He}{i}~3188\AA\ emission, in units of the \hei\ significance: $\epsilon = 5.8\times10^{-19}$~erg~s$^{-1}$~cm$^{-2}$ per spaxel. The measured 1.04\arcsec~$\times$~0.92\arcsec\ seeing is given by the black circle in the lower left, and a 1.0\arcsec\ scale bar is given in the lower right (with the projected physical size of 1.0\arcsec\ at $z = 0.3557$). The red contours are the 3, 5, and 10$\epsilon$ integrated \mgii\ emission contours.  {These contours are not the same as in \autoref{fig:sdss_int}}. The \hei\ and \mgii\ emission are both centered in the same location.}
\label{fig:neb_int}
\end{figure}

\mgii\ emission, in \autoref{fig:sdss_int}, is extended beyond the 1.04\arcsec$\times$0.92\arcsec\ seeing disk (compare the contours to the black circle in the bottom). We fit a 2-dimensional profile to the continuum subtracted \mgii\ profile and measure a FHWM of $1.32\pm0.04$\arcsec$\times1.21\pm0.04$\arcsec\ at 175$^\circ$. In comparison, the nebular \hei~3188~\AA\ in \autoref{fig:neb_int} is not spatially extended, although it does peak at the same location as the \mgii\ emission. Note that the nebular \hei\ emission is faint and the peak of the emission is only detected at the 5$\sigma$ significance level, substantially less than the 21~$\epsilon$ of the \mgii~2796~\AA\ emission. Additionally, we do not detect any statistically significant \mgi\ -- in absorption or emission -- in any of the integrated spectra or individual spaxels. 

\begin{figure}
\includegraphics[width = 0.5\textwidth]{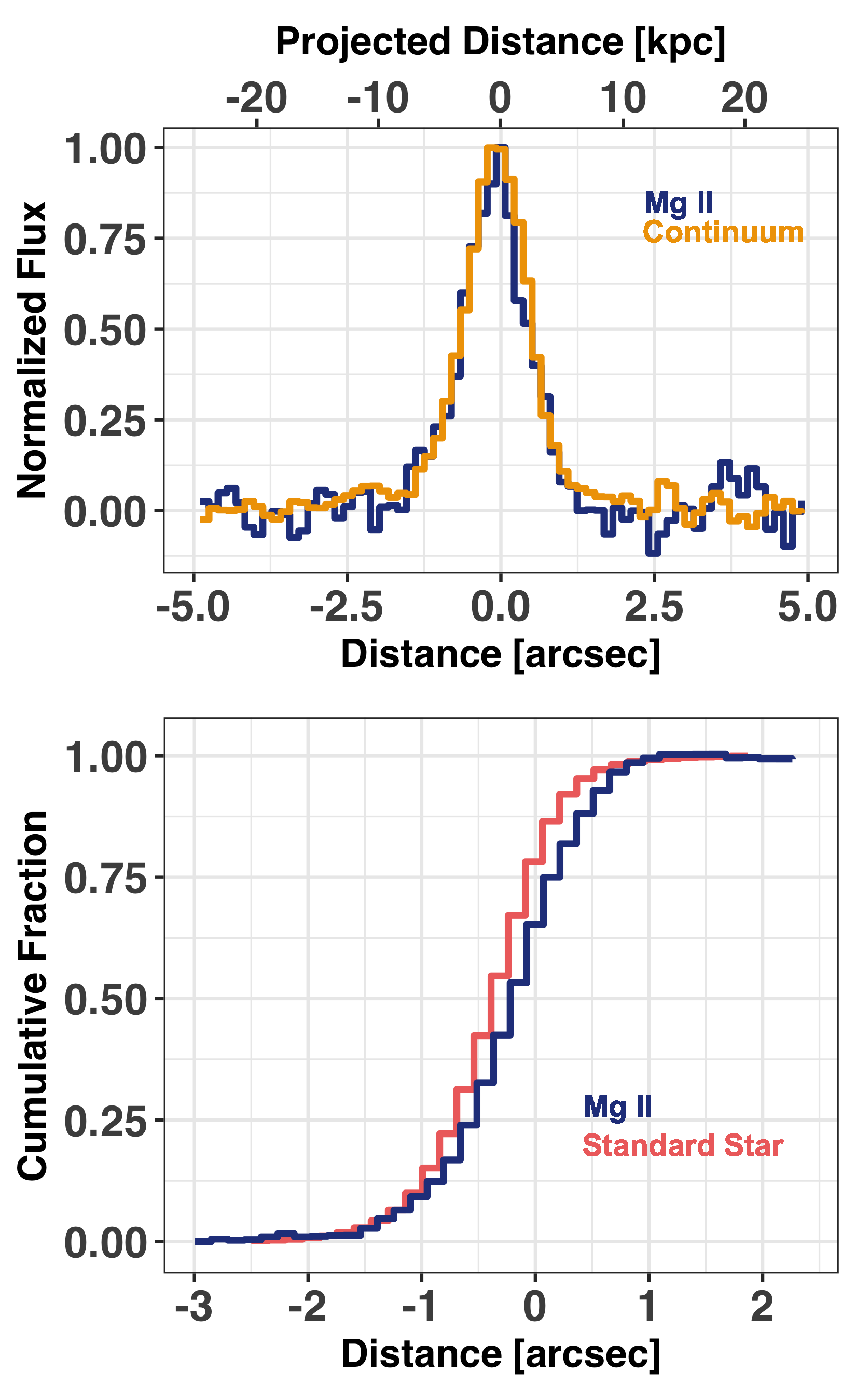}
\caption{\textit{Upper Panel:} The surface brightness profiles of the continuum-subtracted \mgii~2796+2803~\AA\ emission (from \autoref{fig:sdss_int}; blue line) and the stellar continuum (gold line).  {Both surface brightness profiles are extracted from the same vertical line through the brightest \mgii\ emission spaxel}. The \mgii\ and stellar continuum have similar light profiles. The upper x-axis shows the projected physical size at $z= 0.3557$. \textit{Lower Panel:} The cumulative distribution of the continuum-subtracted \mgii\ emission. The pink line compares the \mgii\ cumulative distribution to the standard star cumulative distribution. While the \mgii\ is slightly spatially extended, the 3.0\arcsec diameter SDSS fibers captures all of the \mgii\ emission.}
\label{fig:spatial_comp}
\end{figure}

\begin{table}
\caption{Fitted emission line properties of four spatially distinct regions within the KCWI observations. Each spectrum was extracted from a region with size equal to the measured seeing, separated from the brightest emission peak, and distinct from the other  {three} regions ( {see the inset in \autoref{fig:sdss_int})}. The first column gives the region location relative to the main \mgii\ emission peak, the second and third columns give the fitted intrinsic velocity width, corrected for the spectral resolution ($\sigma_\text{int}$), of the \mgii~2803~\AA\ and \mgii~2796~\AA\ lines, respectively. The forth column is the doublet ratio $R = F_{2796}/F_{2803}$.  { The fifth column gives the inferred column density of neutral hydrogen and column six gives the resultant relative escape fraction. The relative escape fraction is calculated using the \mgii\ flux ratio and the ISM metallicity, but ignores the effects of dust attenuation (see \autoref{eq:fesc_hi}). }  }
\begin{tabular}{lccccc}
Region & $\sigma_{2803}$ & $\sigma_{2796}$ & $R$ & \nhi\ & $f_{\rm esc}^{\rm rel}$ \\
 & [\kmsp] & [\kmsp] &  & [10$^{16}$~cm$^{-2}$] & [\%]\\
\hline
Left& $79\pm3$ & $108 \pm 2$ & $1.5\pm0.2$ & $6\pm3$ & $69\pm13$ \\
Right & $80 \pm 3$ & $81\pm2$ & $1.7\pm0.2$ & $3\pm2$ & $83\pm 10$\\
Upper & $85 \pm 4$ & $108\pm3$ & $1.8\pm0.2$ & $2\pm 2$ & $88\pm10$\\
Lower & $65 \pm 3$ & $96 \pm 3$ & $2.1\pm0.3$ & $<2$ & $>88$\\
\end{tabular}
\label{tab:spatially_distinct}
\end{table}

\begin{figure*}
\includegraphics[width = \textwidth]{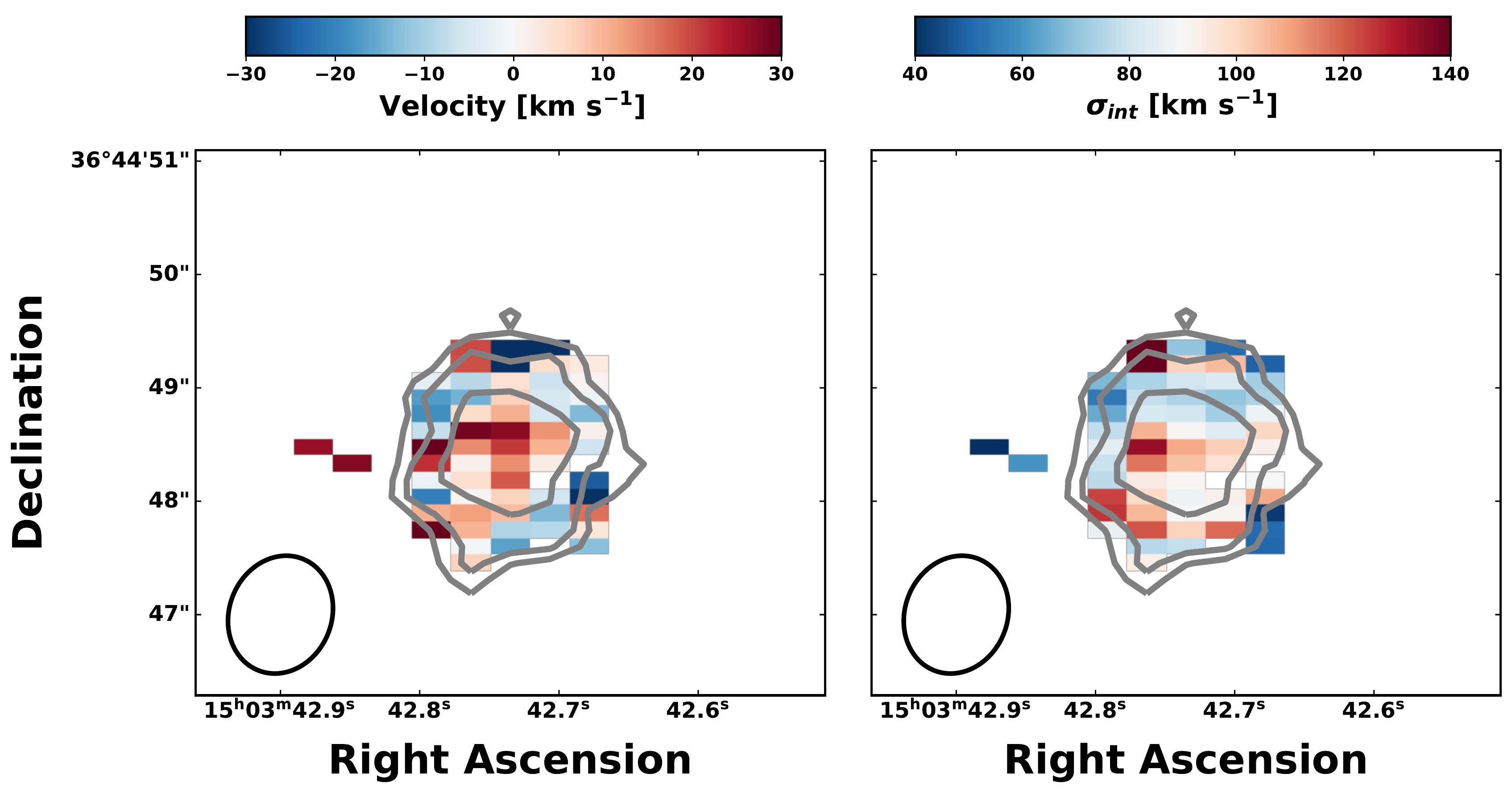}
\caption{\textit{Left Panel:} The spatial distribution of the fitted \mgii~2796~\AA\ velocity centroid. There is no obvious coherent velocity distribution and the mean velocity of the individual spaxels is $3\pm15$~\kms.  \textit{Right Panel:} The spatial distribution of the estimated intrinsic velocity width ($\sigma_\text{int}$) of the  {\mgii~2796~\AA\ emission line}.  The gray contours are the 2, 5, and 10$\epsilon$ significance levels of the integrated \mgii\ emission. The 1.04\arcsec$\times$~0.92\arcsec\ seeing disk is included in the lower left of each panel.}
\label{fig:velocity}
\end{figure*}

The red contours in \autoref{fig:sdss_int} demonstrate that the  stellar continuum closely follows the shape of the \mgii\ emission. To better quantify the spatial relation between the \mgii\ and the stellar continuum, \autoref{fig:spatial_comp} shows their surface brightness profiles. The \mgii\ emission peaks at the same location as the stellar continuum, both have similar FWHMs, and both are roughly symmetric. Both the \mgii\ and stellar continuum are concentrated within a projected physical distance of $\pm5$~kpc. The cumulative \mgii\ flux distribution, shown in the bottom panel of \autoref{fig:spatial_comp}, illustrates that the \mgii\ is highly centrally concentrated, but is slightly extended beyond the 1.04\arcsec$\times$~0.92\arcsec\ seeing disk of the standard star (pink line).

\subsection{Spatially-resolved Emission Properties}
\label{spat_prop}

Using the spaxel-by-spaxel emission line fits, we can explore the spatial variation of the \mgii\ emission. Large-scale galactic outflows from compact galaxies have been observed with \mgii\ emission out to 100~kpc \citep{rubin13, rupke19, burchett20}.  Similarly, if the gas within J1503 rotates then we would expect to observe coherent spatial velocity gradients \citep{Micheva19}.  {We do not find strong spatial variations in the \mgii~2796~\AA\ velocity (left panel of \autoref{fig:velocity}, the 2803~\AA\ velocity does not show spatial trends either)}. The \mgii\ emission does not show signatures of coherent rotation nor ordered structure. The \mgii\ velocities range between $-31$ and $+46$ with a median of $3\pm15$~\kmsp. Similarly, the right panel of \autoref{fig:velocity} shows  {the \mgii~2796~\AA\ intrinsic velocity width ($\sigma_{2796}$), measured to have a mean value of $88 \pm 24$~\kmsp.} The median $v$ and $\sigma_{\rm int}$ values from all of the spaxels are similar to the values determined by fitting the spatially-integrated profile (\autoref{tab:lines}). This indicates that the spatially-integrated value is representative of the median spaxel-by-spaxel values. 

\begin{figure*}
\includegraphics[width = \textwidth]{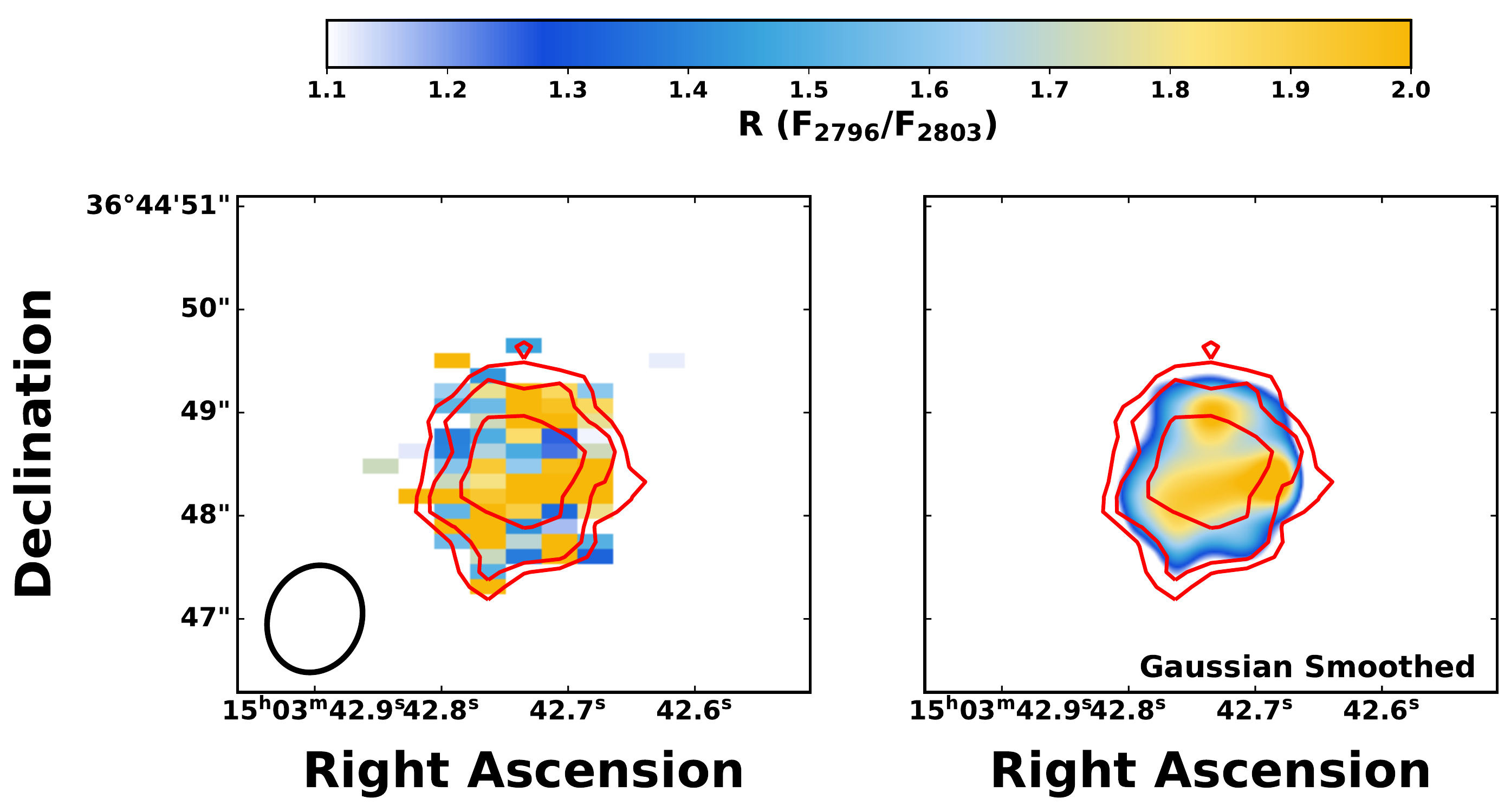}
\caption{\textit{Left Panel: }Spatial map of the \mgii\ doublet flux ratio, $R = F_{2796}/F_{2803}$. There is  spatial variation: the upper left regions have lower $R$ values on average than the lower right regions. The contours show the 2, 5, and 10$\epsilon$ flux levels of the integrated \mgii\ emission. The 1.04\arcsec$\times$~0.92\arcsec\ seeing disk is included as a circle in the lower left. \textit{Right Panel:}  {The same $R$ spatial map as in the left, but convolved with a Gaussian the size of the seeing disk. Note that edge effects artificially cause the outer regions to appear to have lower $R$ values.} }
\label{fig:doub}
\end{figure*}

The final parameter that we study spatially is the \mgii\ doublet flux ratio, $R  = F_{2796}/F_{2803}$. As described below, $R$ traces the \mgii\ optical depth and the spatially-resolved emission maps the Mg$^+$ column density distribution.  The left panel of \autoref{fig:doub} shows that $R$ varies from 0.8 to 2.7 with a median of $1.7\pm0.4$. The median value of the spatial distribution agrees with the value measured from the integrated profile.  {To test the seeing impact and spatially correlated noise on the $R$ distribution, the right panel of \autoref{fig:doub} shows the $R$ distribution convolved with a 2-dimensional Gaussian that has the same parameters as the fitted seeing. While edge effects of the convolution make the outer regions appear at lower $R$ than physical, there are distinct $R$ variations in the inner regions of J1503. Even smoothed to the seeing resolution, there is significant structure in the observed $R$ values.}

\begin{figure}
\includegraphics[width = 0.5\textwidth]{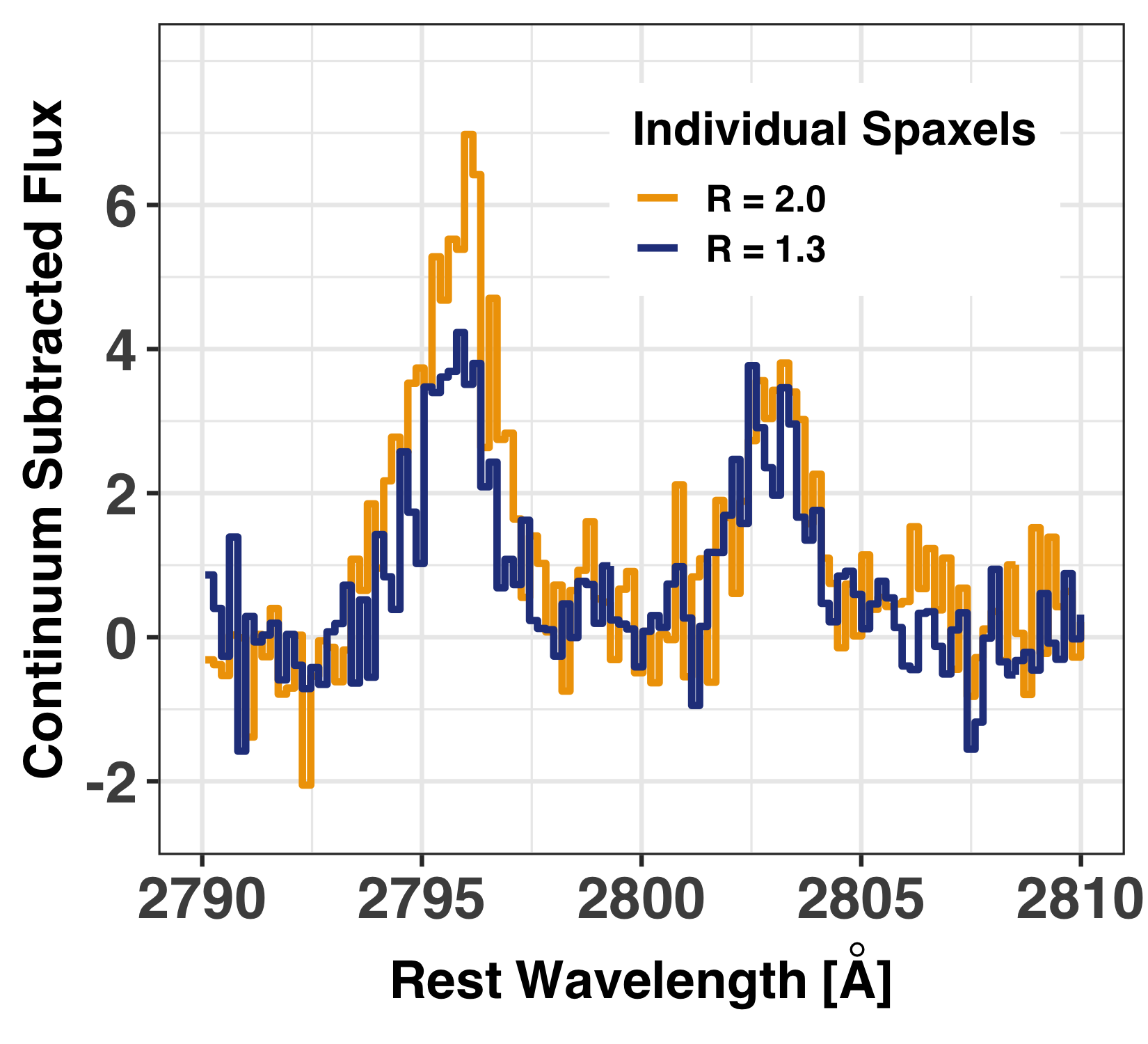}
\caption{Two individual spaxel {s} separated by more than the seeing. Each spaxel shows a different flux ratio of the 2796 and 2803~\AA\ emission feature ($R = F_{2796}/F_{2803}$). These two individual spatial locations within the galaxy have different $R$ values: $2.0\pm0.3$ ( {gold}) and $1.3\pm0.3$ ( {blue}).}
\label{fig:ind}
\end{figure}

\begin{figure}
\includegraphics[width = 0.5\textwidth]{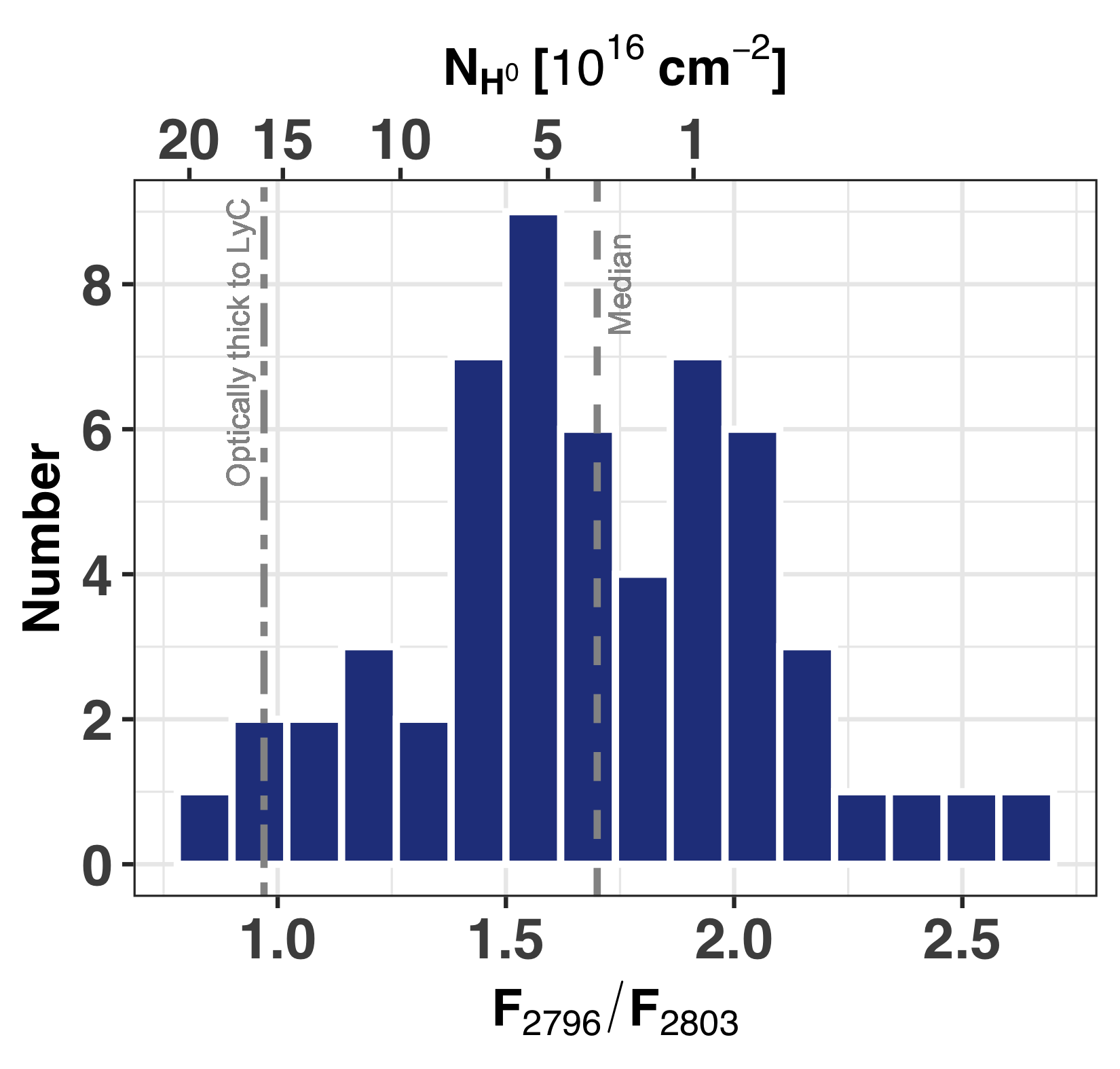}
\caption{Histogram of the \mgii\ flux ratio, $R=F_{2796}/F_{2803}$, for all of the S/N~$> 2$ spaxels within J1503.  {High} $R$ regions imply lower \hi\ column densities, as marked by the upper x-axis, assuming the gas-phase metallicity of J1503  {and a constant dust depletion} (see \autoref{eq:nh1_r}). The dashed line marks the median of the $R$ distribution (1.7) and the dot-dashed line is the column density where \hi\ becomes optically thick to ionizing photons. Note that the typical per spaxel uncertainty on $R$ is 0.3 (two bins), and the distribution could be uniformly distributed. Spaxels with $R > 2$ have large errors such that they are statistically consistent with 2.}
\label{fig:doub_hist}
\end{figure}

\autoref{fig:ind} shows the \mgii\ profiles for two spatially-distinct spaxels within J1503. The \mgii\ emission line profiles change from spaxel-to-spaxel, and equivalently from location-to-location, within the galaxy. The \mgii~2803~\AA\ line has similar strength in both spaxels, but the \mgii~2796~\AA\ is stronger for the blue profile. This leads to $R = 2$ for the blue profile and $R = 1.3$ for the gold profile. \autoref{fig:doub_hist} compresses the individual spaxels from the spatial $R$ map of \autoref{fig:doub} into a histogram. The median value, the dashed gray line, splits the distribution in half, but the distribution does not peak there. Rather, the $R$ distribution is broad with a possible slight double-peaked distribution: one near $R = 2$ and one near $R = 1.5$. However, the median measured error on individual pixels is 0.3, precluding a definitive confirmation of this morphology. Thus, this distribution could be equally-well explained as a uniform distribution.

To further explore the impact of spatial resolution and seeing on the spatial variation of $R$, we extracted four spectra from regions with an aperture size equal to the seeing disk and spatially distinct from each other. These regions are above, below, to the left, and to the right of the bright \mgii\ peak in \autoref{fig:sdss_int}, but do not include the peak. Fitting these four spatially distinct regions finds that the lower and the left quadrants have the largest $R$ differences of $2.1\pm0.3$ and $1.5\pm0.2$, respectively (see \autoref{tab:spatially_distinct}). This is seen in both panels of \autoref{fig:doub}: the lower portion is largely gold, while the spaxels in the left quadrant  are largely  blue.

\subsection{Averaging Based on the Doublet Ratio}
\label{spat_r}
In \autoref{theory} and \autoref{highz}, we emphasize the utility of $R$ to determine the Mg$^+$ and \hi\ optical depths and column densities. To explore the relation between the \mgii\ emission lines and optical depth at high S/N, we determine the average spectra of high and low $R$ regions by averaging all individual spaxels with a measured $R > 1.7$ and all spaxels with $R < 1.7$. We call these averaged spectra the High and Low $R$ spectra, respectively. 
\begin{figure*}
    \centering
    \includegraphics[width = \textwidth]{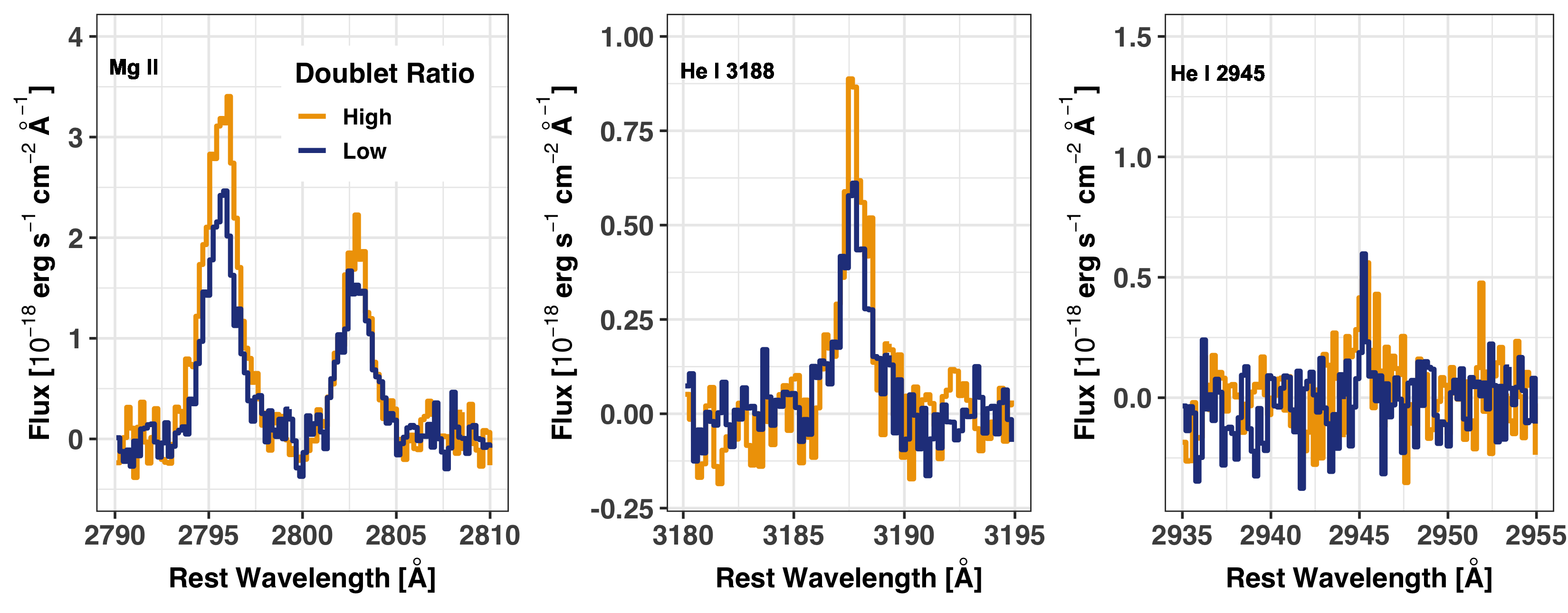}
    \caption{Mean continuum subtracted emission profiles of all spaxels split based upon whether the \mgii\ 2796 to 2803\AA\ flux ratio ($R = F_{2796}/F_{2803}$) is greater than 1.7 (High; blue line) or less than 1.7 (Low; gold line). The three panels are different nebular emission lines: \mgii, \hei~3188~\AA, and \hei~2945\AA\ (from left to right). High $R$ spaxels have a stronger \mgii~2796~\AA\ integrated flux, but nearly the same \mgii~2803~\AA\ flux. The High $R$ composite also has  {marginally} stronger \hei~3188~\AA\ emission. }
    \label{fig:rat_comp}
\end{figure*}

The left panel of \autoref{fig:rat_comp} shows the mean \mgii\ emission line profiles of the High (blue) and Low (gold) $R$ regions. The \mgii\ emission properties, listed in \autoref{tab:rat_comp}, show statistical variation between the two composites. By construction, the High $R$ composite has $R$ values that are consistent with 2, while the Low $R$ composite is 1.43. In the Low $R$ composite, the 2803 and 2796~\AA\ line widths are statistically similar, while the 2796~\AA\ line is significantly broader than the 2803~\AA\ line in regions with High $R$. We do not observe statistically significant differences in the \mgii\ velocity centroids. 

The middle and right panels of \autoref{fig:rat_comp} show two \hei\ emission lines, 3188 and 2945~\AA, respectively. \hei~3188~\AA\ is strongly detected in both composites, with the Low $R$ having weaker average \hei\ than the High $R$ composite. The \hei~2945~\AA\ line is weak in both composites, but moderately stronger in the High $R$ composite. 


\begin{table}
\caption{Fitted line profile properties of two composites of individual spaxels below (Low $R$; Column 2) and above (High $R$; Column 3) the median $R$ value of 1.7. Parameters above the second horizontal line are derived from the \mgii\ profiles, while parameters below it are derived using a combination of the \mgii, the \hei~3188~\AA, and the \hei~2945~\AA\ emission lines (the \hei\ emission flux ratio, the gas-phase attenuation, the inferred relative escape fraction, and the inferred absolute escape fraction).}
\begin{center}
\begin{tabular}{lccc}
Property & Low $R$ & High $R$ \\
\hline
v$_{2796}$~[\kmsp] & $6\pm4$ & $6 \pm 4$ \\
v$_{2803}$~[\kmsp] & $9\pm4$ & $7\pm 4$\\
$\sigma_{2796}$~[\kmsp] & $90\pm3$ & $97 \pm 4$\\
$\sigma_{2803}$~[\kmsp] & $95\pm3$ & $82 \pm 4$ \\
F$_{2796}$/F$_{2803}$ & $1.43 \pm 0.07$ & $1.96 \pm 0.08$  \\
\hline
F$_{3188}$/F$_{2945}$ & $4.26 \pm 1.07$ & $2.17 \pm 0.30$   \\
E$($B-V$)$~[mag] &  3 & 0.13  \\ 
$f_{\rm esc}^{\rm rel}$(LyC) & $63\pm4$ & $97 \pm 6$ \\ 
\fescc &  0  & 19\\
\end{tabular}
\label{tab:rat_comp}
\end{center}
\end{table}

\section{Neutral gas within and surrounding a galaxy that emits ionizing photons}
\label{hi_dist}
The \mgii\ emission from J1503 is not spatially extended beyond the stellar continuum and is centrally concentrated with a FWHM of 1.32\arcsec~$\times$~1.21\arcsec\ (6--6.6~kpc in physical units; \autoref{fig:sdss_int}), considerably smaller than the KCWI field of view of 8.4\arcsec~$\times$~20.4\arcsec.   {As shown in \autoref{theory}, \mgii\ emission is closely connected to the neutral gas properties in the galaxy. The combination of the small spatial distribution and the relationship between \mgii\ and neutral gas implies that the bulk of the dense metal-bearing neutral gas is not extended beyond the 13~kpc of the stellar continuum (\autoref{fig:spatial_comp})}. Since ionizing photons are observed to escape from this galaxy, it is perhaps not surprising that the circum-galactic medium of J1503 is devoid of neutral hydrogen: there must be a paucity of gas along the line-of-sight for ionizing photons to escape. 

However, a diminished neutral hydrogen circum-galactic medium is in stark contrast to galaxies observed at moderate \citep{steidel11, ostlin14, hayes14, cantalupo, erb18, cai} and high-redshift \citep{ouchi, sobral} with \lya\ emission. At $z \sim 3$, studies find \lya\ halos around Lyman Break Galaxies that are upwards of 15 times larger than the stellar continuum \citep{steidel11, wisotzki}. Some of these galaxies have since been confirmed to emit ionizing photons \citep{steidel18}. However, these galaxies are either more massive or more spatially extended than J1503. 

Without ordered rotational motion (\autoref{fig:velocity}), there is not a stable disk within J1503 and the gaseous dynamics are completely dispersion dominated. Given the low stellar mass and compact size of the galaxy (log(M$_\ast$/M$_\odot) = 8.22$), it is perhaps not surprising that the rotational velocities are small (or not observed), although the 5~kpc spatial resolution likely inhibits detection of rotational motion. However, the observed $\sigma_\text{int} \sim 50-80$~\kms is much larger than typically found in local galaxies with similar stellar mass \citep[typically $\sim10$~\kmsp;][]{walter08, hunter, ott, pardy}. For instance, Blue Compact Dwarfs have resolvable rotation curves with $v_{\rm rot}$ up to 92~\kmsp, but with velocity dispersions of only 5--10~\kms \citep{vanzee98, vanzee01}. Extreme $\sigma_{\rm int}/v$ values have been observed in some local extremely compact star-forming galaxies \citep{Micheva19} and are reminiscent of the dispersion dominated systems found at $z \sim 1-6$ \citep{genzel11, swinbank12,leethocawalit16, swinbank17, smit18, girard18, girard}. 

The kinematic properties suggest that J1503 did not form as a rotationally supported disk. Rather, intense gravitational instabilities, possibly generated by a large accretion event, could have formed this local LyC emitter \citep{genzel11, swinbank12, girard18}. {This rapid assembly of J1503 is strengthened by the very young observed stellar age from both the FUV stellar continuum and the large observed H$\beta$ equivalent width \citep{izotov16b}.}  Future observations of the stellar kinematics, which cannot be probed by the current observations, may help to explain the formation of sources of ionizing photons at high-redshift.

\section{Expected Mg~\textsc{ii} Emission Properties}
\label{theory}

\begin{table}
\caption{The inferred escape fractions ($f_{\rm esc}$). The upper section gives the \fescc\ measured directly from the LyC observations \citep{izotov16b}. The middle section gives the \mgii~2803~\AA\ escape fraction calculated using the photoionization models of \citet{henry18}. We also give the \fescc\ inferred by applying the FUV dust attenuation to \fescl. The bottom section gives the values inferred from the spatially integrated \mgii\ spectrum, including: the \mgii~2803~\AA\ optical depth ($\tau_{2803}$), the Mg$^+$ column density ($N_{\rm Mg^{+}}$), the relative LyC escape fraction (not accounting for dust), and the absolute LyC escape fractions.  }
\begin{center}
\begin{tabular}{lccc}
Escape Fraction & Value \\
\hline
\textbf{Measured Directly} \\
\fescc & $5.8\pm0.6$ \\
\hline
\textbf{Inferred from photoionization models} \\
\fescl & $95\pm14$ \\
\fescc & $8\pm1$ \\
\hline 
\textbf{Inferred from the \mgii\ doublet ratio} \\
$\tau_{2803}$ & $0.16\pm0.07$ \\
log($N_{\rm Mg^+}$ [cm$^{-2}$]) & $10.9\pm0.2$ \\
\fescl &  $85\pm7$ \\
$f_{\rm esc}^{\rm rel}\left({\rm LyC}\right)$ & $80 \pm 7$ \\ 
$f_{\rm esc} \left({\rm LyC}\right)$ & $5.9\pm0.4$ \\
\end{tabular}
\label{tab:esc}
\end{center}
\end{table}

\subsection{Producing \ion{Mg}{ii} Emission}
In \autoref{spatial}, we found that the total \mgii\ emission from J1503 has a similar spatial distribution as the stellar continuum (\autoref{fig:sdss_int} and \autoref{fig:spatial_comp}) and the nebular \hei\ emission (\autoref{fig:em_comp}). We do not observe \mgi\ or \mgii\ absorption in any spaxel, nor in any composites. Each of the \mgii\ emission lines are well-fit by single Gaussians (\autoref{fig:mg2_fit}). All of these observations suggest that the resonant \mgii\ doublet traces neutral gas that is not strongly impacted by resonant absorption. In other words, the Mg$^+$ gas is optically thin. Meanwhile, the \mgii\ doublet emission ratio, $R = F_{2796}/F_{2803}$, has a broad range between 0.8--2.7 (\autoref{fig:doub_hist}). In this section we explore the physical origin and implication of  $R$. 

\ion{Mg}{ii}~2796 and 2803~\AA\ are strong resonance lines produced when Mg$^+$ electrons transition between the $^2$P$^{3/2, 1/2}$ upper levels (hereafter refereed to as levels 2), respectively, and a common $^{2}$S$^{1/2}$ ground state (level 1). Each transition can be treated as a two level system, where electrons in the ground state (level 1) can be collisionally excited by free electrons or radiatively excited by photons into either of the excited states (levels 2) with 4.4~eV of energy. These resonant transitions have large Einstein A coefficients (A$_{21} \sim 2.6\times10^8$~s$^{-1}$), such that downward transitions exclusively occur through spontaneous decay unless the electron densities are greater than 10$^8$~cm$^{-3}$. Thus, the rate that Mg$^+$ electrons are excited from level 1 in level 2 is given as 
\begin{equation}
\begin{split}
    \frac{dn_{2}}{dt} &= \text{Collisions up} + \text{Absorption Up} - \text{Spontaneous Decay} \\
    &=C_{\rm coll} n_{\rm e} n_1 + J B_{12} n_{1} - A_{21} n_{2} , 
    \end{split}
\end{equation}
{where $C_{\rm coll}$ is the collisional rate coefficient, $n_1$ and $n_2$ are the density of the Mg$^+$ electrons in the ground state and excited states, $n_{\rm e}$ is the electron density, $J$ is the mean radiation field, and B$_{12}$ is the absorption coefficient.} 

The observed resonant \mgii\ emission lines do not show prominent absorption signatures. This leads to the hypothesis that collisions dominate the Mg$^+$ excitation within J1503 ($C_{\rm coll}n_{\rm e} > J B_{12}$; see \autoref{cloudy}). When collisions dominate the excitation of the Mg$^+$ gas (or in the optically-thin limit), the intrinsic flux ratio of the two \mgii\ emission lines is the ratio of their emissivities, $j$ (see \autoref{plane-parallel}). The intrinsic emissivity ratio of the 2796 to 2803~\AA\ emission lines is
\begin{equation}
    R_{\rm int} = \frac{{F}_{\rm 2796, int}}{{F}_{\rm 2803, int}} = \frac{j_{2796}}{j_{2803}}  = \frac{C_{2796}}{C_{2803}} =  \frac{g_{2796}}{g_{2803}} = 2 ,
    \label{eq:frat}
\end{equation}
{where $C$ and $g$ are the collisional rate coefficient and quantum degeneracy factors ($g = 2J+1$) of each upper level
\citep[see \autoref{cloudy};][]{mendoza81, sigut}. When the Mg$^+$ excitation is dominated by collisional excitation the intrinsic $R$ will be constant and equal to 2. This $R = 2$ is confirmed by \textsc{cloudy} photoionization modeling \citep[see \autoref{cloudy};][]{henry18}. If the excitation is dominated instead by photon absorption, the emission flux ratio would be set by the ratio of the Einstein A values instead of the $g$ values. Since both \mgii\ lines have similar A$_{21}$ values, this would lead to $R$ values near 1 instead of 2 \citep[see the flux values in table 4 from the radiative transfer model in][]{prochaska2011}.}

However, we observe $R$ values that vary between 0.8--2.7. The values above 2 are consistent with 2 at the 1$\sigma$ significance level, but 40\% of the spaxels are statistically less than two and greater than 1. This departure from a constant $R = 2$ observed in \autoref{fig:doub} and \autoref{fig:doub_hist} suggests that collisional excitation creates intrinsic emission that is then incident on a small foreground screen of \mgii\ gas that imprints optical depth variations onto the $R$ ratio. Intervening Mg$^{+}$ gas absorbs and scatters the intrinsic light, removing a portion of the emission, and decreasing the transmitted light. In turn, these optical depth effects reduce $R$ from the intrinsic value of 2 to the observed $R$ values.

{Broadly speaking, there are two optical depth geometrical effects that could impact the \mgii\ emission: a uniform column density of Mg$^+$ gas (\autoref{thick}) or a porous distribution of optically thick Mg$^+$ (\autoref{partial}). Reality likely resides in a combination of the two scenarios, where lower column density Mg$^{+}$ channels exist between relatively higher column density Mg$^+$ regions (\autoref{combine}). The low column density gas regions likely transmit the \mgii\ and LyC emission, while the high column density regions absorb and scatter the \mgii. In the next three sub-sections we explore each scenario, the implied physical properties, and their impact on the observed \ion{Mg}{ii} emission lines.}

\subsection{\ion{Mg}{ii} escape through optically thin gas}
\label{thick}

\begin{figure}
    \centering
    \includegraphics[width = 0.5\textwidth]{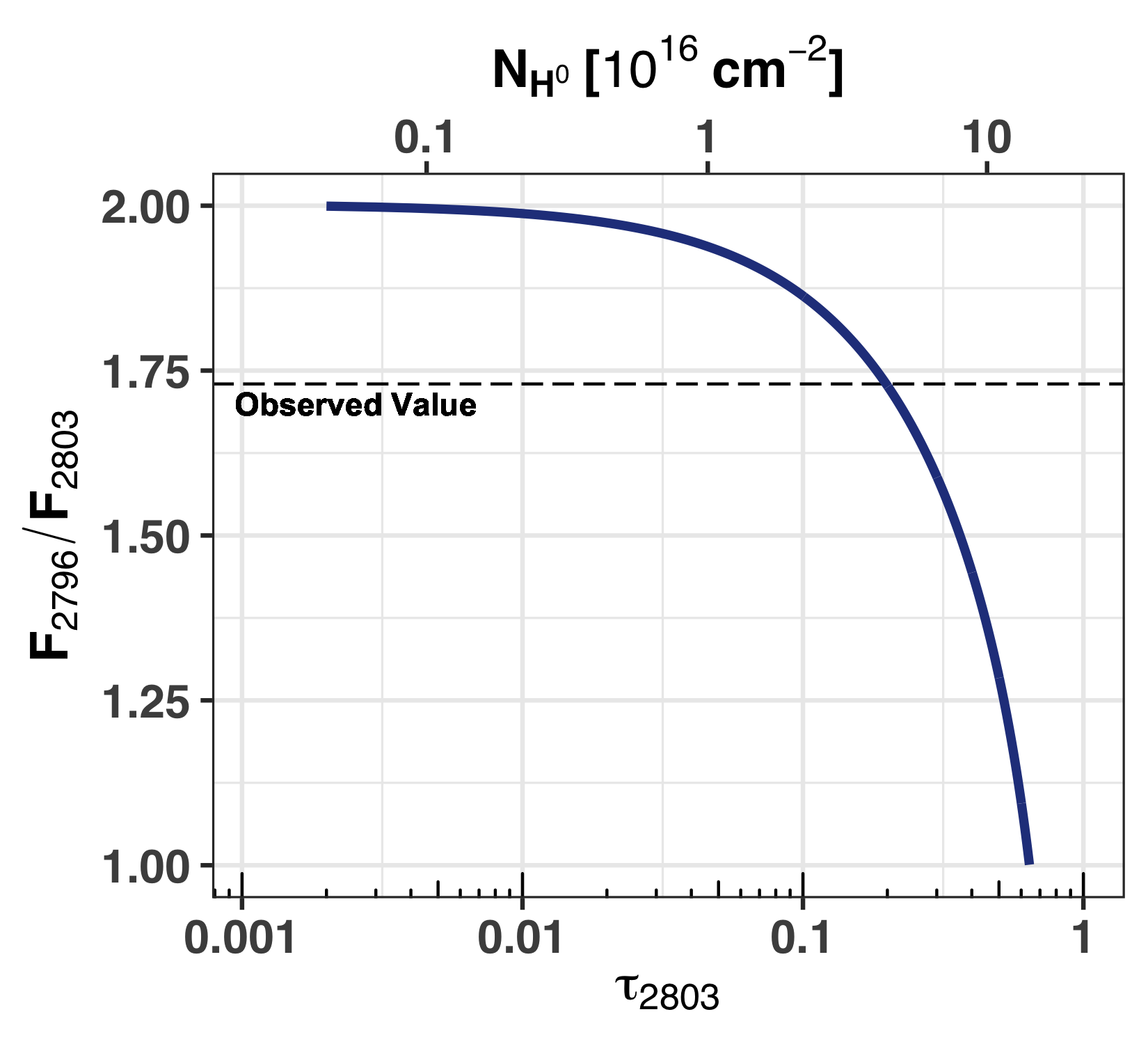}
    \caption{The change of the \mgii\ doublet flux ratio, $R = F_{2796}/F_{2803}$, with the optical depth of the \mgii~2803~\AA\ line ($\tau_{2803}$) for a simple synthetic line profile that assumes that the nebular \mgii\ emission is much brighter than the continuum emission (\autoref{thick}). The observed $R=1-2$ values from J1503 correspond to $\tau_{2803}$ values of 0.01--0.6, along the rapidly changing portion of the curve. The $R$ of optically thin \mgii\ emission strongly varies with \hi\ column densities (upper x-axis, assuming the metallicity of J1503 and a typical dust depletion). \mgii\ emission with $R > 1$ corresponds to \hi\ column densities less than 10$^{17.2}$~cm$^{-2}$. Optically-thin \mgii\ emission ($\tau_{2803} < 1$) sensitively probes neutral gas column densities that allow ionizing photons to escape galaxies.}
    \label{fig:cf1}
\end{figure}

The first scenario envisions a uniform distribution of dustless, low column density Mg$^{+}$ gas residing in the foreground along the line-of-sight to a background continuum plus \mgii\ emission source (see \autoref{highz} for the impact of dust). A small fraction of the incident light excites the Mg$^+$ electrons into excited states. The Mg$^+$ electrons then de-excite and emit \mgii\ photons in random directions, predominately not along the line of sight. Since the \mgii\ profile of J1503 indicates that the Mg$^+$ is optically thin, these re-emitted \mgii\ photons propagate through the Mg$^+$ gas without being reabsorbed and are removed from along the line of sight to the observer. The transmitted \mgii\ flux decreases, with the observed flux ($F_{\rm obs}$) given as
\begin{equation}
    F_{\rm obs} =  \left( F_{\rm int} + F_{\rm cont} \right) e^{-\tau} \approx  F_{\rm int} e^{-\tau} ,
    \label{eq:tau}
\end{equation}
where $F_{\rm cont}$ is the background continuum and $F_{\rm int}$ is the background intrinsic \mgii\ collisionally-excited nebular flux. Here we have approximated the intrinsic flux as being dominated by the \mgii\ emission rather than the continuum flux (see \autoref{continuum}). $\tau$ is the optical depth of the transition integrated over the line profile, defined as
\begin{equation}
    \tau = \frac{\pi e^2}{m_{\rm e} c^2} f \lambda N_{\rm Mg^+} ,
    \label{eq:optical_depth}
\end{equation}
where $e$ and $m_{\rm e}$ are the electron charge and mass, $c$ is the speed of light, $\lambda$ is the restframe wavelength, and $f$ is oscillator strength of the transition. The $\tau$ ratio of the two \mgii\ transitions is equal to the ratio of the $f$ values.  $f_{2796}$ is twice that of the \mgii~2803~\AA\ transition, such that
\begin{equation}
    \tau_{2796} = 2 \tau_{2803} .
\end{equation}
Thus, if the intrinsic two-to-one \mgii\ emission escapes through optically-thin Mg$^+$ gas, $R$ is a function of $\tau_{2803}$ as
\begin{equation}
    R  = \frac{F_{\rm 2796, obs}}{F_{\rm 2803, obs}}= \frac{F_{\rm 2796, int}}{F_{\rm 2803, int}}e^{-\tau_{2796}+\tau_{2803}} = 2 e^{-\tau_{2803}} . 
    \label{eq:r_tau}
\end{equation}
$R$ varies depending on $\tau_{2803}$ of the foreground \mgii, and, through \autoref{eq:optical_depth}, the Mg$^+$ column density along the line of sight. 

To illustrate how $R$ varies with $\tau_{2803}$ we created mock line profiles in three steps: (1) created a flat unity continuum level (consistent with the O-star dominated continuum spectrum of J1503), (2) added a \mgii\ emission profile to the continuum with an $R_{\rm int} = 2$, and (3) multiplied the resultant profiles by an absorption doublet profile with $\tau_{2803}$ (and $\tau_{2796} = 2 \tau_{2803}$). We assumed that both the emission and absorption profiles have the same velocities and widths. We then calculated $R$ the same way we did for the data. This process was repeated for a large range of $\tau_{2803}$ values to derive a distribution of $R$ values.

\autoref{fig:cf1} shows the change in $R$ with $\tau_{2803}$ for these mock spectra. As $\tau_{2803} \rightarrow 0$ the flux ratio tends to the intrinsic value of $R = 2$ because there is no Mg$^+$ gas along the line of sight to absorb \mgii\ photons. As $\tau_{2803}$ increases, Mg$^+$ removes a factor of $e^{\tau_{2803}}$ more flux from the 2796~\AA\ line than from the 2803~\AA\ line (\autoref{eq:r_tau}), resulting in a declining $R$ with increasing $\tau_{2803}$.  {Realistically observable $R$ values between $1-2$ probe  $\tau_{2803}$ values between $0.6-0.01$, where extremely high-quality observations are required to estimate very low $\tau_{2803}$ values}. This is the observed $R$ parameter space from J1503. 

\begin{figure}
    \centering
    \includegraphics[width = 0.5\textwidth]{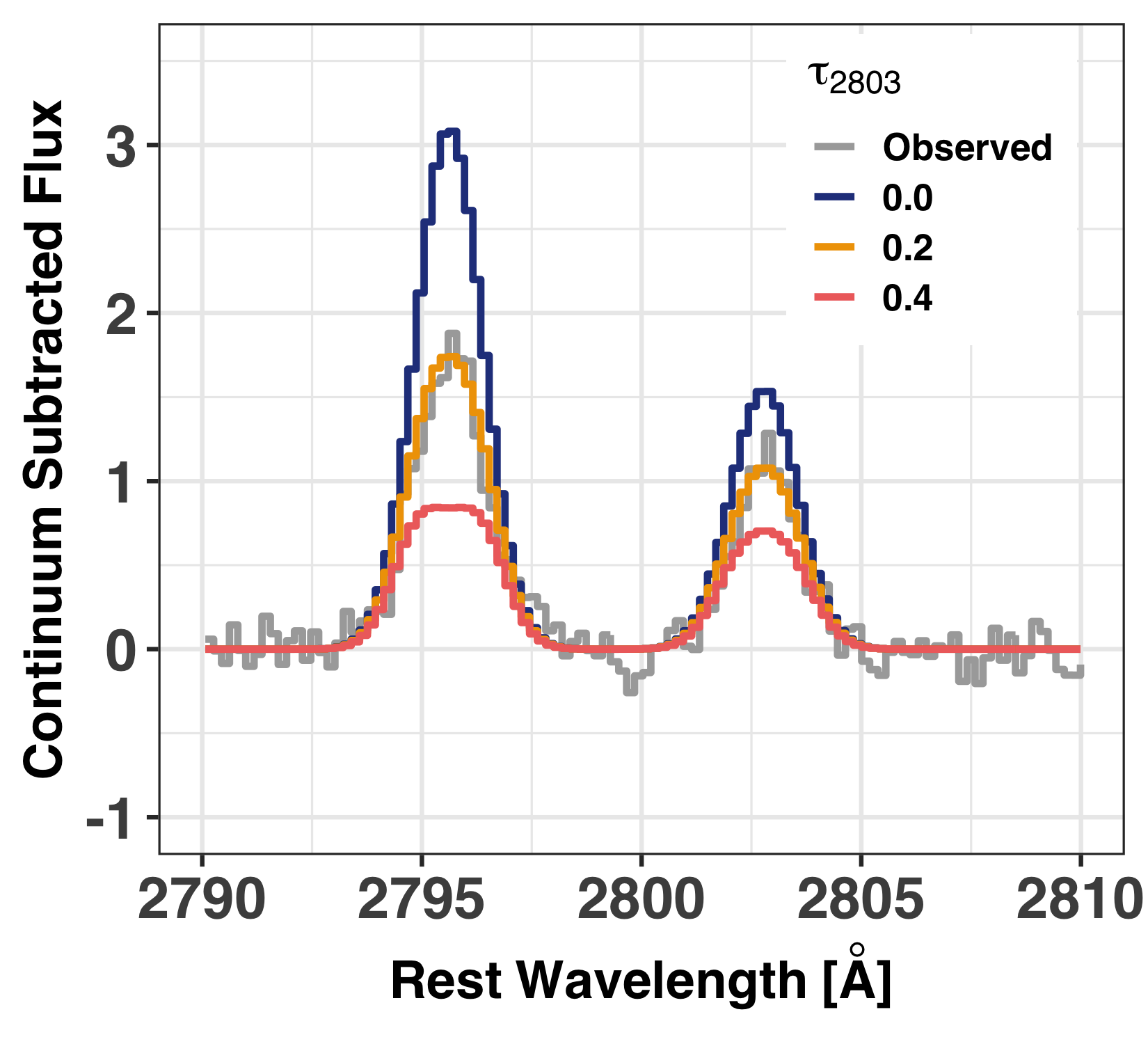}
    \caption{Simulated \mgii\ doublet emission profiles for three different \mgii~2803~\AA\ optical depths ($\tau_{2803}$): 0 (blue), 0.2 (gold), and 0.4 (red). At $\tau_{2803} = 0$, the line profiles have a flux ratio of 2. As $\tau_{2803}$ increases, the \mgii~2796~\AA\ line weakens by a factor of $e^{\tau_{2803}}$ more than the \mgii~2803~\AA\ line because the oscillator strength of the \mgii~2796~\AA\ is twice that of the \mgii~2803~\AA\ transition. The spatially integrated \mgii\ profile from J1503 is included in gray and is consistent with the $\tau_{2803} = 0.2$ model.  }
    \label{fig:mock_comp}
\end{figure}

Mock spectra, with parameters tailored to the J1503 observations (\autoref{tab:lines}), are shown in \autoref{fig:mock_comp}. At $\tau_{2803} = 0$ (dark blue line), the \mgii\ profile is the intrinsic emission profile with a stronger \mgii~2796~\AA\ line than the \mgii~2803~\AA\ line and $R = 2$. As $\tau_{2803}$ increases the 2796~\AA\ profile decreases in strength until at $\tau_{2803} \sim 0.4-0.6$ the two \mgii\ lines have similar strength (red line). In between $\tau_{2803}= 0$ and $0.4$, the 2796~\AA\ line is still stronger than the 2803~\AA\ transition (gold line), but the relative strength of the two lines has noticeably decreased. 

The observed spatially integrated \mgii\ profile is included as a grey line in \autoref{fig:mock_comp}. The mock profiles explain a few of the key observations. First, a small, optically-thin $\tau_{2803}$ of 0.2 (gold line) reproduces the spatially-integrated emission profile. The optically thin nature of \mgii\ is consistent with the observations that the \mgii\ emission is not spatially extended beyond the stellar or nebular emission (\autoref{fig:em_comp} and \autoref{fig:spatial_comp}) and that the \mgii\ line profiles are well-fit by single Gaussians centered at zero-velocity ( {\autoref{fig:mg2_fit}}). Second, the individual pixels (\autoref{fig:ind}) and the $R$ composites (\autoref{fig:rat_comp}) show that decreasing the 2796~\AA\ flux drives the $R$ from 2 to $\sim1.3$, while the \mgii~2803~\AA\ line remains nearly constant. The mock spectra explain this shift because the 2796~\AA\ transition has a larger oscillator strength and is reduced by a factor of $e^{\tau_{2803}}$ more than the 2803~\AA\ transition.

Optical depth variations describe the foreground neutral gas properties in J1503 through which ionizing photons must pass through to escape the galaxy. First, \autoref{eq:r_tau} directly infers $\tau_{2803}$ from the observed $R$ as
\begin{equation}
    \tau_{2803} = -\ln\left(R/2\right) .
    \label{eq:tau_r}
\end{equation}
Using the average value of $R = 1.7$ from the spatially-integrated line profile, $\tau_{2803} = 0.16$ over the entire galaxy of J1503 (\autoref{tab:esc}). At these low optical depths, most of the $F_{\rm int}$ escapes the optically thin gas. Ignoring scattering because there are no observational signatures of either scattering or absorption, the amount of \mgii\ emission that escapes J1503 can be inferred from \autoref{eq:tau} as the escape fraction as
\begin{equation}
    f_{\rm esc}\left({\rm \ion{Mg}{ii}}\right) = \frac{F_{\rm obs}}{F_{\rm int}} = e^{-\tau}. \label{eq:esc_def}
\end{equation}
A $\tau_{2803} =0.16$ leads to $f_{\rm esc} \left(2803\right) = 85\pm7$\% (\autoref{tab:esc}). 

Photoionization models can also predict $F_{\rm int}$ in \autoref{eq:esc_def} by using multiple strong optical emission lines that are set by the ionization structure, metallicity, density, and ionizing continua of a hypothetical nebula \citep{erb12, guseva13}. \citet{henry18} used \textsc{cloudy} photoionization models \citep{ferland} to determine a relationship between the extinction-corrected [\ion{O}{iii}]~5007~\AA\ and [\ion{O}{ii}]~3727~\AA\ emission lines and $F_{\rm int}$(\ion{Mg}{ii}) (their Eq.~2). 
Comparing $F_{\rm int}$ to the observed SDSS \mgii~2803~\AA\ line fluxes (\autoref{eq:esc_def}) suggests that J1503 transmits $95\pm14$\% of the \mgii~2803~\AA\ photons (\autoref{tab:esc}). This is consistent with the \fescm\ estimated with $R$ above. 

From \autoref{eq:optical_depth}, the total Mg$^+$ column density, $N_{{\rm Mg}^+}$, can be estimated in terms of $\tau_{2803}$ as
\begin{equation}
\begin{split}
        { N}_{{\rm Mg}^{+}} &= 3.8\times10^{14}~{\rm cm}^{-2}~\text{\AA} \frac{\tau_{2803}}{f \lambda}\\ 
        & = 4.4 \times 10^{11}~{\rm cm}^{-2}~\tau_{2803},
\end{split}
\end{equation}
where we used the oscillator strength ($f=0.303$ for the 2803~\AA\ transition) and the restframe wavelength (in \AA). This is cast in terms of $R$ using \autoref{eq:tau_r} as
\begin{equation}
N_{{\rm Mg}^{+}} = -4.4 \times 10^{11}~{\rm cm}^{-2}~\ln\left(R/2\right).
\label{eq:nmg_r}
\end{equation}
\autoref{eq:nmg_r} implies that the spatially-integrated Mg$^+$ column density in J1503 is \nmg~=~$7\pm3\times10^{10}$~cm$^{-2}$. 

The main goal of this paper is to relate the \mgii\ emission to the LyC escape through the \hi\ column density. The Mg$^0$ and Mg$^+$ ionic phases overlap with \hi\ and their column densities can be converted into \hi\ column densities using the Mg/H abundance of the galaxy. There are two complications with this: (1) a rather uncertain fraction of Mg is depleted onto dust ($\delta_{\rm Mg}$) and (2) the Mg/H abundance must be determined. First, we assume that $\delta_{\rm Mg} = 27$\%, or that 27\% of Mg is in the gas phase in the warm neutral medium \citep{jenkins}. The Mg depletion factor has appreciable scatter, but a general trend is not observed with metallicity and the depletion only depends on the ionization for the most highly ionized galaxies \citep[][]{guseva19}. Secondly, while the Mg/H abundance is challenging to observe, the O/H abundance is readily observable from multiple optical oxygen emission lines, some of which are temperature sensitive (e.g., [\ion{O}{iii}]~4363~\AA). Since both oxygen and magnesium are $\alpha$ elements that are primarily produced by core-collapse supernova \citep{johnson19}, the Mg/O value should not appreciably vary (except for dust depletion differences). Thus, we estimate the Mg/H abundance using the observed O/H abundance (see \autoref{tab:props}) and a solar O/Mg abundance of O/Mg~$ = 12.3$ \citep{asplund}. The \hi\ column density can then be approximated as:
\begin{equation}
\begin{split}
    {\rm N}_{{\rm H}^0} &= \frac{\rm H}{\rm O} \frac{\rm O}{ \rm Mg} \frac{1}{\delta_{\rm Mg}} \left({\rm N}_{\rm Mg^{+}} + {\rm N}_{\rm Mg^{0}}\right) \approx 46 \frac{\text{H}}{\text{O}} {\rm N}_{\rm Mg^{+}} \\
    &\approx -2\times10^{13}~{\rm cm}^{-2}~\frac{\rm H}{\rm O} \ln\left(R/2\right) , 
\end{split}
\end{equation}
where we neglected the contribution of Mg$^0$ because the \mgi~2852~\AA\ line is undetected in all our observations (individual spaxels or composites). Using the O/H of J1503, the assumed dust depletion, and \autoref{eq:nmg_r}, the neutral hydrogen column density is expressed as
\begin{equation}
\begin{split}
    {\rm N}_{{\rm H}^0} &= 5.1\times10^5~{\rm cm}^{-2}~{\rm N}_{{\rm Mg}^+} \\
    &= -2.2 \times 10^{17}~{\rm cm}^{-2} \ln\left(R/2\right) .
    \label{eq:nh1_r}
   \end{split}
\end{equation}
Thus, the \mgii\ doublet ratio and the gas-phase metallicity can infer the \hi\ column density. \autoref{eq:nh1_r} enables an unprecedented study of one of the chief sinks of ionizing photons within LyC emitters: the neutral hydrogen column density. The integrated \hi\ average column density over the entire galaxy of J1503 is \nhi~$=4\pm1\times10^{16}$~cm$^{-2}$. 

Inverting \autoref{eq:nh1_r} provides guidelines for which $R$ values correspond to neutral gas that is optically thin to ionizing photons, assuming the nebular metallicity of J1503 and that dust destroys a negligible amount of ionizing photons (we refer to this as the relative escape fraction, $f_{\rm esc}^{\rm rel}$, see \autoref{highz} for the importance of dust). Galaxies become optically thin to ionizing radiation at \hi\ column densities less than 10$^{17.2}$~cm$^{-2}$, which corresponds to an $R > 0.97$ in J1503.

The metallicity crucially impacts the interpretation of whether \mgii\ traces LyC escape. At a lower metallicity of 12+log(O/H) = 7.5, $R > 1.6$ corresponds to a LyC emitter because there are fewer Mg atoms per hydrogen atom. Conversely, at solar metallicities \citep[12+log(O/H) = 8.69; ][]{asplund} galaxies  {with $R > 0$} will be optically thin to ionizing radiation. Regardless of the gas-phase metallicity, $R\sim 2$ and \mgii\ line profiles without signatures of scattering likely indicate \hi\ gas that is optically thin to ionizing radiation. 

While this simple optical depth framework provides intuition for the propagation of ionizing photons through low column density neutral gas, some previous observations require further exploration. Namely, the Lyman Series of J1503 is strong \citep{gazanges, gazagnes20}, implying a significant \hi\ column density. Strong saturated Lyman Series absorption is not just found in J1503, but other confirmed LyC emitting galaxies \citep{steidel18}.

\subsection{\mgii\ escape through a clumpy geometry}
\label{partial}
A possible solution is to have highly clumped neutral gas within the galaxy \citep{heckman2011, zackrisson13}. \mgii\ (and LyC) photons that encounter the dense clouds are absorbed and are not transmitted along the line of sight. Meanwhile, the channels between the clumps have zero column density and allow the photons to pass through these channels without being absorbed. Either large-scale gaseous instabilities \citep{koki} or massive star feedback \citep{jaskot19} could inject the energy and momentum required to redistribute the optically-thick gas and create evacuated channels for the photons to pass through.

Envision that a fraction of the total area along the line of sight (called the covering fraction; $C_f$) is covered by high column density gas and 1-$C_f$ of the area is completely free of gas. Therefore, a fraction, 1-$C_f$, of the background radiation escapes the gas without being absorbed by Mg$^+$, while the complement, $C_f$, of the light is absorbed by high column density Mg$+$ with an optical depth of $\tau \gg 1$. The radiative transfer equation (\autoref{eq:tau}; still neglecting dust) then takes the form of
\begin{equation}
    F_{\rm obs} = F_{\rm int} \times \left[\left(1-C_f\right) + C_f e^{-\tau}\right] =  F_{\rm int} \times \left(1-C_f\right) ,
    \label{eq:rad_pf}
\end{equation}
where the simplistic geometry assumes that the gaseous regions are optically thick ($\tau \gg 1$). In this idealized scenario, all of the \mgii\ photons that originate in optically thick regions are destroyed, either by dust or through scattering out of the line of sight. The observed \mgii\ doublet flux ratio,  {which traces lines of sight through the empty channels,} becomes simply
\begin{equation}
    R = \frac{F_{\rm 2796, obs}}{F_{\rm 2803, obs}} = \frac{F_{\rm 2796, int}}{F_{\rm 2803, int}} = 2 ,
\end{equation}
because both transitions have the same $C_f$ at $\tau \gg 1$. The extremely clumpy scenario predicts that $R$ does not vary from the intrinsic $R$. Rather, any \mgii\ emission that escapes the neutral gas has the flux ratio equal to the intrinsic ratio. However, the observations of J1503 demonstrate a variation of $R$ values.  Thus, neither of these overly-simplified physical scenarios match all of the observations.

\subsection{\mgii\ escapes through low column density channels surrounded by high column density regions}
\label{combine}
Both of the proposed scenarios above reproduce portions of the observations, but neither satisfies all the constraints. A slight modification of the extremely clumpy scenario is that the channels that photons pass through have a low column density medium, with $\tau_{\rm chan}$, while the optically-thick regions, with $\tau_{\rm thick}$, absorb all of the photons incident upon them \citep{gazagnes20}.  This modifies the radiative transfer equation in \autoref{eq:rad_pf} to become
\begin{equation}
    F_{\rm obs} = F_{\rm int} \times \left[ \left(1-C_f\right)e^{-\tau_{\rm chan}} + C_f e^{-\tau_{\rm thick}} \right] . \label{eq:rad_chan}
\end{equation}
If $\tau_{\rm thick} \gg 1$, the \mgii\ doublet flux ratio becomes
\begin{equation}
        \frac{F_{\rm 2796, obs}}{F_{\rm 2803, obs}} = \frac{F_{\rm 2796, int}}{F_{\rm 2803, int}} \frac{\left(1-C_f\right)e^{-2\tau_{\rm 2803, chan}}}{\left(1-C_f\right)e^{-\tau_{\rm 2803, chan}}} \label{eq:low-den}
\end{equation}
and the observed $R$ becomes
\begin{equation}
    R = 2 e^{-\tau_{\rm 2803, chan}}.\label{eq:tau_chan}
\end{equation}
This equation is nearly identical to the $R$-relation found in the optically thin scenario (\autoref{eq:r_tau}) but with the important physical clarification that the \mgii\ flux ratio is entirely determined by the optical depth of the low column density channels. All of the derivations in \autoref{thick} that relate \mgii\ emission to the \hi\ properties hold, but have the crucial physical distinction that they only correspond to the gas within the low column density channels. 

Is this scenario of low column density gas in between higher column density clouds consistent with the Lyman Series observations? \citet{gazanges} and \citet{gazagnes20} used the observed Ly$\beta$, Ly$\gamma$, Ly$\delta$, and Ly5 absorption lines of J1503 to determine that the Lyman series is saturated, but the fitted Lyman Series covering fraction is $0.72\pm0.06$. Ly$\beta$ is the strongest transition fit by \citet{gazagnes20}, and Ly$\beta$ saturates at \nhi~$\gtrsim 8\times 10^{15}$~cm$^{-2}$. In other words, the \citet{gazagnes20} observations indicate that 72\% of the area of the stellar continuum is covered by neutral gas with \nhi~$\gtrsim 8\times 10^{15}$~cm$^{-2}$. Using \autoref{eq:nh1_r}, we find that 67\% of the stellar continuum in J1503 is covered by \mgii\ gas with  $R < 1.928$ (the $R$ that corresponds to $8\times 10^{15}$~cm$^{-2}$), similar to the observed Lyman Series covering fraction. This suggests that the \mgii\ emission can serve as a proxy of the neutral gas properties.

Likewise, the covering fraction of optically thick Mg$^+$ gas can be estimated by combining \autoref{eq:esc_def} and \autoref{eq:rad_chan} as 
\begin{equation}
    \text{\fescm} = \left(1-C_f(\text{\mgii})\right)e^{-\tau_{\rm chan}} .
\end{equation}
If \fescm\ is measured using photoionization models \citep[such as from ][]{henry18} and $\tau_{\rm chan}$ is estimated using $R$ (which is independent of $C_f$; see \autoref{eq:tau_chan}), the \mgii\ covering fraction can be approximated as
\begin{equation}
    C_f(\text{\mgii)} = 1- \text{\fescm} e^{\tau_{\rm chan}} . \label{eq:cf}
\end{equation}
Using the spatially-integrated values (\autoref{tab:esc}), this leads to a $C_f(\text{\mgii)} = -0.11\pm0.05$. At a 2$\sigma$ significance level, this $C_f(\text{\mgii)}$ is a non-physical $C_f$ that is less than zero. This implies that either \fescm\ is over-estimated or $\tau_{\rm chan}$ is slightly under-estimated. The mis-estimate could possibly arise from either the effects of  {absorption subsequently followed by re-emission (or scattering, which is not accounted for here)} or issues with the photoionization models. Regardless, the low inferred $C_f$ agrees with the fact that we do not observe any spaxels with $R < 0.7$, which corresponds to $\tau_{\rm chan} = 1$, and that $C_f(\text{\mgii)}\approx0$. Thus, a small $C_f(\text{\mgii)}$  can be inferred from both the spatially resolved and spatially-unresolved observations.

Meanwhile, $C_f(\text{H}^0)$ is required to estimate the LyC escape fraction. The $C_f$ inferred from the Lyman Series or the metal absorption lines does \textit{not} properly quantify the covering of \hi\ related to LyC escape because these transitions become optically thick at different column densities than the LyC. \autoref{eq:rad_chan} defines $C_f$ as the fraction of the total sight lines with $\tau \gg 1$. For LyC emission, this can be recast in terms of the fraction of sightlines that intersect \hi\ gas with \nhi~$> 10^{17.2}$~cm$^{-2}$. \autoref{fig:doub_hist} indicates that no observed spaxel has \mgii\ emission that suggests $\tau_\text{\hi} \gg 1$ and only 1.6\% of the spaxels have $R < 0.97$ and  $\tau_\text{\hi}\gtrsim1$. This implies that, at least for J1503, $C_f(\text{\hi})\approx0$ for gas that is optically thick to the LyC. 

The neutral gas covering fraction can be approximated using $\tau_{\rm chan}$ and \fescm, if the abundance is such that \mgii\ becomes optically thick at the same \nhi\ as the Lyman Break (10$^{17.2}$~cm$^{-2}$). For J1503, $R>1.0$ corresponds to optically thin neutral hydrogen and $R > 0.7$ corresponds to optically thin \mgii. Thus, $C_f(\text{\mgii)}$ provides an upper limit for $C_f(\text{\hi})$ in J1503. 

Meanwhile, at a slightly higher 12+log(O/H)~=~8.1, the \mgii\ covering fraction exactly equals the \hi\ covering fraction. If the \mgii\ covering fraction differs from the \hi\ covering fraction, $C_f(\text{\mgii)}$ can only provide limits for $C_f(\text{\hi})$. If $C_f(\text{\mgii)}$ is large, then the neutral gas covering fraction must be accounted for when estimating the \fescc. Conversely, if $C_f(\text{\mgii)}$ is small, then the covering fraction can be approximated as zero. 

This is \textit{not} to say that LyC passes through the low column density channels unabsorbed, rather the LyC optical depth is less than 1. The strong correlations between the $C_f$ of the Lyman Series and \fescc\ and the \lya\ properties \citep{steidel18, gazagnes20} could arise because the \nhi\ distribution within a galaxy is rather broad (\autoref{fig:doub_hist}). The LyC escapes through the lowest column density gas, and the fraction of \hi\ that is optically thin to LyC depends on the median and shape of the \nhi\ distribution. Thus, if \nhi\ is low on average, fewer total sight lines would have optically thick Lyman Series absorption lines and the $C_f$ will be lower. 

\begin{figure}
    \centering
    \includegraphics[width = 0.5\textwidth]{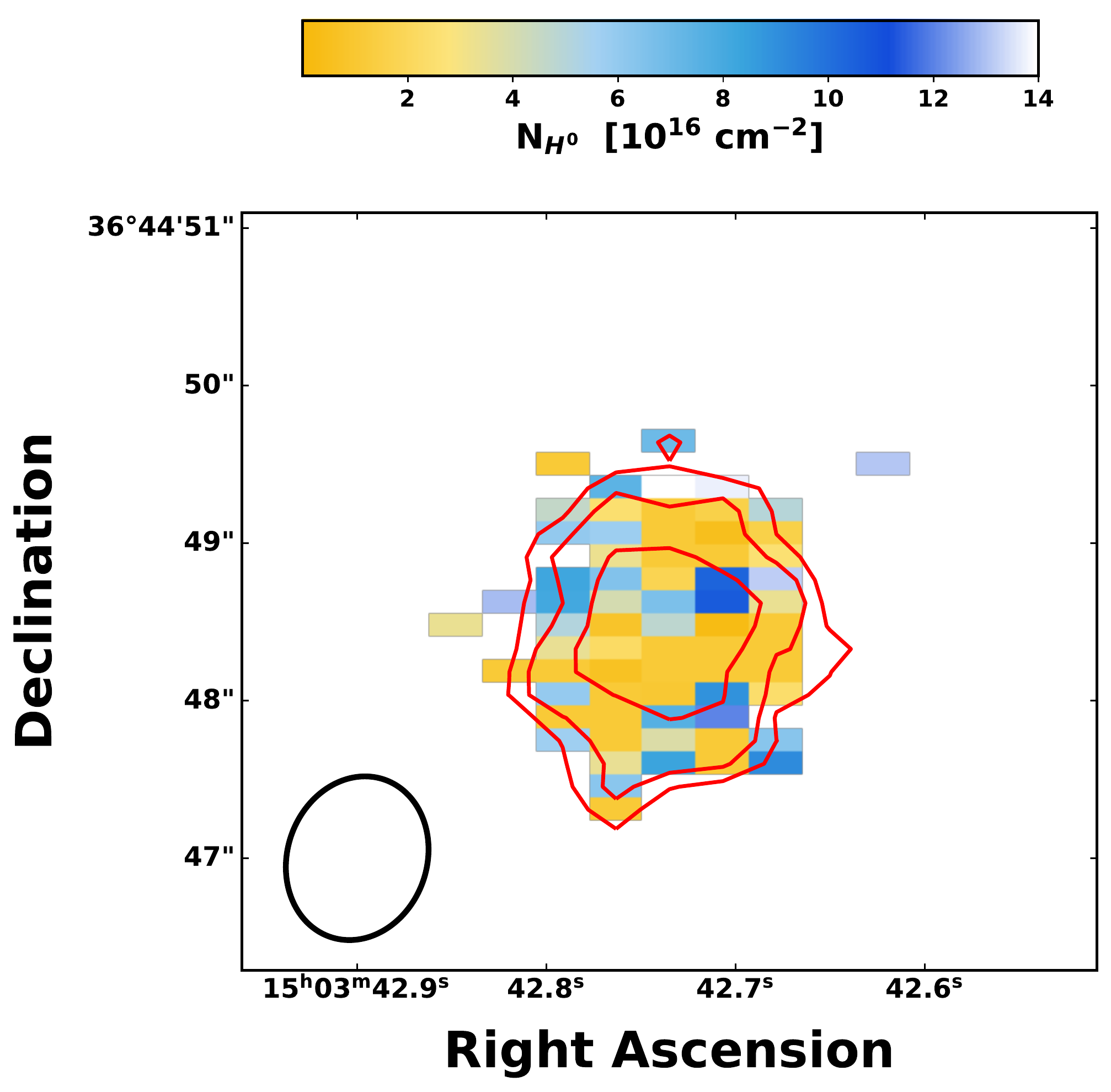}
    \caption{The spatial distribution of the \hi\ column density in J1503 calculated using the \mgii\ doublet flux ratio and assuming a constant nebular metallicity and dust depletion (\autoref{eq:nh1_r}). The contours show the 2, 5, and 10$\epsilon$ contours of the integrated \mgii\ emission.  {Gold} regions correspond to low \hi\ column density and are patchy over the face of the galaxy.  The low column density regions are likely the channels through which the ionizing photons escape J1503.} 
    \label{fig:nhi}
\end{figure}

Thus, the \mgii\ observations suggest that J1503 emits ionizing photons due to a combination of the two proposed scenarios: the \hi\ column density is low (a spatially averaged value of $4\times10^{16}$~cm$^{-2}$), but the column density \textit{also} has a broad spatial distribution. \autoref{fig:nhi} shows the spatial map of \nhi\ and depicts a factor of more than 10 variation in the \hi\ column density. This clumpy \hi\ distribution suggests that ionizing photons escape through the regions of relatively low \hi\ column density (the light blue regions in \autoref{fig:nhi}) with sizes smaller than, or on the order of, the 5~kpc spatial resolution of the observations.

\section{An ideal Lyman Continuum tracer}
\label{highz}
A chief goal of this study is to provide an indirect method for future observations to infer the absolute escape fraction of ionizing photons, \fescc, from galaxies within the Epoch of Reionization. Above we proposed the combination of the \mgii\ doublet and nebular metallicities as an ideal tracer of the \hi\ column density. The combination of \nhi\ and dust attenuation then provides an estimate of the absolute LyC escape fraction. In this section we walk through the process of inferring \fescc\ from the \mgii\ observations of J1503 (\autoref{esc_1503}), and how \fescc\ spatially varies (\autoref{spatial_lyc}). We then calculate \fescc\ using literature \mgii\ observations of 6 other known LyC emitters (\autoref{esc_samp}). In \autoref{future}, we comment on the possibility of observing \mgii\ with JWST and ELTs to constrain the sources of cosmic reionization.

\subsection{The LyC escape fraction of J1503}
\label{esc_1503}
Ionizing radiation is absorbed by two main sinks: \hi\ gas and dust. Breaking up the contributions of the two simplifies the determination of the absolute escape fraction, or \fescc. The fraction of ionizing photons that pass through a dustless neutral medium, sometimes referred to as the relative escape fraction, is
\begin{equation}
\begin{split}
    f_{\rm esc}^{\rm rel} &= e^{-N_{{\rm \textsc{H}}^{0}} \sigma_{\rm phot}} \\ &\approx e^{2.2\times10^{17}~{\rm cm}^{-2 }\sigma_{\rm phot} \ln{R/2}} ,
    \label{eq:fesc_hi}
    \end{split}
\end{equation}
where $\sigma_{\rm phot} = 6.3\times10^{-18}$~cm$^{2}$ is the photoionization cross-section of \hi\ at 912~\AA. Using  \nhi~$=3.6\times10^{16}$~cm$^{-2}$, the average column density inferred from the spatially-integrated \mgii\ emission from \autoref{eq:nh1_r}, 80\% of the intrinsic ionizing photons pass through the dustless neutral gas in J1503. For the metallicity of J1503, $f_{\rm esc}^{\rm rel}$ and \fescl\ are statistically consistent with each other (\autoref{tab:esc}). However, this will not always be true and will depend on the metallicity of the gas. 

Dust is the second sink of ionizing photons. The fraction of ionizing photons at 912~\AA\ that pass through a dusty medium without being absorbed is given as
\begin{equation}
    f_{\rm esc}^{\rm Dust} = 10^{-0.4 {\rm E(B-V) k(912)}} .
\end{equation}
\citet{gazagnes20} used the \citet{reddy16} attenuation law, with k(912)~$=$~12.87, to fit theoretical stellar continuum models to the observed FUV stellar continuum features of J1503. The FUV stellar continuum is best-fit with E(B-V) = 0.22~mag.  {This can be compared to the E(B-V)~$= 0.09$ that \citet{izotov16b} calculated  for J1503 using the Balmer emission lines. However, the \citet{izotov16b} value was determined using the \citet{cardelli} attenuation law with $R_V = 2.3$, whereas we used the \citet{reddy16} law.  If we calculate E(B-V) using the observed H$\alpha$/H$\beta$, H$\gamma$/H$\beta$ and H$\delta$/H$\beta$ flux ratios from \citet{izotov16b} and the \citet{reddy16} attenuation law, we estimate E(B-V)$ = 0.14, 0.21, 0.23$~mag, respectively. Thus, the stellar and nebular attenuations are similar when similar attenuation laws are used.} This relatively large inferred E(B-V) means that dust removes a large fraction of the ionizing photons; only 7\% of the ionizing photons escape the galaxy without being absorbed by dust ($f_{\rm esc}^{\rm Dust}$). As stressed by \citet{chisholm18}, dust is the major sink of ionizing photons in J1503. 

The total, or absolute, \fescc\ is equal to the product of the relative escape fraction and the fraction unabsorbed by dust such that
\begin{equation}
    f_{\rm esc}\left({\rm LyC}\right)  = e^{-N_{{\rm \textsc{H}}^0} \sigma_{\rm ph}} \times 10^{-0.4 \text{E(B-V) k(912)}} .
    \label{eq:fescc}
\end{equation}
This leads to an absolute \fescc~$=5.9\pm0.4$\% from the spatially-integrated observations. This is consistent with the \fescc\ directly observed from the LyC of  $5.8\pm0.6$\% \citep{izotov16b}. Thus, the escape fraction of ionizing photons from J1503 can be recovered using the \mgii\ doublet flux ratio and the dust attenuation. 

Note that the covering fraction does \textit{not} appear in the estimation of \fescc, even though we assumed a highly clumpy, non-uniform geometry. As discussed in \autoref{combine}, this is because hardly any of the observed spaxels have \mgii\ emission properties that indicate  optically thick \hi. This effectively means that $C_f(\text{\hi})~=~0$. The \mgii\ covering fraction (\autoref{eq:cf}) from spatially unresolved observations can be used as a guide for whether the neutral covering fraction must be included, but only if the abundance is close to 12+log(O/H)~=~8.1, Mg$^+$ is the dominant neutral ionization state, and the dust depletion is similar to our assumptions (see \autoref{combine}).

Further, the \mgii\ emission profile itself indicates that the individual spaxels have near zero covering fractions of optically thick gas because there are no  {strong} indications of absorption or scattering,  { with only minor, <3$\sigma$ significant, deviations from the single Gaussian fit } (\autoref{fig:mg2_fit}, \autoref{fig:ind}, and \autoref{fig:rat_comp}). If there was both optically thin and optically thick gas along the line of sight, a complex profile, with absorption and emission, would be observed. Such complex \mgii\ profiles have been observed in galaxy spectra at moderate redshifts \citep{erb12, guseva13, martin13, rubin13, finley, feltre18, burchett20}, but are not observed from J1503. Further, at larger \mgii\ optical depths, radiative transfer effects will impact the resultant \mgii\ emission profiles: shifting the velocities of emission lines away from zero velocity, broadening the lines, decreasing \fescm, and leading to non-Gaussian profiles \citep{henry18}. Any of these more complex profiles would strongly suggest that either $\tau_{2803}$ or $C_f(\mgii)$ are large, indicating that $R$ \textit{cannot} be used to robustly determine $\tau_{2803}$ and \fescc.

\subsection{The LyC escape fraction in different regions of J1503}
\label{spatial_lyc} 
In \autoref{resolved} we found that the \mgii\ doublet flux ratio, $R$, varies spatially in J1503. Combined with the hypothesis that ionizing photons escape through low column density channels in a denser medium (\autoref{combine}), this spatial $R$ variation suggests that $f_{\rm esc}^{\rm rel}$ varies from location-to-location within J1503. Here we first explore the spatial variation of the higher S/N composites and then the individual spaxel-by-spaxel variations. 

The \nhi\ values in the High and Low $R$ composites are $4\pm8\times10^{15}$ and $7\pm1\times10^{16}$~cm$^{-2}$ (\autoref{fig:rat_comp} and \autoref{tab:rat_comp}). This is a factor of 18 difference (with a 4$\sigma$ separation) in \hi\ column density between the two composites that translates to relative escape fractions of 97 and 63\%, respectively, if we assume that there is a constant metallicity and dust depletion. Similarly, the four spatially distinct integrated regions (\autoref{tab:spatially_distinct}) have relative escape fractions between $>88$ and 67\%. While this difference in the relative escape fraction is significant, it does not fully explain total the spatial variation of the absolute \fescc.

Dust attenuation can also strongly vary spatially, especially if the dust column density is related to the total neutral column density through a dust-to-gas ratio. In \autoref{esc_1503}, we used stellar fits to the FUV stellar continuum to determine the dust reddening for spatially integrated FUV observations. Another common way to measure the dust attenuation is the ratio of recombination emission lines that have intrinsic intensity ratios set by atomic physics. The Balmer emission lines are the most typical way to infer the reddening from emission lines, but our narrow wavelength regime does not contain Balmer lines. However, the \hei~3188~\AA\ and \hei~2945~\AA\ lines are both \hei\ recombination lines whose intrinsic ratios are set by atomic physics, if the electron temperature and density are constrained (\autoref{tab:props}). The \hei\ lines are not ideal to trace the attenuation because their intrinsic strength varies more with temperature and density than Balmer lines, and future observations will explore this limitation. The intrinsic \hei~3188~\AA\ to \hei~2945~\AA\ flux ratio, for the observed temperatures and densities \citep{benjamin99}, is
\begin{equation}
    \frac{F_{\rm 3188, int}}{F_{\rm 2945, int}} = 2.1 .
\end{equation}
The \hei~2945~\AA\ emission in the High and Low $R$ composites are weak, but the \hei\ flux ratio is different at the $1.5\sigma$ significance levels in the two stacks (\autoref{tab:rat_comp}). The flux ratio for the High $R$ stack is nearly 2.1, while the Low $R$ stack has a larger \hei\ flux ratio. This rough comparison implies that the High $R$ stack has less dust attentuation than the Low $R$ stack. 

We estimate E(B-V) values for both composites by assuming the \citet{reddy16} reddening law and comparing the observed and intrinsic $F_{3188}/{F_{2945}}$ (\autoref{tab:esc}). This suggests that the Low $R$ composite has higher \nhi\ regions and more dust attenuation. When we determine the absolute escape fraction of the Low $R$ regions with these inferred dust attenuations, we find that it is zero. The combination of the higher dust attenuation and \hi\ column density in these more optically thick regions means that all of the ionizing photons are absorbed. Meanwhile, the High $R$ composite has both lower dust attenuation and lower \nhi, such that 19\% of the ionizing photons escape the galaxy. This suggests that all of the ionizing photons that escape J1503 come from the High $R$ regions.

\begin{figure}
    \centering
    \includegraphics[width = 0.5\textwidth]{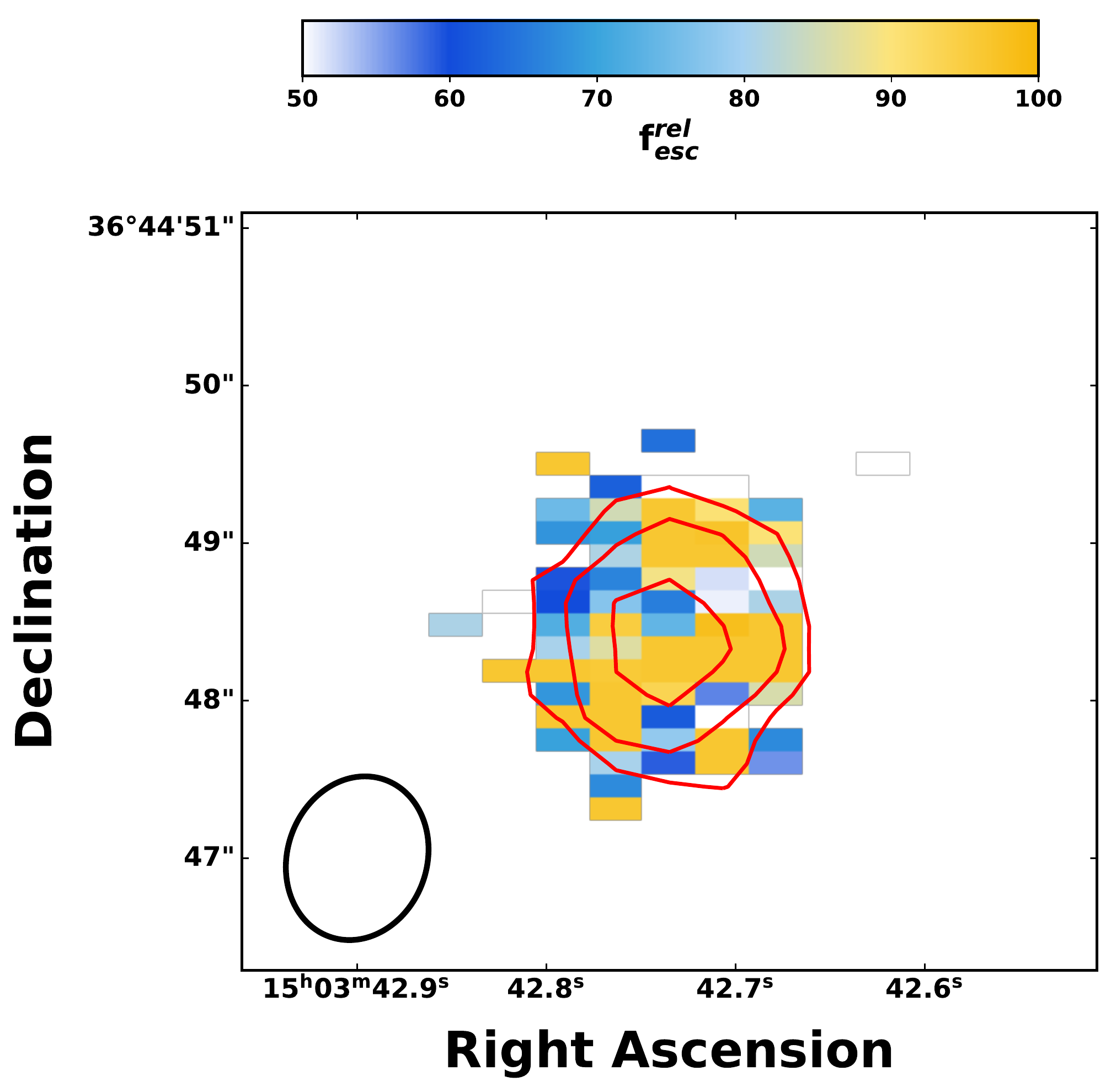}
    \caption{The spatial variation of the relative escape fraction of ionizing photons, or the escape fraction without accounting for dust. The red contours are the 2, 5, and 10$\epsilon$ stellar continuum flux. There is a factor of two difference between regions of high (gold) and low (blue) relative escape fractions. The LyC escape fraction varies spatially across the galaxy.  The black circle gives the observed seeing.}
    \label{fig:fesc_rel}
\end{figure}

Another way to demonstrate the spatial variation is to consider $f_{\rm esc}^{\rm rel}$. \autoref{fig:fesc_rel} shows that $f_{\rm esc}^{\rm rel}$ varies spatially by a factor of two, where gold regions have relative escape fraction near 100\% and blue regions are as low as a 44\%. The stellar continuum (red contours in \autoref{fig:fesc_rel}) weighted relative escape fraction is 83\%, consistent with the relative escape fraction of 80\% calculated from the spatially-integrated \mgii\ line profiles (\autoref{esc_1503}). Thus, the spatially resolved escape fraction for J1503 is similar to the spatially-integrated value. 

The \mgii\ observations suggest that ionizing photons escape from J1503 through regions of low \hi\ column density that are embedded within regions of slightly higher, but still relatively low, \hi\ column density. As stressed in \autoref{combine}, the escape fraction determined by the \mgii\ emission lines is only through these low column density channels. The variations in the \nhi\ and dust attenuation in these low column density channels imprint spatial variations on the escape fraction of ionizing photons (\autoref{fig:fesc_rel}). This introduces a crucial orientation effect for the interpretation of the observed escape fractions: whether the observer is along a line of sight to low column density dust-free channel or high column density dusty region determines whether LyC is observed \citep{zackrisson13, Micheva, Micheva19, fletcher19}. 

\subsection{\mgii\ reproduces the LyC escape fraction}
\label{esc_samp}

\begin{figure}
\includegraphics[width = 0.5\textwidth]{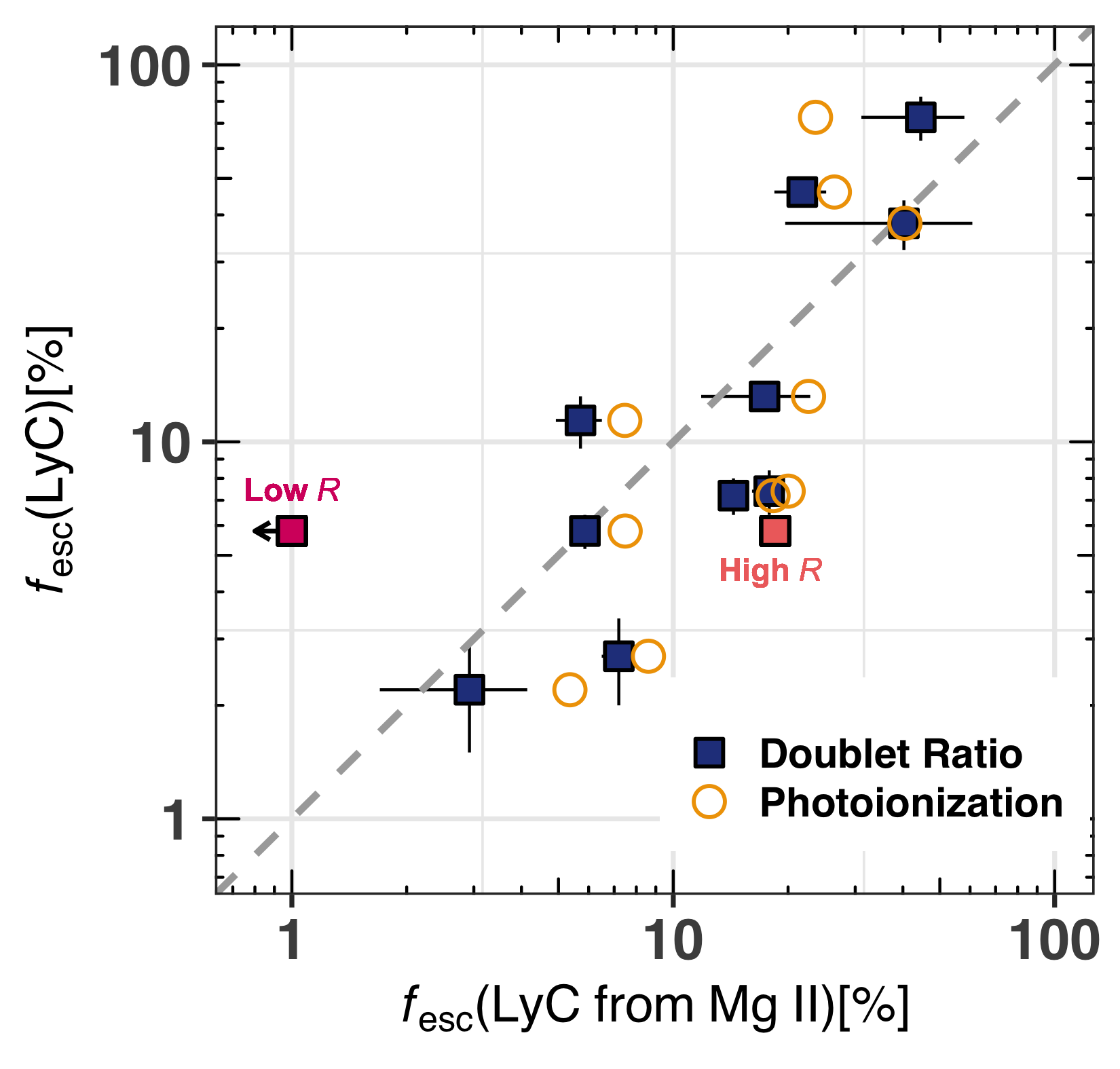}
\caption{The directly observed absolute LyC escape fraction (\fescc) versus the escape fraction predicted using the \mgii\ emission and the FUV stellar continuum attenuation (see \autoref{esc_1503})  {for 10 confirmed LyC emitters from the literature \citep[listed in \autoref{tab:leaker_samp}; ][]{izotov16b, izotov16a, izotov18a, izotov18b, guseva20}} . We predicted the escape fraction in two ways: using the \mgii\ doublet flux ratio (blue squares) and the photoionization models from \citet{henry18} (gold circles). The predicted LyC escape fractions from both methods accurately reproduce \fescc\ (see the gray dashed one-to-one line). This suggests that the \mgii\ emission, nebular metallicity, and dust continuum attenuation can be used to predict \fescc\ of high-redshift galaxies. The red points are the LyC escape fraction observed from composites of spaxels with Low and High $R$, demonstrating that all of the ionizing photons in J1503 escape through regions with High $R$. Regions with Low $R$ have \fescc~$\sim$~0.} 
\label{fig:mg2_esc}
\end{figure}

\begin{table*}
    \caption{Properties of LyC emitting galaxies from \citet{izotov16b}, \citet{izotov16a}, \citet{izotov18a}, and \citet{izotov18b} used to infer the LyC escape fractions using the \mgii\ emission doublet. The first column gives the galaxy's name. The second column gives the nebular oxygen abundance.  {The third column gives the \mgii\ doublet flux ratio ($R$) from the SDSS \citep{izotov16a, izotov16b, izotov18a, izotov18b}, from \citet{guseva20} (for J1154+2443, J1011+1947, J1442-0209, J0925+1403, and J0901+2119), or the KCWI observations presented here (for J1503+3644).} The forth column gives the attenuation of the FUV stellar continuum \citep{gazagnes20}. The fifth column gives the  [\ion{O}{iii}]/[\ion{O}{ii}] flux ratio (O$_{32}$)  {that has been corrected for internal extinction using the value from the Balmer ratio \citep{izotov16b, izotov16a, izotov18a, izotov18b, guseva20}}. The sixth column gives the \mgii~2803~\AA\ to [\ion{O}{iii}]~5007~\AA\ flux ratio  {corrected for internal attenuation with the nebular Balmer-line extinction correction}. Columns 7-9 are the directly observed absolute LyC escape fraction \citep[\fescc; ][]{izotov16b, izotov16a, izotov18a, izotov18b}, the LyC escape fraction inferred from the \mgii\ doublet flux ratio ($f_{\rm esc}$(LyC from \mgii); \autoref{eq:fescc}), and the LyC escape fraction inferred using the photoionization models of \citet{henry18}  (using columns 5 and 6 and their Eq.~2; $f_{\rm esc}(\rm phot)$). Both $f_{\rm esc}$(LyC from \mgii) and  $f_{\rm esc}(\rm phot)$ have been corrected with the same  {FUV} dust attenuation  {from \citet{gazagnes20}}.}
    \begin{tabular}{ccccccccc}
       Name & 12+log(O/H) & $F_{2796}/F_{2803}$ & E(B-V) & O$_{32}$ & log(Mg/[\ion{O}{iii}]) & \fescc & $f_{\rm esc} \left({\rm LyC~from~\mgii}\right)$ & $f_{\rm esc} \left({\rm phot}\right)$ \\
       (1) & (2) & (3) & (4) & (5) & (6) & (7) & (8) & (9) \\
       \hline
       J1243+4646 & 7.89 & $2.5\pm0.8$  & 0.10 & 13.5 & -1.44 & 73 & $45\pm13$ & 24  \\
       { J1154+2443} & 7.75 & $1.9\pm0.3$ & 0.12 & 11.5 &  -1.93 & 46 & $14\pm3.5$ & 26 \\
       J1256+4504 & 7.87 & $2.0\pm1.0$ & 0.08 & 16.3 & -1.92 & 38 & $40\pm21$ & 40 \\
       J1152+3400 & 8.00 & $1.9 \pm 0.6$ & 0.14 & 5.4 & -1.91 & 13 & $17\pm5.5$& 23 \\
        {J1011+1947} &  7.97 & $1.8\pm0.2$ & 0.23  & 27.1  & -2.39 & 11  & $5.7 \pm 0.8$  & 7.4 \\
        {J1442-0209} & 7.99 & $1.9 \pm 0.2$ & 0.14 & 6.7 & -1.70 & 7.4 & $24\pm17$& 20 \\        {J0925+1403} & 8.12 & $2.0\pm0.2$ & 0.16 &5.0  &-1.92 & 7.2  & $14\pm1.0$ & 18 \\
       J1503+3644 & 7.95 & $1.7\pm0.1$ & 0.22 & 4.9 & -1.52 & 5.8 & $5.9\pm0.4$ & 7.5 \\
        {J0901+2119} & 8.05  & $2.0\pm0.2$ & 0.22  & 8.0  & -1.96 & 2.7  & $7.2\pm0.7$  & 7.4 \\
       J1248+4259 & 7.64 & $1.7\pm0.7$ & 0.25 & 11.8 & -1.95 & 2.2 & $2.9\pm1.2$ & 5.4 \\
       \end{tabular}
    \label{tab:leaker_samp}
\end{table*}
J1503 is just a single example of a LyC emitting galaxy with \mgii\ emission. Of the 11 galaxies with confirmed LyC emission in the \citet{izotov16b}, { \citet{izotov16a},} \citet{izotov18a}, and \citet{izotov18b} samples,  {five} are at redshifts $z > 0.34$ where \mgii\ is redshifted into the SDSS DR15 wavelength regime.  { For four of the five SDSS spectra, we take the literature extinction-corrected $R$ and metallicity measurements, while we downloaded the SDSS data for J1243+4646 and measured the extinction-corrected \mgii\ values ourselves (\autoref{tab:leaker_samp}). An additional five have been observed with \textsc{xshooter} on the VLT and have strong \mgii\ detections \citep{guseva20}. Thus, we have a sample of 10 galaxies with both \mgii\ emission and LyC detections.} 

We use the HST/COS FUV stellar continuum observations to determine the dust attenuation following the methods \citet{chisholm19} and published in \citet{gazagnes20}. These measurements estimate the absolute LyC escape fraction using the \mgii\ doublet and dust attenuation the same way we did in \autoref{esc_1503}. The inferred \fescc\ from the \mgii, the blue squares in \autoref{fig:mg2_esc}, strongly follow the \fescc\ directly observed from the LyC  {with a 3$\sigma$ statistical significance (p-value < 0.001; Pearson's correlation coefficient of 0.88)}.

Similarly, photoionization models from \citet{henry18} and the FUV attenuation can predict \fescc\ in a manner that is complementary to $R$ (gold circles). We use the dust corrected [\ion{O}{iii}]~5007~\AA\ and [\ion{O}{ii}]~3727~\AA\ values from \citet{izotov16b}, \citet{izotov16a}, \citet{izotov18a}, \citet{izotov18b},  {and \citet{guseva20}} to estimate the intrinsic \mgii~2803~\AA\ emission using the photoionization models of \citet{henry18} (their Eq. 2). We then compare the observed \mgii\ emission fluxes to the expected intrinsic \mgii\ emission to determine \fescl\ (\autoref{eq:esc_def}). We then dust correct the \fescl\ ratios in the same exact way as we did when calculating \fescc\ using $R$. The only difference between the blue squares and gold circles in \autoref{fig:mg2_esc} is the determination of $F_{\rm int}$ (either with $R$ or photoionization models).  Comparing the two methods indicates that both reasonably reproduce the absolute LyC escape fraction. While a small sample-size, the sample of \mgii\ and LyC emitters spans the full observed range of \fescc\ from 2-73\%, suggesting that the \mgii\ doublet and the dust attenuation can infer the LyC escape fraction over a wide \fescc.

\subsection{The \mgii\ Doublet May Determine the Escape Fraction of EoR Galaxies}
\label{future}

J1503 is a local LyC emitter with a $z \sim 0.3$. The \mgii\ observations presented here provide a template to constrain the escape fraction of ionizing photons from galaxies within the Epoch of Reionization. In the next 5--10 years the advent of the \textit{James Webb Space Telescope} (JWST) and 20--40~m class Extremely Large Telescopes (ELTs) will provide the necessary collecting area and reduced FIR backgrounds to definitively measure the \mgii\ doublet ratio, the dust attenuation, and the nebular metallicities of these distant galaxies. The methods presented in \autoref{esc_samp} can then determine \fescc\ of galaxies within the Epoch of Reionization. These inferences will be the first  {indirect} constraints on the LyC escape fraction of star-forming galaxies during the Epoch of Reionization, testing crucial galaxy and AGN formation and evolution models.

Is it feasible to empirically infer the escape fractions of ionizing photons using \mgii\ and the next generation of telescopes? Recent observational campaigns with HST have found galaxies within the EoR with F160W (mean wavelength of 1.5~$\mu$m)  magnitudes near 25~mag \citep{livermore17, oesch18}. When the Spitzer imaging is included, the Spectral Energy Distributions require mean  [\ion{O}{iii}]+H$\beta$ equivalent widths near 650~\AA\ and extreme values up to 2600~\AA\ \citep{labbe, smit15, debarros}. Assuming that the stellar continuum is flat in $F_\nu$ and that the H$\beta$ equivalent width is 0.2 times the combined H$\beta$+[\ion{O}{iii}] equivalent width (for a mean H$\beta$ equivalent of 215\AA, consistent with J1503), the average (extreme) of these $z \sim 7$ galaxies would have observed H$\beta$ fluxes of $2\times10^{-18}$ ($8\times10^{-18}$)~erg~s$^{-1}$~cm$^{-2}$. With $F$(\mgii~2796)~$=1/3F$(H$\beta$) from J1503, we expect these bright $z\sim7$ galaxies to have $F$(\mgii~2796)~=~$6\times10^{-19}$ ($3\times10^{-18}$)~erg~s$^{-1}$~cm$^{-2}$. Note that \citet{guseva13} find galaxies in the local Universe with \mgii/H$\beta$ flux ratios as high as 0.6. J1503 could be a conservative estimate. 

The JWST NIRSpec Exposure Time Calculator predicts that a 10~hr observation would detect \mgii~2796~\AA\ with a S/N of 5 (17) per pixel with a resolving power of $\sim2000$, or $\sim150$~\kmsp, at 2.2~$\mu$m using the G235H grating. These spectral resolutions are required to resolve the \mgii\ emission properties and rule out possible impacts of absorption and scattering. Combining the \mgii\ observations with a measured dust attenuation and metallicity will determine the absolute escape fraction of galaxies within the Epoch of Reionzation. A 10~hr integration is large, but the Multi-Object Spectroscopy capabilities of JWST enables for \fescc\ to be simultaneously inferred for multiple $z \sim 6-10$ galaxies. Similarly, future ELTs will have extreme collecting areas coupled with the next generation of Adaptive Optics. This will enable \mgii\ observations in the K-band out to $z \sim 7.5$. From the ground, these observations will determine the $R$ values for galaxies within the Epoch of Reionization which can be combined with restframe FUV continuum observations in the observed Y and H-bands to determine the escape fractions of galaxies within the Epoch of Reionization.

\section{Summary}
Here we presented Keck Cosmic Web Imager (KCWI) spatially-resolved spectroscopic observations of the \mgii~2796, 2803~\AA\ doublet from  {J1503+3644}.  {J1503+3644} is a $z\sim 0.36$ Lyman Continuum emitter with a spatially averaged 6\% escape fraction of ionizing photons. \mgii\ is a resonant transition with an ionization potential of 15~eV that overlaps in ionization state with \hi. Thus, \mgii\ provides an opportunity to determine the properties of the neutral gas that ionizing photons must propagate through to escape the galaxy. These observations are a test-case for indirectly determining the escape fraction of ionizing photons from galaxies within the Epoch of Reionization.

The resonant \mgii\ emission is not spatially-extended beyond the observed stellar continuum (\autoref{fig:sdss_int} and \autoref{fig:spatial_comp}). We find that both the spatially integrated and spatially-resolved \mgii\ line profiles are centered at zero-velocity, well-fit by a single Gaussian  {(\autoref{fig:mg2_fit})}, and marginally broader than the observed nebular \hei\ emission (\autoref{fig:em_comp}). Further, we do not detect any \mgi\ emission or absorption (\autoref{fig:sdss_full}); Mg$^+$ is the only observed ionization state overlapping with \hi. The strong emission profiles  suggest that the resonant \mgii\ emission lines trace the neutral gas in this Lyman Continuum emitter and are not significantly impacted by absorption or scattering. 

We observe spatial variations of the \mgii\ emission doublet $R = F_{2796}/F_{2800}$. The median $R=1.7$, with a spatial range of 0.8--2.7 (\autoref{fig:doub_hist}). To emphasize the spatial variation, we integrated four spatially-distinct regions (e.g. separated by more than the observed seeing) within the galaxy and measured statistically different $R$ values from region-to-region (\autoref{tab:spatially_distinct}). We then created composites of High and Low $R$ spectra by averaging all spaxels above and below the median $R$. High $R$ regions have $R = 1.96\pm0.08$, while lower $R$ spaxels have ratios of $1.43\pm0.07$ (\autoref{fig:rat_comp}). 

In \autoref{theory}, we discussed the physical origin of the $R$ variation. In the optically thin limit, the \mgii\ doublet is collisionally excited and has a constant flux ratio of $R= 2$.  As the \mgii\ optical depth increases, the \mgii~2796~\AA\ line is reduced more than the \mgii~2803~\AA\ transition because it has a larger oscillator strength (\autoref{fig:mock_comp}). Thus, the \mgii\ doublet becomes optically-thin as $R$ approaches 2. As $R$ decreases the transition becomes  more optically thick. This qualitatively describes the observed $R$ spatial variations (\autoref{fig:cf1}).

Using the observed gas-phase metallicity, the $R$ values are inversely related to the \hi\ column density (\nhi; \autoref{eq:nh1_r}), such that higher $R$ corresponds to lower \nhi. Optically thin \mgii\ profiles correspond to regions with low \hi\ column densities that transmit ionizing photons (10$^{15-17}$~cm$^{-2}$, depending on the ISM metallicity; \autoref{fig:cf1}). This ensures that \mgii\ emission sensitively probes the gaseous conditions required for ionizing photons to escape star-forming galaxies. 

The \nhi\ inferred from $R$ in most spaxels (98.4\%) of  {J1503+3644} is optically thin to ionizing photons, with a median value of \nhi~$=~4\pm1\times10^{16}$~cm$^{-2}$ (\autoref{fig:doub_hist}). However, we also observe spaxel-to-spaxel variations in \nhi.  There is a factor of 18 difference between High $R$ regions (which have low \nhi) and Low $R$ regions (which have high \nhi).

The relative escape fraction, or without accounting for dust attenuation, varies from spaxel-to-spaxel by more than a factor of 2 (\autoref{fig:fesc_rel}). To test the spatial variation of the absolute LyC escape fraction, we approximated the dust attenuation in the High and Low $R$ composites using \hei\ emission lines. Regions with high $R$ have lower \nhi\ and lower dust attenuation, while regions with lower $R$ have higher \nhi\ and higher dust attenuation. The escape fraction of Low $R$ regions is 0\%, while High $R$ regions emit 19\% of the ionizing photons. Finally, we used the \nhi\ inferred from the \mgii\ emission and the dust reddening from the FUV stellar continuum to estimate the spatially-integrated absolute LyC escape fraction. The estimated value, $5.9\pm0.4$\%, is consistent with the value directly observed. 

The combination of \mgii\ emission and FUV dust attenuation accurately estimates the LyC escape fraction for the seven LyC emitters from the literature with \mgii\ observations (\autoref{fig:mg2_esc}). The \mgii\ flux is 10-35\% of the H$\beta$ flux, which will be observable from redshift $6-10$ galaxies with $\sim10$~hr integrations with the \textit{James Webb Space Telescope} and Extremely Large Telescopes (\autoref{future}). Combining these observations with dust attenuations and metallicities will empirically determine the escape fraction of galaxies within the Epoch of Reionization to discriminate whether star-forming galaxies emitted a sufficient number of ionizing to reionize the early universe.

\section*{Acknowledgements}
The authors wish to recognize and acknowledge the very significant cultural role and reverence that the summit of Maunakea has always had within the indigenous Hawaiian community.  We are most fortunate to have the opportunity to conduct observations from this mountain.

Support for this work was provided by NASA through the NASA Hubble Fellowship grant \#51432 awarded by the Space Telescope Science Institute, which is operated by the Association of Universities for Research in Astronomy, Inc., for NASA, under contract NAS5-26555.

The authors thank the referee for an extremely helpful and detailed report that dramatically improved the quality of the final manuscript. 

JC thanks Brant Robertson for conversations that greatly improved the scope of this paper.

\section*{Data Availability}
The data underlying this article will be shared on request to the corresponding author.

\label{lastpage}
\bsp	
\bibliographystyle{mnras}
\bibliography{mg2}

\begin{thebibliography}{}
\makeatletter
\relax
\def\mn@urlcharsother{\let\do\@makeother \do\$\do\&\do\#\do\^\do\_\do\%\do\~}
\def\mn@doi{\begingroup\mn@urlcharsother \@ifnextchar [ {\mn@doi@}
  {\mn@doi@[]}}
\def\mn@doi@[#1]#2{\def\@tempa{#1}\ifx\@tempa\@empty \href
  {http://dx.doi.org/#2} {doi:#2}\else \href {http://dx.doi.org/#2} {#1}\fi
  \endgroup}
\def\mn@eprint#1#2{\mn@eprint@#1:#2::\@nil}
\def\mn@eprint@arXiv#1{\href {http://arxiv.org/abs/#1} {{\tt arXiv:#1}}}
\def\mn@eprint@dblp#1{\href {http://dblp.uni-trier.de/rec/bibtex/#1.xml}
  {dblp:#1}}
\def\mn@eprint@#1:#2:#3:#4\@nil{\def\@tempa {#1}\def\@tempb {#2}\def\@tempc
  {#3}\ifx \@tempc \@empty \let \@tempc \@tempb \let \@tempb \@tempa \fi \ifx
  \@tempb \@empty \def\@tempb {arXiv}\fi \@ifundefined
  {mn@eprint@\@tempb}{\@tempb:\@tempc}{\expandafter \expandafter \csname
  mn@eprint@\@tempb\endcsname \expandafter{\@tempc}}}

\bibitem[\protect\citeauthoryear{{Alam} et~al.,}{{Alam} et~al.}{2015}]{sdss}
{Alam} S.,  et~al., 2015, \mn@doi [\apjs] {10.1088/0067-0049/219/1/12}, \href
  {http://adsabs.harvard.edu/abs/2015ApJS..219...12A} {219, 12}

\bibitem[\protect\citeauthoryear{{Alexandroff}, {Heckman}, {Borthakur},
  {Overzier}  \& {Leitherer}}{{Alexandroff} et~al.}{2015}]{alexandroff}
{Alexandroff} R.~M.,  {Heckman} T.~M.,  {Borthakur} S.,  {Overzier} R.,
  {Leitherer} C.,  2015, \mn@doi [\apj] {10.1088/0004-637X/810/2/104}, \href
  {http://adsabs.harvard.edu/abs/2015ApJ...810..104A} {810, 104}

\bibitem[\protect\citeauthoryear{{Asplund}, {Grevesse}, {Sauval}  \&
  {Scott}}{{Asplund} et~al.}{2009}]{asplund}
{Asplund} M.,  {Grevesse} N.,  {Sauval} A.~J.,   {Scott} P.,  2009, \mn@doi
  [\araa] {10.1146/annurev.astro.46.060407.145222}, \href
  {http://adsabs.harvard.edu/abs/2009ARA%26A..47..481A} {47, 481}

\bibitem[\protect\citeauthoryear{{Ba{\~n}ados} et~al.,}{{Ba{\~n}ados}
  et~al.}{2018}]{banados}
{Ba{\~n}ados} E.,  et~al., 2018, \mn@doi [\nat] {10.1038/nature25180}, \href
  {https://ui.adsabs.harvard.edu/abs/2018Natur.553..473B} {553, 473}

\bibitem[\protect\citeauthoryear{{Becker} et~al.,}{{Becker}
  et~al.}{2001}]{becker}
{Becker} R.~H.,  et~al., 2001, \mn@doi [\aj] {10.1086/324231}, \href
  {http://adsabs.harvard.edu/abs/2001AJ....122.2850B} {122, 2850}

\bibitem[\protect\citeauthoryear{{Benjamin}, {Skillman}  \& {Smits}}{{Benjamin}
  et~al.}{1999}]{benjamin99}
{Benjamin} R.~A.,  {Skillman} E.~D.,   {Smits} D.~P.,  1999, \mn@doi [\apj]
  {10.1086/306923}, \href
  {https://ui.adsabs.harvard.edu/abs/1999ApJ...514..307B} {514, 307}

\bibitem[\protect\citeauthoryear{{Berg}, {Chisholm}, {Erb}, {Pogge}, {Henry}
  \& {Olivier}}{{Berg} et~al.}{2019}]{berg19}
{Berg} D.~A.,  {Chisholm} J.,  {Erb} D.~K.,  {Pogge} R.,  {Henry} A.,
  {Olivier} G.~M.,  2019, \mn@doi [\apjl] {10.3847/2041-8213/ab21dc}, \href
  {https://ui.adsabs.harvard.edu/abs/2019ApJ...878L...3B} {878, L3}

\bibitem[\protect\citeauthoryear{{Borthakur}, {Heckman}, {Leitherer}  \&
  {Overzier}}{{Borthakur} et~al.}{2014}]{borthakur}
{Borthakur} S.,  {Heckman} T.~M.,  {Leitherer} C.,   {Overzier} R.~A.,  2014,
  \mn@doi [Science] {10.1126/science.1254214}, \href
  {http://adsabs.harvard.edu/abs/2014Sci...346..216B} {346, 216}

\bibitem[\protect\citeauthoryear{{Burchett}, {Rubin}, {Prochaska}, {Coil},
  {Rickards Vaught}  \& {Hennawi}}{{Burchett} et~al.}{2020}]{burchett20}
{Burchett} J.~N.,  {Rubin} K. H.~R.,  {Prochaska} J.~X.,  {Coil} A.~L.,
  {Rickards Vaught} R.,   {Hennawi} J.~F.,  2020, arXiv e-prints, \href
  {https://ui.adsabs.harvard.edu/abs/2020arXiv200503017B} {p. arXiv:2005.03017}

\bibitem[\protect\citeauthoryear{{Cai} et~al.,}{{Cai} et~al.}{2019}]{cai}
{Cai} Z.,  et~al., 2019, \mn@doi [\apjs] {10.3847/1538-4365/ab4796}, \href
  {https://ui.adsabs.harvard.edu/abs/2019ApJS..245...23C} {245, 23}

\bibitem[\protect\citeauthoryear{{Cantalupo}, {Arrigoni-Battaia}, {Prochaska},
  {Hennawi}  \& {Madau}}{{Cantalupo} et~al.}{2014}]{cantalupo}
{Cantalupo} S.,  {Arrigoni-Battaia} F.,  {Prochaska} J.~X.,  {Hennawi} J.~F.,
  {Madau} P.,  2014, \mn@doi [\nat] {10.1038/nature12898}, \href
  {https://ui.adsabs.harvard.edu/abs/2014Natur.506...63C} {506, 63}

\bibitem[\protect\citeauthoryear{{Cardelli}, {Clayton}  \& {Mathis}}{{Cardelli}
  et~al.}{1989}]{cardelli}
{Cardelli} J.~A.,  {Clayton} G.~C.,   {Mathis} J.~S.,  1989, \mn@doi [\apj]
  {10.1086/167900}, \href {http://adsabs.harvard.edu/abs/1989ApJ...345..245C}
  {345, 245}

\bibitem[\protect\citeauthoryear{{Chisholm} et~al.,}{{Chisholm}
  et~al.}{2018}]{chisholm18}
{Chisholm} J.,  et~al., 2018, \mn@doi [\aap] {10.1051/0004-6361/201832758},
  \href {http://adsabs.harvard.edu/abs/2018A%26A...616A..30C} {616, A30}

\bibitem[\protect\citeauthoryear{{Chisholm}, {Rigby}, {Bayliss}, {Berg},
  {Dahle}, {Gladders}  \& {Sharon}}{{Chisholm} et~al.}{2019}]{chisholm19}
{Chisholm} J.,  {Rigby} J.~R.,  {Bayliss} M.,  {Berg} D.~A.,  {Dahle} H.,
  {Gladders} M.,   {Sharon} K.,  2019, \mn@doi [\apj]
  {10.3847/1538-4357/ab3104}, \href
  {https://ui.adsabs.harvard.edu/abs/2019ApJ...882..182C} {882, 182}

\bibitem[\protect\citeauthoryear{{De Barros}, {Oesch}, {Labb{\'e}}, {Stefanon},
  {Gonz{\'a}lez}, {Smit}, {Bouwens}  \& {Illingworth}}{{De Barros}
  et~al.}{2019}]{debarros}
{De Barros} S.,  {Oesch} P.~A.,  {Labb{\'e}} I.,  {Stefanon} M.,
  {Gonz{\'a}lez} V.,  {Smit} R.,  {Bouwens} R.~J.,   {Illingworth} G.~D.,
  2019, \mn@doi [\mnras] {10.1093/mnras/stz940}, \href
  {https://ui.adsabs.harvard.edu/abs/2019MNRAS.tmp..907D} {}

\bibitem[\protect\citeauthoryear{{Dijkstra}, {Haiman}  \& {Spaans}}{{Dijkstra}
  et~al.}{2006}]{dijkstra}
{Dijkstra} M.,  {Haiman} Z.,   {Spaans} M.,  2006, \mn@doi [\apj]
  {10.1086/506243}, \href {http://adsabs.harvard.edu/abs/2006ApJ...649...14D}
  {649, 14}

\bibitem[\protect\citeauthoryear{{Erb}, {Quider}, {Henry}  \& {Martin}}{{Erb}
  et~al.}{2012}]{erb12}
{Erb} D.~K.,  {Quider} A.~M.,  {Henry} A.~L.,   {Martin} C.~L.,  2012, \mn@doi
  [\apj] {10.1088/0004-637X/759/1/26}, \href
  {http://adsabs.harvard.edu/abs/2012ApJ...759...26E} {759, 26}

\bibitem[\protect\citeauthoryear{{Erb}, {Steidel}  \& {Chen}}{{Erb}
  et~al.}{2018}]{erb18}
{Erb} D.~K.,  {Steidel} C.~C.,   {Chen} Y.,  2018, \mn@doi [\apjl]
  {10.3847/2041-8213/aacff6}, \href
  {https://ui.adsabs.harvard.edu/abs/2018ApJ...862L..10E} {862, L10}

\bibitem[\protect\citeauthoryear{{Fan} et~al.,}{{Fan} et~al.}{2006}]{fan2006}
{Fan} X.,  et~al., 2006, \mn@doi [\aj] {10.1086/504836}, \href
  {http://adsabs.harvard.edu/abs/2006AJ....132..117F} {132, 117}

\bibitem[\protect\citeauthoryear{{Feltre} et~al.,}{{Feltre}
  et~al.}{2018}]{feltre18}
{Feltre} A.,  et~al., 2018, \mn@doi [\aap] {10.1051/0004-6361/201833281}, \href
  {https://ui.adsabs.harvard.edu/abs/2018A%26A...617A..62F} {617, A62}

\bibitem[\protect\citeauthoryear{{Ferland} et~al.,}{{Ferland}
  et~al.}{2017}]{ferland}
{Ferland} G.~J.,  et~al., 2017, \rmxaa, \href
  {https://ui.adsabs.harvard.edu/abs/2017RMxAA..53..385F} {53, 385}

\bibitem[\protect\citeauthoryear{{Finkelstein} et~al.,}{{Finkelstein}
  et~al.}{2019}]{finkelstein19}
{Finkelstein} S.~L.,  et~al., 2019, \mn@doi [\apj] {10.3847/1538-4357/ab1ea8},
  \href {https://ui.adsabs.harvard.edu/abs/2019ApJ...879...36F} {879, 36}

\bibitem[\protect\citeauthoryear{{Finley} et~al.,}{{Finley}
  et~al.}{2017}]{finley}
{Finley} H.,  et~al., 2017, \mn@doi [\aap] {10.1051/0004-6361/201731499}, \href
  {https://ui.adsabs.harvard.edu/abs/2017A&A...608A...7F} {608, A7}

\bibitem[\protect\citeauthoryear{{Fletcher}, {Tang}, {Robertson}, {Nakajima},
  {Ellis}, {Stark}  \& {Inoue}}{{Fletcher} et~al.}{2019}]{fletcher19}
{Fletcher} T.~J.,  {Tang} M.,  {Robertson} B.~E.,  {Nakajima} K.,  {Ellis}
  R.~S.,  {Stark} D.~P.,   {Inoue} A.,  2019, \mn@doi [\apj]
  {10.3847/1538-4357/ab2045}, \href
  {https://ui.adsabs.harvard.edu/abs/2019ApJ...878...87F} {878, 87}

\bibitem[\protect\citeauthoryear{{Gazagnes}, {Chisholm}, {Schaerer},
  {Verhamme}, {Rigby}  \& {Bayliss}}{{Gazagnes} et~al.}{2018}]{gazanges}
{Gazagnes} S.,  {Chisholm} J.,  {Schaerer} D.,  {Verhamme} A.,  {Rigby} J.~R.,
   {Bayliss} M.,  2018, \mn@doi [\aap] {10.1051/0004-6361/201832759}, \href
  {http://adsabs.harvard.edu/abs/2018A%26A...616A..29G} {616, A29}

\bibitem[\protect\citeauthoryear{{Gazagnes}, {Chisholm}, {Schaerer}, {Verhamme}
   \& {Izotov}}{{Gazagnes} et~al.}{2020}]{gazagnes20}
{Gazagnes} S.,  {Chisholm} J.,  {Schaerer} D.,  {Verhamme} A.,   {Izotov} Y.,
  2020, \mn@doi [\aap] {10.1051/0004-6361/202038096}, \href
  {https://ui.adsabs.harvard.edu/abs/2020A&A...639A..85G} {639, A85}

\bibitem[\protect\citeauthoryear{{Genzel} et~al.,}{{Genzel}
  et~al.}{2011}]{genzel11}
{Genzel} R.,  et~al., 2011, \mn@doi [\apj] {10.1088/0004-637X/733/2/101}, \href
  {https://ui.adsabs.harvard.edu/abs/2011ApJ...733..101G} {733, 101}

\bibitem[\protect\citeauthoryear{{Giallongo} et~al.,}{{Giallongo}
  et~al.}{2015}]{Giallongo15}
{Giallongo} E.,  et~al., 2015, \mn@doi [\aap] {10.1051/0004-6361/201425334},
  \href {http://adsabs.harvard.edu/abs/2015A\%26A...578A..83G} {578, A83}

\bibitem[\protect\citeauthoryear{{Girard} et~al.,}{{Girard}
  et~al.}{2018}]{girard18}
{Girard} M.,  et~al., 2018, \mn@doi [\aap] {10.1051/0004-6361/201731988}, \href
  {https://ui.adsabs.harvard.edu/abs/2018A&A...613A..72G} {613, A72}

\bibitem[\protect\citeauthoryear{{Girard}, {Dessauges-Zavadsky}, {Combes},
  {Chisholm}, {Patr{\'\i}cio}, {Richard}  \& {Schaerer}}{{Girard}
  et~al.}{2019}]{girard}
{Girard} M.,  {Dessauges-Zavadsky} M.,  {Combes} F.,  {Chisholm} J.,
  {Patr{\'\i}cio} V.,  {Richard} J.,   {Schaerer} D.,  2019, \mn@doi [\aap]
  {10.1051/0004-6361/201935896}, \href
  {https://ui.adsabs.harvard.edu/abs/2019A&A...631A..91G} {631, A91}

\bibitem[\protect\citeauthoryear{{Grazian} et~al.,}{{Grazian}
  et~al.}{2016}]{Grazian16}
{Grazian} A.,  et~al., 2016, \mn@doi [\aap] {10.1051/0004-6361/201526396},
  \href {http://cdsads.u-strasbg.fr/abs/2016A\%26A...585A..48G} {585, A48}

\bibitem[\protect\citeauthoryear{{Green} et~al.,}{{Green} et~al.}{2012}]{cos}
{Green} J.~C.,  et~al., 2012, \mn@doi [\apj] {10.1088/0004-637X/744/1/60},
  \href {http://adsabs.harvard.edu/abs/2012ApJ...744...60G} {744, 60}

\bibitem[\protect\citeauthoryear{{Green} et~al.,}{{Green}
  et~al.}{2015}]{green15}
{Green} G.~M.,  et~al., 2015, \mn@doi [\apj] {10.1088/0004-637X/810/1/25},
  \href {http://adsabs.harvard.edu/abs/2015ApJ...810...25G} {810, 25}

\bibitem[\protect\citeauthoryear{{Grimes} et~al.,}{{Grimes}
  et~al.}{2009}]{grimes09}
{Grimes} J.~P.,  et~al., 2009, \mn@doi [\apjs] {10.1088/0067-0049/181/1/272},
  \href {http://adsabs.harvard.edu/abs/2009ApJS..181..272G} {181, 272}

\bibitem[\protect\citeauthoryear{{Gronke}, {Bull}  \& {Dijkstra}}{{Gronke}
  et~al.}{2015}]{gronke}
{Gronke} M.,  {Bull} P.,   {Dijkstra} M.,  2015, \mn@doi [\apj]
  {10.1088/0004-637X/812/2/123}, \href
  {https://ui.adsabs.harvard.edu/abs/2015ApJ...812..123G} {812, 123}

\bibitem[\protect\citeauthoryear{{Guseva}, {Izotov}, {Fricke}  \&
  {Henkel}}{{Guseva} et~al.}{2013}]{guseva13}
{Guseva} N.~G.,  {Izotov} Y.~I.,  {Fricke} K.~J.,   {Henkel} C.,  2013, \mn@doi
  [\aap] {10.1051/0004-6361/201221010}, \href
  {https://ui.adsabs.harvard.edu/abs/2013A%26A...555A..90G} {555, A90}

\bibitem[\protect\citeauthoryear{{Guseva}, {Izotov}, {Fricke}  \&
  {Henkel}}{{Guseva} et~al.}{2019}]{guseva19}
{Guseva} N.~G.,  {Izotov} Y.~I.,  {Fricke} K.~J.,   {Henkel} C.,  2019, \mn@doi
  [\aap] {10.1051/0004-6361/201834935}, \href
  {https://ui.adsabs.harvard.edu/abs/2019A%26A...624A..21G} {624, A21}

\bibitem[\protect\citeauthoryear{{Guseva} et~al.,}{{Guseva}
  et~al.}{2020}]{guseva20}
{Guseva} N.~G.,  et~al., 2020, arXiv e-prints, \href
  {https://ui.adsabs.harvard.edu/abs/2020arXiv200711977G} {p. arXiv:2007.11977}

\bibitem[\protect\citeauthoryear{{Hayes} et~al.,}{{Hayes}
  et~al.}{2014}]{hayes14}
{Hayes} M.,  et~al., 2014, \mn@doi [\apj] {10.1088/0004-637X/782/1/6}, \href
  {http://adsabs.harvard.edu/abs/2014ApJ...782....6H} {782, 6}

\bibitem[\protect\citeauthoryear{{Heckman} et~al.,}{{Heckman}
  et~al.}{2011}]{heckman2011}
{Heckman} T.~M.,  et~al., 2011, \mn@doi [\apj] {10.1088/0004-637X/730/1/5},
  \href {http://adsabs.harvard.edu/abs/2011ApJ...730....5H} {730, 5}

\bibitem[\protect\citeauthoryear{{Henry}, {Scarlata}, {Martin}  \&
  {Erb}}{{Henry} et~al.}{2015}]{henry}
{Henry} A.,  {Scarlata} C.,  {Martin} C.~L.,   {Erb} D.,  2015, \mn@doi [\apj]
  {10.1088/0004-637X/809/1/19}, \href
  {http://adsabs.harvard.edu/abs/2015ApJ...809...19H} {809, 19}

\bibitem[\protect\citeauthoryear{{Henry}, {Berg}, {Scarlata}, {Verhamme}  \&
  {Erb}}{{Henry} et~al.}{2018}]{henry18}
{Henry} A.,  {Berg} D.~A.,  {Scarlata} C.,  {Verhamme} A.,   {Erb} D.,  2018,
  \mn@doi [\apj] {10.3847/1538-4357/aab099}, \href
  {https://ui.adsabs.harvard.edu/abs/2018ApJ...855...96H} {855, 96}

\bibitem[\protect\citeauthoryear{{Hopkins}, {Hernquist}, {Cox}  \& {Kere{\v
  s}}}{{Hopkins} et~al.}{2008}]{hopkins08}
{Hopkins} P.~F.,  {Hernquist} L.,  {Cox} T.~J.,   {Kere{\v s}} D.,  2008,
  \mn@doi [\apjs] {10.1086/524362}, \href
  {http://adsabs.harvard.edu/abs/2008ApJS..175..356H} {175, 356}

\bibitem[\protect\citeauthoryear{{Hunter} et~al.,}{{Hunter}
  et~al.}{2012}]{hunter}
{Hunter} D.~A.,  et~al., 2012, \mn@doi [\aj] {10.1088/0004-6256/144/5/134},
  \href {https://ui.adsabs.harvard.edu/abs/2012AJ....144..134H} {144, 134}

\bibitem[\protect\citeauthoryear{{Izotov}, {Schaerer}, {Thuan}, {Worseck},
  {Guseva}, {Orlitov{\'a}}  \& {Verhamme}}{{Izotov} et~al.}{2016a}]{izotov16b}
{Izotov} Y.~I.,  {Schaerer} D.,  {Thuan} T.~X.,  {Worseck} G.,  {Guseva} N.~G.,
   {Orlitov{\'a}} I.,   {Verhamme} A.,  2016a, \mn@doi [\mnras]
  {10.1093/mnras/stw1205}, \href
  {http://adsabs.harvard.edu/abs/2016MNRAS.461.3683I} {461, 3683}

\bibitem[\protect\citeauthoryear{{Izotov}, {Orlitov{\'a}}, {Schaerer}, {Thuan},
  {Verhamme}, {Guseva}  \& {Worseck}}{{Izotov} et~al.}{2016b}]{izotov16a}
{Izotov} Y.~I.,  {Orlitov{\'a}} I.,  {Schaerer} D.,  {Thuan} T.~X.,  {Verhamme}
  A.,  {Guseva} N.~G.,   {Worseck} G.,  2016b, \mn@doi [\nat]
  {10.1038/nature16456}, \href
  {http://adsabs.harvard.edu/abs/2016Natur.529..178I} {529, 178}

\bibitem[\protect\citeauthoryear{{Izotov}, {Schaerer}, {Worseck}, {Guseva},
  {Thuan}, {Verhamme}, {Orlitov{\'a}}  \& {Fricke}}{{Izotov}
  et~al.}{2018a}]{izotov18a}
{Izotov} Y.~I.,  {Schaerer} D.,  {Worseck} G.,  {Guseva} N.~G.,  {Thuan} T.~X.,
   {Verhamme} A.,  {Orlitov{\'a}} I.,   {Fricke} K.~J.,  2018a, \mn@doi
  [\mnras] {10.1093/mnras/stx3115}, \href
  {http://adsabs.harvard.edu/abs/2018MNRAS.474.4514I} {474, 4514}

\bibitem[\protect\citeauthoryear{{Izotov}, {Worseck}, {Schaerer}, {Guseva},
  {Thuan}, {Fricke}  \& {Orlitov{\'a}}}{{Izotov} et~al.}{2018b}]{izotov18b}
{Izotov} Y.~I.,  {Worseck} G.,  {Schaerer} D.,  {Guseva} N.~G.,  {Thuan} T.~X.,
   {Fricke} A. V.,   {Orlitov{\'a}} I.,  2018b, \mn@doi [\mnras]
  {10.1093/mnras/sty1378}, \href
  {http://adsabs.harvard.edu/abs/2018MNRAS.478.4851I} {478, 4851}

\bibitem[\protect\citeauthoryear{{Jaskot}, {Dowd}, {Oey}, {Scarlata}  \&
  {McKinney}}{{Jaskot} et~al.}{2019}]{jaskot19}
{Jaskot} A.~E.,  {Dowd} T.,  {Oey} M.~S.,  {Scarlata} C.,   {McKinney} J.,
  2019, \mn@doi [\apj] {10.3847/1538-4357/ab3d3b}, \href
  {https://ui.adsabs.harvard.edu/abs/2019ApJ...885...96J} {885, 96}

\bibitem[\protect\citeauthoryear{{Jenkins}}{{Jenkins}}{2009}]{jenkins}
{Jenkins} E.~B.,  2009, \mn@doi [\apj] {10.1088/0004-637X/700/2/1299}, \href
  {http://adsabs.harvard.edu/abs/2009ApJ...700.1299J} {700, 1299}

\bibitem[\protect\citeauthoryear{{Johnson}}{{Johnson}}{2019}]{johnson19}
{Johnson} J.~A.,  2019, \mn@doi [Science] {10.1126/science.aau9540}, \href
  {https://ui.adsabs.harvard.edu/abs/2019Sci...363..474J} {363, 474}

\bibitem[\protect\citeauthoryear{{Kakiichi} \& {Gronke}}{{Kakiichi} \&
  {Gronke}}{2019}]{koki}
{Kakiichi} K.,  {Gronke} M.,  2019, arXiv e-prints, \href
  {https://ui.adsabs.harvard.edu/abs/2019arXiv190502480K} {p. arXiv:1905.02480}

\bibitem[\protect\citeauthoryear{Kramida, {Yu.~Ralchenko}, Reader  \& {and NIST
  ASD Team}}{Kramida et~al.}{2018}]{nist}
Kramida A.,  {Yu.~Ralchenko} Reader J.,   {and NIST ASD Team} 2018, {NIST
  Atomic Spectra Database (ver. 5.6.1), [Online]. Available:
  {\tt{https://physics.nist.gov/asd}} [2019, September 3]. National Institute
  of Standards and Technology, Gaithersburg, MD.}

\bibitem[\protect\citeauthoryear{{Labb{\'e}} et~al.,}{{Labb{\'e}}
  et~al.}{2013}]{labbe}
{Labb{\'e}} I.,  et~al., 2013, \mn@doi [\apjl] {10.1088/2041-8205/777/2/L19},
  \href {https://ui.adsabs.harvard.edu/abs/2013ApJ...777L..19L} {777, L19}

\bibitem[\protect\citeauthoryear{{Leethochawalit}, {Jones}, {Ellis}, {Stark},
  {Richard}, {Zitrin}  \& {Auger}}{{Leethochawalit}
  et~al.}{2016}]{leethocawalit16}
{Leethochawalit} N.,  {Jones} T.~A.,  {Ellis} R.~S.,  {Stark} D.~P.,  {Richard}
  J.,  {Zitrin} A.,   {Auger} M.,  2016, \mn@doi [\apj]
  {10.3847/0004-637X/820/2/84}, \href
  {https://ui.adsabs.harvard.edu/abs/2016ApJ...820...84L} {820, 84}

\bibitem[\protect\citeauthoryear{{Leitet}, {Bergvall}, {Piskunov}  \&
  {Andersson}}{{Leitet} et~al.}{2011}]{leitet11}
{Leitet} E.,  {Bergvall} N.,  {Piskunov} N.,   {Andersson} B.-G.,  2011,
  \mn@doi [\aap] {10.1051/0004-6361/201015654}, \href
  {http://adsabs.harvard.edu/abs/2011A%26A...532A.107L} {532, A107}

\bibitem[\protect\citeauthoryear{{Leitherer}, {Ekstr{\"o}m}, {Meynet},
  {Schaerer}, {Agienko}  \& {Levesque}}{{Leitherer} et~al.}{2014}]{leitherer14}
{Leitherer} C.,  {Ekstr{\"o}m} S.,  {Meynet} G.,  {Schaerer} D.,  {Agienko}
  K.~B.,   {Levesque} E.~M.,  2014, \mn@doi [\apjs]
  {10.1088/0067-0049/212/1/14}, \href
  {http://adsabs.harvard.edu/abs/2014ApJS..212...14L} {212, 14}

\bibitem[\protect\citeauthoryear{{Leitherer}, {Hernandez}, {Lee}  \&
  {Oey}}{{Leitherer} et~al.}{2016}]{leitherer16}
{Leitherer} C.,  {Hernandez} S.,  {Lee} J.~C.,   {Oey} M.~S.,  2016, \mn@doi
  [\apj] {10.3847/0004-637X/823/1/64}, \href
  {http://adsabs.harvard.edu/abs/2016ApJ...823...64L} {823, 64}

\bibitem[\protect\citeauthoryear{{Livermore}, {Finkelstein}  \&
  {Lotz}}{{Livermore} et~al.}{2017}]{livermore17}
{Livermore} R.~C.,  {Finkelstein} S.~L.,   {Lotz} J.~M.,  2017, \mn@doi [\apj]
  {10.3847/1538-4357/835/2/113}, \href
  {http://adsabs.harvard.edu/abs/2017ApJ...835..113L} {835, 113}

\bibitem[\protect\citeauthoryear{{Markwardt}}{{Markwardt}}{2009}]{mpfit}
{Markwardt} C.~B.,  2009, in {Bohlender} D.~A.,  {Durand} D.,   {Dowler} P.,
  eds,  Astronomical Society of the Pacific Conference Series Vol. 411,
  Astronomical Data Analysis Software and Systems XVIII. p.~251 (\mn@eprint
  {arXiv} {0902.2850})

\bibitem[\protect\citeauthoryear{{Martin}, {Shapley}, {Coil}, {Kornei},
  {Bundy}, {Weiner}, {Noeske}  \& {Schiminovich}}{{Martin}
  et~al.}{2012}]{martin12}
{Martin} C.~L.,  {Shapley} A.~E.,  {Coil} A.~L.,  {Kornei} K.~A.,  {Bundy} K.,
  {Weiner} B.~J.,  {Noeske} K.~G.,   {Schiminovich} D.,  2012, \mn@doi [\apj]
  {10.1088/0004-637X/760/2/127}, \href
  {http://adsabs.harvard.edu/abs/2012ApJ...760..127M} {760, 127}

\bibitem[\protect\citeauthoryear{{Martin}, {Shapley}, {Coil}, {Kornei},
  {Murray}  \& {Pancoast}}{{Martin} et~al.}{2013}]{martin13}
{Martin} C.~L.,  {Shapley} A.~E.,  {Coil} A.~L.,  {Kornei} K.~A.,  {Murray} N.,
    {Pancoast} A.,  2013, \mn@doi [\apj] {10.1088/0004-637X/770/1/41}, \href
  {https://ui.adsabs.harvard.edu/abs/2013ApJ...770...41M} {770, 41}

\bibitem[\protect\citeauthoryear{{Matsuoka} et~al.,}{{Matsuoka}
  et~al.}{2018}]{matsuoka18}
{Matsuoka} Y.,  et~al., 2018, \mn@doi [\apj] {10.3847/1538-4357/aaee7a}, \href
  {https://ui.adsabs.harvard.edu/abs/2018ApJ...869..150M} {869, 150}

\bibitem[\protect\citeauthoryear{{Matsuoka} et~al.,}{{Matsuoka}
  et~al.}{2019}]{matsuoka19}
{Matsuoka} Y.,  et~al., 2019, \mn@doi [\apjl] {10.3847/2041-8213/ab0216}, \href
  {https://ui.adsabs.harvard.edu/abs/2019ApJ...872L...2M} {872, L2}

\bibitem[\protect\citeauthoryear{{McKinney}, {Jaskot}, {Oey}, {Yun}, {Dowd}  \&
  {Lowenthal}}{{McKinney} et~al.}{2019}]{Mckinney}
{McKinney} J.~H.,  {Jaskot} A.~E.,  {Oey} M.~S.,  {Yun} M.~S.,  {Dowd} T.,
  {Lowenthal} J.~D.,  2019, \mn@doi [\apj] {10.3847/1538-4357/ab08eb}, \href
  {https://ui.adsabs.harvard.edu/abs/2019ApJ...874...52M} {874, 52}

\bibitem[\protect\citeauthoryear{{McMullin}, {Waters}, {Schiebel}, {Young}  \&
  {Golap}}{{McMullin} et~al.}{2007}]{casa}
{McMullin} J.~P.,  {Waters} B.,  {Schiebel} D.,  {Young} W.,   {Golap} K.,
  2007, in {Shaw} R.~A.,  {Hill} F.,   {Bell} D.~J.,  eds,  Astronomical
  Society of the Pacific Conference Series Vol. 376, Astronomical Data Analysis
  Software and Systems XVI. p.~127

\bibitem[\protect\citeauthoryear{{Mendoza}}{{Mendoza}}{1981}]{mendoza81}
{Mendoza} C.,  1981, \mn@doi [Journal of Physics B Atomic Molecular Physics]
  {10.1088/0022-3700/14/14/018}, \href
  {https://ui.adsabs.harvard.edu/abs/1981JPhB...14.2465M} {14, 2465}

\bibitem[\protect\citeauthoryear{{Michel-Dansac}, {Blaizot}, {Garel},
  {Verhamme}, {Kimm}  \& {Trebitsch}}{{Michel-Dansac}
  et~al.}{2020}]{Michel-Dansac}
{Michel-Dansac} L.,  {Blaizot} J.,  {Garel} T.,  {Verhamme} A.,  {Kimm} T.,
  {Trebitsch} M.,  2020, \mn@doi [\aap] {10.1051/0004-6361/201834961}, \href
  {https://ui.adsabs.harvard.edu/abs/2020A&A...635A.154M} {635, A154}

\bibitem[\protect\citeauthoryear{{Micheva}, {Oey}, {Keenan}, {Jaskot}  \&
  {James}}{{Micheva} et~al.}{2018}]{Micheva}
{Micheva} G.,  {Oey} M.~S.,  {Keenan} R.~P.,  {Jaskot} A.~E.,   {James} B.~L.,
  2018, \mn@doi [\apj] {10.3847/1538-4357/aae372}, \href
  {https://ui.adsabs.harvard.edu/abs/2018ApJ...867....2M} {867, 2}

\bibitem[\protect\citeauthoryear{{Micheva}, {Christian Herenz}, {Roth},
  {{\"O}stlin}  \& {Girichidis}}{{Micheva} et~al.}{2019}]{Micheva19}
{Micheva} G.,  {Christian Herenz} E.,  {Roth} M.~M.,  {{\"O}stlin} G.,
  {Girichidis} P.,  2019, \mn@doi [\aap] {10.1051/0004-6361/201834838}, \href
  {https://ui.adsabs.harvard.edu/abs/2019A%26A...623A.145M} {623, A145}

\bibitem[\protect\citeauthoryear{{Morrissey} et~al.,}{{Morrissey}
  et~al.}{2018}]{kcwi}
{Morrissey} P.,  et~al., 2018, \mn@doi [\apj] {10.3847/1538-4357/aad597}, \href
  {https://ui.adsabs.harvard.edu/abs/2018ApJ...864...93M} {864, 93}

\bibitem[\protect\citeauthoryear{{Naidu}, {Forrest}, {Oesch}, {Tran}  \&
  {Holden}}{{Naidu} et~al.}{2018}]{naidu18}
{Naidu} R.~P.,  {Forrest} B.,  {Oesch} P.~A.,  {Tran} K.-V.~H.,   {Holden}
  B.~P.,  2018, \mn@doi [\mnras] {10.1093/mnras/sty961}, \href
  {https://ui.adsabs.harvard.edu/abs/2018MNRAS.478..791N} {478, 791}

\bibitem[\protect\citeauthoryear{{Naidu}, {Tacchella}, {Mason}, {Bose}, {Oesch}
   \& {Conroy}}{{Naidu} et~al.}{2020}]{naidu19}
{Naidu} R.~P.,  {Tacchella} S.,  {Mason} C.~A.,  {Bose} S.,  {Oesch} P.~A.,
  {Conroy} C.,  2020, \mn@doi [\apj] {10.3847/1538-4357/ab7cc9}, \href
  {https://ui.adsabs.harvard.edu/abs/2020ApJ...892..109N} {892, 109}

\bibitem[\protect\citeauthoryear{{Neufeld}}{{Neufeld}}{1990}]{neufeld}
{Neufeld} D.~A.,  1990, \mn@doi [\apj] {10.1086/168375}, \href
  {https://ui.adsabs.harvard.edu/abs/1990ApJ...350..216N} {350, 216}

\bibitem[\protect\citeauthoryear{{Oesch}, {Bouwens}, {Illingworth}, {Labb{\'e}}
   \& {Stefanon}}{{Oesch} et~al.}{2018}]{oesch18}
{Oesch} P.~A.,  {Bouwens} R.~J.,  {Illingworth} G.~D.,  {Labb{\'e}} I.,
  {Stefanon} M.,  2018, \mn@doi [\apj] {10.3847/1538-4357/aab03f}, \href
  {http://adsabs.harvard.edu/abs/2018ApJ...855..105O} {855, 105}

\bibitem[\protect\citeauthoryear{{Onoue} et~al.,}{{Onoue} et~al.}{2017}]{onoue}
{Onoue} M.,  et~al., 2017, \mn@doi [\apjl] {10.3847/2041-8213/aa8cc6}, \href
  {http://adsabs.harvard.edu/abs/2017ApJ...847L..15O} {847, L15}

\bibitem[\protect\citeauthoryear{{Orlitov{\'a}}, {Verhamme}, {Henry},
  {Scarlata}, {Jaskot}, {Oey}  \& {Schaerer}}{{Orlitov{\'a}}
  et~al.}{2018}]{orlitova}
{Orlitov{\'a}} I.,  {Verhamme} A.,  {Henry} A.,  {Scarlata} C.,  {Jaskot} A.,
  {Oey} M.~S.,   {Schaerer} D.,  2018, \mn@doi [\aap]
  {10.1051/0004-6361/201732478}, \href
  {https://ui.adsabs.harvard.edu/abs/2018A%26A...616A..60O} {616, A60}

\bibitem[\protect\citeauthoryear{{{\"O}stlin} et~al.,}{{{\"O}stlin}
  et~al.}{2014}]{ostlin14}
{{\"O}stlin} G.,  et~al., 2014, \mn@doi [\apj] {10.1088/0004-637X/797/1/11},
  \href {http://adsabs.harvard.edu/abs/2014ApJ...797...11O} {797, 11}

\bibitem[\protect\citeauthoryear{{Ott} et~al.,}{{Ott} et~al.}{2012}]{ott}
{Ott} J.,  et~al., 2012, \mn@doi [\aj] {10.1088/0004-6256/144/4/123}, \href
  {https://ui.adsabs.harvard.edu/abs/2012AJ....144..123O} {144, 123}

\bibitem[\protect\citeauthoryear{{Ouchi} et~al.,}{{Ouchi}
  et~al.}{2009a}]{ouchi}
{Ouchi} M.,  et~al., 2009a, \mn@doi [\apj] {10.1088/0004-637X/696/2/1164},
  \href {https://ui.adsabs.harvard.edu/abs/2009ApJ...696.1164O} {696, 1164}

\bibitem[\protect\citeauthoryear{{Ouchi} et~al.,}{{Ouchi}
  et~al.}{2009b}]{ouchi09}
{Ouchi} M.,  et~al., 2009b, \mn@doi [\apj] {10.1088/0004-637X/706/2/1136},
  \href {http://adsabs.harvard.edu/abs/2009ApJ...706.1136O} {706, 1136}

\bibitem[\protect\citeauthoryear{{Pardy} et~al.,}{{Pardy} et~al.}{2014}]{pardy}
{Pardy} S.~A.,  et~al., 2014, \mn@doi [\apj] {10.1088/0004-637X/794/2/101},
  \href {http://adsabs.harvard.edu/abs/2014ApJ...794..101P} {794, 101}

\bibitem[\protect\citeauthoryear{{Prochaska}, {Kasen}  \& {Rubin}}{{Prochaska}
  et~al.}{2011}]{prochaska2011}
{Prochaska} J.~X.,  {Kasen} D.,   {Rubin} K.,  2011, \mn@doi [\apj]
  {10.1088/0004-637X/734/1/24}, \href
  {http://adsabs.harvard.edu/abs/2011ApJ...734...24P} {734, 24}

\bibitem[\protect\citeauthoryear{{Reddy}, {Steidel}, {Pettini}  \&
  {Bogosavljevi{\'c}}}{{Reddy} et~al.}{2016}]{reddy16}
{Reddy} N.~A.,  {Steidel} C.~C.,  {Pettini} M.,   {Bogosavljevi{\'c}} M.,
  2016, \mn@doi [\apj] {10.3847/0004-637X/828/2/107}, \href
  {http://adsabs.harvard.edu/abs/2016ApJ...828..107R} {828, 107}

\bibitem[\protect\citeauthoryear{{Ricci}, {Marchesi}, {Shankar}, {La Franca}
  \& {Civano}}{{Ricci} et~al.}{2017}]{ricci}
{Ricci} F.,  {Marchesi} S.,  {Shankar} F.,  {La Franca} F.,   {Civano} F.,
  2017, \mn@doi [\mnras] {10.1093/mnras/stw2909}, \href
  {http://adsabs.harvard.edu/abs/2017MNRAS.465.1915R} {465, 1915}

\bibitem[\protect\citeauthoryear{{Rivera-Thorsen} et~al.,}{{Rivera-Thorsen}
  et~al.}{2019}]{rivera19}
{Rivera-Thorsen} T.~E.,  et~al., 2019, \mn@doi [Science]
  {10.1126/science.aaw0978}, \href
  {https://ui.adsabs.harvard.edu/abs/2019Sci...366..738R} {366, 738}

\bibitem[\protect\citeauthoryear{{Robertson} et~al.,}{{Robertson}
  et~al.}{2013}]{robertson13}
{Robertson} B.~E.,  et~al., 2013, \mn@doi [\apj] {10.1088/0004-637X/768/1/71},
  \href {http://adsabs.harvard.edu/abs/2013ApJ...768...71R} {768, 71}

\bibitem[\protect\citeauthoryear{{Robertson}, {Ellis}, {Furlanetto}  \&
  {Dunlop}}{{Robertson} et~al.}{2015}]{robertson15}
{Robertson} B.~E.,  {Ellis} R.~S.,  {Furlanetto} S.~R.,   {Dunlop} J.~S.,
  2015, \mn@doi [\apjl] {10.1088/2041-8205/802/2/L19}, \href
  {http://adsabs.harvard.edu/abs/2015ApJ...802L..19R} {802, L19}

\bibitem[\protect\citeauthoryear{{Rubin}, {Prochaska}, {Koo}, {Phillips},
  {Martin}  \& {Winstrom}}{{Rubin} et~al.}{2014}]{rubin13}
{Rubin} K.~H.~R.,  {Prochaska} J.~X.,  {Koo} D.~C.,  {Phillips} A.~C.,
  {Martin} C.~L.,   {Winstrom} L.~O.,  2014, \mn@doi [\apj]
  {10.1088/0004-637X/794/2/156}, \href
  {http://adsabs.harvard.edu/abs/2014ApJ...794..156R} {794, 156}

\bibitem[\protect\citeauthoryear{{Rupke} et~al.,}{{Rupke}
  et~al.}{2019}]{rupke19}
{Rupke} D. S.~N.,  et~al., 2019, \mn@doi [\nat] {10.1038/s41586-019-1686-1},
  \href {https://ui.adsabs.harvard.edu/abs/2019Natur.574..643R} {574, 643}

\bibitem[\protect\citeauthoryear{{Scarlata} \& {Panagia}}{{Scarlata} \&
  {Panagia}}{2015}]{scarlata}
{Scarlata} C.,  {Panagia} N.,  2015, \mn@doi [\apj]
  {10.1088/0004-637X/801/1/43}, \href
  {http://adsabs.harvard.edu/abs/2015ApJ...801...43S} {801, 43}

\bibitem[\protect\citeauthoryear{{Schaerer} \& {de Barros}}{{Schaerer} \& {de
  Barros}}{2010}]{schaerer10}
{Schaerer} D.,  {de Barros} S.,  2010, \mn@doi [\aap]
  {10.1051/0004-6361/200913946}, \href
  {http://adsabs.harvard.edu/abs/2010A%26A...515A..73S} {515, A73}

\bibitem[\protect\citeauthoryear{{Shapley}, {Steidel}, {Strom},
  {Bogosavljevi{\'c}}, {Reddy}, {Siana}, {Mostardi}  \& {Rudie}}{{Shapley}
  et~al.}{2016}]{shapley16}
{Shapley} A.~E.,  {Steidel} C.~C.,  {Strom} A.~L.,  {Bogosavljevi{\'c}} M.,
  {Reddy} N.~A.,  {Siana} B.,  {Mostardi} R.~E.,   {Rudie} G.~C.,  2016,
  \mn@doi [\apjl] {10.3847/2041-8205/826/2/L24}, \href
  {http://adsabs.harvard.edu/abs/2016ApJ...826L..24S} {826, L24}

\bibitem[\protect\citeauthoryear{{Shen}, {Hopkins}, {Faucher-Gigu{\`e}re},
  {Alexander}, {Richards}, {Ross}  \& {Hickox}}{{Shen} et~al.}{2020}]{shen20}
{Shen} X.,  {Hopkins} P.~F.,  {Faucher-Gigu{\`e}re} C.-A.,  {Alexander} D.~M.,
  {Richards} G.~T.,  {Ross} N.~P.,   {Hickox} R.~C.,  2020, \mn@doi [\mnras]
  {10.1093/mnras/staa1381}, \href
  {https://ui.adsabs.harvard.edu/abs/2020MNRAS.495.3252S} {495, 3252}

\bibitem[\protect\citeauthoryear{{Sigut} \& {Pradhan}}{{Sigut} \&
  {Pradhan}}{1995}]{sigut}
{Sigut} T.~A.~A.,  {Pradhan} A.~K.,  1995, \mn@doi [Journal of Physics B Atomic
  Molecular Physics] {10.1088/0953-4075/28/22/018}, \href
  {https://ui.adsabs.harvard.edu/abs/1995JPhB...28.4879S} {28, 4879}

\bibitem[\protect\citeauthoryear{{Smit} et~al.,}{{Smit} et~al.}{2015}]{smit15}
{Smit} R.,  et~al., 2015, \mn@doi [\apj] {10.1088/0004-637X/801/2/122}, \href
  {https://ui.adsabs.harvard.edu/abs/2015ApJ...801..122S} {801, 122}

\bibitem[\protect\citeauthoryear{{Smit} et~al.,}{{Smit} et~al.}{2018}]{smit18}
{Smit} R.,  et~al., 2018, \mn@doi [\nat] {10.1038/nature24631}, \href
  {https://ui.adsabs.harvard.edu/abs/2018Natur.553..178S} {553, 178}

\bibitem[\protect\citeauthoryear{{Sobral}, {Matthee}, {Darvish}, {Schaerer},
  {Mobasher}, {R{\"o}ttgering}, {Santos}  \& {Hemmati}}{{Sobral}
  et~al.}{2015}]{sobral}
{Sobral} D.,  {Matthee} J.,  {Darvish} B.,  {Schaerer} D.,  {Mobasher} B.,
  {R{\"o}ttgering} H. J.~A.,  {Santos} S.,   {Hemmati} S.,  2015, \mn@doi
  [\apj] {10.1088/0004-637X/808/2/139}, \href
  {https://ui.adsabs.harvard.edu/abs/2015ApJ...808..139S} {808, 139}

\bibitem[\protect\citeauthoryear{{Steidel}, {Bogosavljevi{\'c}}, {Shapley},
  {Kollmeier}, {Reddy}, {Erb}  \& {Pettini}}{{Steidel}
  et~al.}{2011}]{steidel11}
{Steidel} C.~C.,  {Bogosavljevi{\'c}} M.,  {Shapley} A.~E.,  {Kollmeier} J.~A.,
   {Reddy} N.~A.,  {Erb} D.~K.,   {Pettini} M.,  2011, \mn@doi [\apj]
  {10.1088/0004-637X/736/2/160}, \href
  {https://ui.adsabs.harvard.edu/abs/2011ApJ...736..160S} {736, 160}

\bibitem[\protect\citeauthoryear{{Steidel}, {Bogosavljevi{\'c}}, {Shapley},
  {Reddy}, {Rudie}, {Pettini}, {Trainor}  \& {Strom}}{{Steidel}
  et~al.}{2018}]{steidel18}
{Steidel} C.~C.,  {Bogosavljevi{\'c}} M.,  {Shapley} A.~E.,  {Reddy} N.~A.,
  {Rudie} G.~C.,  {Pettini} M.,  {Trainor} R.~F.,   {Strom} A.~L.,  2018,
  \mn@doi [\apj] {10.3847/1538-4357/aaed28}, \href
  {http://adsabs.harvard.edu/abs/2018ApJ...869..123S} {869, 123}

\bibitem[\protect\citeauthoryear{{Swinbank}, {Sobral}, {Smail}, {Geach},
  {Best}, {McCarthy}, {Crain}  \& {Theuns}}{{Swinbank}
  et~al.}{2012}]{swinbank12}
{Swinbank} A.~M.,  {Sobral} D.,  {Smail} I.,  {Geach} J.~E.,  {Best} P.~N.,
  {McCarthy} I.~G.,  {Crain} R.~A.,   {Theuns} T.,  2012, \mn@doi [\mnras]
  {10.1111/j.1365-2966.2012.21774.x}, \href
  {https://ui.adsabs.harvard.edu/abs/2012MNRAS.426..935S} {426, 935}

\bibitem[\protect\citeauthoryear{{Swinbank} et~al.,}{{Swinbank}
  et~al.}{2017}]{swinbank17}
{Swinbank} A.~M.,  et~al., 2017, \mn@doi [\mnras] {10.1093/mnras/stx201}, \href
  {https://ui.adsabs.harvard.edu/abs/2017MNRAS.467.3140S} {467, 3140}

\bibitem[\protect\citeauthoryear{{Tremonti}, {Moustakas}  \&
  {Diamond-Stanic}}{{Tremonti} et~al.}{2007}]{tremonti07}
{Tremonti} C.~A.,  {Moustakas} J.,   {Diamond-Stanic} A. a.~M.,  2007, \mn@doi
  [\apjl] {10.1086/520083}, \href
  {https://ui.adsabs.harvard.edu/abs/2007ApJ...663L..77T} {663, L77}

\bibitem[\protect\citeauthoryear{{Vanzella} et~al.,}{{Vanzella}
  et~al.}{2010}]{vanzella10}
{Vanzella} E.,  et~al., 2010, \mn@doi [\apj] {10.1088/0004-637X/725/1/1011},
  \href {https://ui.adsabs.harvard.edu/abs/2010ApJ...725.1011V} {725, 1011}

\bibitem[\protect\citeauthoryear{{Vanzella} et~al.,}{{Vanzella}
  et~al.}{2016}]{vanzella16}
{Vanzella} E.,  et~al., 2016, \mn@doi [\apj] {10.3847/0004-637X/825/1/41},
  \href {http://adsabs.harvard.edu/abs/2016ApJ...825...41V} {825, 41}

\bibitem[\protect\citeauthoryear{{Verhamme}, {Schaerer}  \&
  {Maselli}}{{Verhamme} et~al.}{2006}]{verhamme}
{Verhamme} A.,  {Schaerer} D.,   {Maselli} A.,  2006, \mn@doi [\aap]
  {10.1051/0004-6361:20065554}, \href
  {http://adsabs.harvard.edu/abs/2006A%26A...460..397V} {460, 397}

\bibitem[\protect\citeauthoryear{{Verhamme}, {Orlitov{\'a}}, {Schaerer}  \&
  {Hayes}}{{Verhamme} et~al.}{2015}]{verhamme15}
{Verhamme} A.,  {Orlitov{\'a}} I.,  {Schaerer} D.,   {Hayes} M.,  2015, \mn@doi
  [\aap] {10.1051/0004-6361/201423978}, \href
  {https://ui.adsabs.harvard.edu/abs/2015A%26A...578A...7V} {578, A7}

\bibitem[\protect\citeauthoryear{{Verhamme}, {Orlitov{\'a}}, {Schaerer},
  {Izotov}, {Worseck}, {Thuan}  \& {Guseva}}{{Verhamme}
  et~al.}{2017}]{verhamme17}
{Verhamme} A.,  {Orlitov{\'a}} I.,  {Schaerer} D.,  {Izotov} Y.,  {Worseck} G.,
   {Thuan} T.~X.,   {Guseva} N.,  2017, \mn@doi [\aap]
  {10.1051/0004-6361/201629264}, \href
  {https://ui.adsabs.harvard.edu/abs/2017A%26A...597A..13V} {597, A13}

\bibitem[\protect\citeauthoryear{{Walter}, {Brinks}, {de Blok}, {Bigiel},
  {Kennicutt}, {Thornley}  \& {Leroy}}{{Walter} et~al.}{2008}]{walter08}
{Walter} F.,  {Brinks} E.,  {de Blok} W.~J.~G.,  {Bigiel} F.,  {Kennicutt}
  Robert~C. J.,  {Thornley} M.~D.,   {Leroy} A.,  2008, \mn@doi [\aj]
  {10.1088/0004-6256/136/6/2563}, \href
  {https://ui.adsabs.harvard.edu/abs/2008AJ....136.2563W} {136, 2563}

\bibitem[\protect\citeauthoryear{{Wang}, {Heckman}, {Leitherer}, {Alexandroff},
  {Borthakur}  \& {Overzier}}{{Wang} et~al.}{2019}]{wang19}
{Wang} B.,  {Heckman} T.~M.,  {Leitherer} C.,  {Alexandroff} R.,  {Borthakur}
  S.,   {Overzier} R.~A.,  2019, \mn@doi [\apj] {10.3847/1538-4357/ab418f},
  \href {https://ui.adsabs.harvard.edu/abs/2019ApJ...885...57W} {885, 57}

\bibitem[\protect\citeauthoryear{{Weiner} et~al.,}{{Weiner}
  et~al.}{2009}]{weiner}
{Weiner} B.~J.,  et~al., 2009, \mn@doi [\apj] {10.1088/0004-637X/692/1/187},
  \href {http://adsabs.harvard.edu/abs/2009ApJ...692..187W} {692, 187}

\bibitem[\protect\citeauthoryear{{Willott} et~al.,}{{Willott}
  et~al.}{2010}]{willott}
{Willott} C.~J.,  et~al., 2010, \mn@doi [\aj] {10.1088/0004-6256/139/3/906},
  \href {http://adsabs.harvard.edu/abs/2010AJ....139..906W} {139, 906}

\bibitem[\protect\citeauthoryear{{Wisotzki} et~al.,}{{Wisotzki}
  et~al.}{2016}]{wisotzki}
{Wisotzki} L.,  et~al., 2016, \mn@doi [\aap] {10.1051/0004-6361/201527384},
  \href {https://ui.adsabs.harvard.edu/abs/2016A&A...587A..98W} {587, A98}

\bibitem[\protect\citeauthoryear{{Worseck} et~al.,}{{Worseck}
  et~al.}{2014}]{worseck14}
{Worseck} G.,  et~al., 2014, \mn@doi [\mnras] {10.1093/mnras/stu1827}, \href
  {http://adsabs.harvard.edu/abs/2014MNRAS.445.1745W} {445, 1745}

\bibitem[\protect\citeauthoryear{{Worseck}, {Prochaska}, {Hennawi}  \&
  {McQuinn}}{{Worseck} et~al.}{2016}]{worseck16}
{Worseck} G.,  {Prochaska} J.~X.,  {Hennawi} J.~F.,   {McQuinn} M.,  2016,
  \mn@doi [\apj] {10.3847/0004-637X/825/2/144}, \href
  {https://ui.adsabs.harvard.edu/abs/2016ApJ...825..144W} {825, 144}

\bibitem[\protect\citeauthoryear{{Zackrisson}, {Inoue}  \&
  {Jensen}}{{Zackrisson} et~al.}{2013}]{zackrisson13}
{Zackrisson} E.,  {Inoue} A.~K.,   {Jensen} H.,  2013, \mn@doi [\apj]
  {10.1088/0004-637X/777/1/39}, \href
  {https://ui.adsabs.harvard.edu/abs/2013ApJ...777...39Z} {777, 39}

\bibitem[\protect\citeauthoryear{{van Zee}, {Skillman}  \& {Salzer}}{{van Zee}
  et~al.}{1998}]{vanzee98}
{van Zee} L.,  {Skillman} E.~D.,   {Salzer} J.~J.,  1998, \mn@doi [\aj]
  {10.1086/300510}, \href
  {https://ui.adsabs.harvard.edu/abs/1998AJ....116.1186V} {116, 1186}

\bibitem[\protect\citeauthoryear{{van Zee}, {Salzer}  \& {Skillman}}{{van Zee}
  et~al.}{2001}]{vanzee01}
{van Zee} L.,  {Salzer} J.~J.,   {Skillman} E.~D.,  2001, \mn@doi [\aj]
  {10.1086/321108}, \href
  {https://ui.adsabs.harvard.edu/abs/2001AJ....122..121V} {122, 121}

\makeatother
\end{thebibliography}

\appendix
\section{The Radiative Transfer Equation with a Uniform Source Function in a Plane-Parallel Geometry} 
\label{plane-parallel}
An implicit assumption made to determine 
the neutral gas properties in \autoref{theory} is that the \mgii\ emission flux ratio is best described as a background source incident upon a foreground screen of Mg$^{+}$ gas. However, there are alternative geometries and scenarios that could complicate the interpretation of the \mgii\ emission lines. Here, we explore the next more complex possibility: a flat continuum source, incident on a parcel of Mg$^{+}$ gas that both absorbs the background continuum and is also collisionally excited to emit \mgii\ photons. In this scenario, the Mg$^+$ gas produces the \mgii\ photons along the path length of the observations.  

In this situation, the radiative transfer equation becomes
\begin{equation}
    \frac{d{\rm I}}{d\tau} = -I + S
    \label{eq:S}
\end{equation}
Where $I$ is the intensity of the light, $\tau$ is the optical depth, and $S$ is the source function of the parcel of gas. The source function is defined as the ratio of the emissivity (emission added per volume by the Mg$^+$ gas; $j$) to the absorption coefficient ($\alpha$) as 
\begin{equation} 
    S = \frac{j}{\alpha} .\label{eq:s}
\end{equation}
$\alpha$ is defined as
\begin{equation}
    \alpha = \frac{d\tau}{dl}
\end{equation}
where $l$ is the total path length of a ray traced through the Mg$^+$ gas.

Solving the radiative transfer equation to account for the source function requires assumptions about the geometry of the Mg$^+$ gas as well as the variation of the source function. If we assume that the source function is spatially uniform ($j$ and $\alpha$ are constant and $\alpha = \tau/l$), and that the geometry is plane parallel, then \autoref{eq:S} can be solved to be
\begin{equation}
    {I}_{\rm obs} = {I}_{\rm int} e^{-\tau} + S \left(1-e^{-\tau}\right) 
\end{equation}
Which describes the flux ratio of the \text{observed} \mgii\ emission lines as
\begin{equation}
    R =  \frac{F_{2796}}{ F_{2803}} = \frac{ F_{\rm int} e^{-\tau_{2796}} +S_{2796} \left(1-e^{-\tau_{2796}}\right)}{F_{\rm int} e^{-\tau_{2803}} +S_{2803} \left(1-e^{-\tau_{2803}}\right)} .
    \label{eq:rad_w_source}
\end{equation}

We can make two approximations to understand how \autoref{eq:rad_w_source} relates to the relations given in \autoref{theory}. First, as $\tau \rightarrow 0 $ we can expand $e^{-\tau}$ as a Taylor Series such that $e^{-\tau} \approx 1 - \tau$. This is to approximate the low optical depth observed from the \mgii\ emission lines. Second, we assume that $S \gg F_{\rm int}$, or that the \mgii\ emission is stronger than the background continuum (\autoref{continuum} below discusses this assumption). In this approximation, \autoref{eq:rad_w_source} becomes
\begin{equation}
    R \approx \frac{S_{2796} \tau_{2796}}{S_{2803} \tau_{2803}} \label{eq:r_ws}
\end{equation}
If we substitute the definitions of $S$ (\autoref{eq:s}) back into \autoref{eq:r_ws} we find that 
\begin{equation}
    R = \frac{\tau_{2796} \left(j_{2796}/\alpha_{2796}\right) }{\tau_{2803} \left(j_{2803}/\alpha_{2803}\right)} .
\end{equation}
However, $\alpha = \tau/l$ and
\begin{equation}
    R =  \frac{j_{2796}}{j_{2803}} .
\end{equation}
As the optical depth goes to zero, the ratio of the \mgii\ emission lines only depends on the ratio of the emissivities. Thus, even in a plane-parallel geometry with a constant source function, the intrinsic emission ratio of the \mgii\ doublet is set by the ratio of the two emissivities.

\section{The impact of absorption on the emissivities}
\label{cloudy}

One of the assumption that we made in \autoref{theory} is that collisions dominate the excitation of the Mg$^+$ electrons. However, there are other excitation possibilities: photo-excitation (also called photon pumping) or radiative cascades due to recombination. To test the dominant excitation source for J1503, we created \textsc{cloudy} photoionization models \citep[using V17.01;][]{ferland} with a 3~Myr, 0.2~Z$_\odot$, \textsc{starburst99} stellar continuum model as the continuum source \citet{leitherer14}, a fixed hydrogen density of 280~cm$^{-3}$ (measured from the \ion{S}{ii} lines), and the measured metallicity \citep[12+log(O/H) = 7.95; ][]{izotov16b}. We then varied the ionization parameter to give a range in [\ion{O}{iii}]/[\ion{O}{ii}], or, equivalently, the ionization parameter (ratio of the photon density to gas density). 

\begin{figure}
\includegraphics[width = 0.5\textwidth]{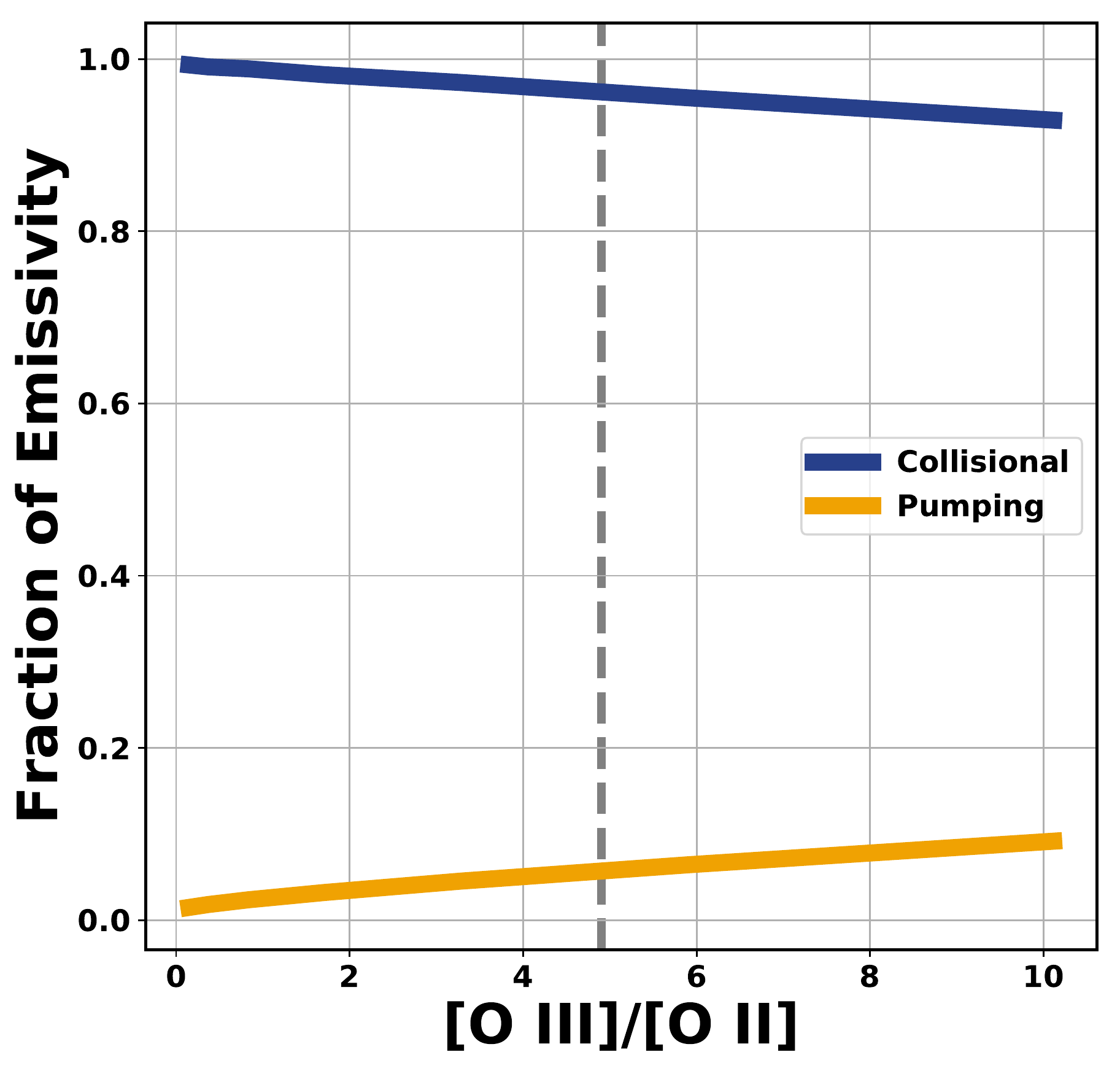}
\caption{Plot of the fraction of the total \mgii~2796~\AA\ emissivity contributed by collisional excitation (blue) and photon pumping (gold) versus the [\ion{O}{iii}]~5007~\AA/[\ion{O}{ii}]~3727~\AA\ emission ratio from \textsc{cloudy} photoionization models. The dashed gray line shows the observed [\ion{O}{iii}]~5007~\AA/[\ion{O}{ii}]~3727~\AA\ flux ratio in J1503. The total \mgii\ emissivity from galaxies with conditions similar to J1503 is dominated by collisional excitation ($>95$\%).}
\label{fig:em_test}
\end{figure}

\autoref{fig:em_test} uses the \textsc{cloudy} models to decompose the fractional impact of the total emissivity due to collisions (blue) and photon pumping of the \mgii~2796~\AA\ transition. \textsc{cloudy} does not predict other excitation mechanisms (e.g. recombination) to contribute to the \mgii\ emissivity. This shows that the \mgii\ emissivity is overwhelmingly ($>95$\%) dominated by collisional excitation at the observed [\ion{O}{iii}]/[\ion{O}{ii}] values of J1503 (gray line), and photon pumping does not largely contribute to the emissivity until much higher photon densities or lower gas densities (characterized by log(U)). These \textsc{cloudy} models also confirm that $R_{\rm int} = 2$, when collisions dominate the emissivity. This is the assumed $R_{\rm int}$ to determine \fescm\ throughout the paper (\autoref{eq:frat}). 

We caution that the \textsc{cloudy} documentation warns that the inferred emission fluxes of resonant lines dominated by photon pumping will not include line-of-sight absorption. Therefore, \textsc{cloudy} will over-estimate the total contribution of photon pumping to observed resonant emission line fluxes because, in reality, absorption will decrease the transmitted flux. The contribution of photon pumping to the \mgii\ emissivity will depend on the geometry of the situation (Section 16.44.3 of Hazy v17). In the geometry where the line of sight to the observer includes the continuum source, the absorption will reduce the pumping contribution to the emissivity, with the resultant line profile becoming a combination of emission plus absorption.

\section{The impact of the background continuum on the  optical depth}
\label{continuum}
In \autoref{theory}, we solved the radiative transfer equation to determine the \mgii\ optical depth and column density. These \mgii\ properties were combined  with the gas-phase metallicity to determine \nhi. One simplifying assumption that we made was that the \mgii\ emission was much brighter than the continuum, such that \autoref{eq:tau} became
\begin{equation}
    F_{\rm obs} = \left( F_{\rm int} + F_{\rm cont}\right) e^{-\tau} \approx F_{\rm int} e^{-\tau} . \label{eq:rad1}
\end{equation}
However, this simplification is not necessary if the continuum can be accurately measured. If the continuum emission (F$_{\rm cont}$), and the flux of both \mgii\ emission lines is measured, \autoref{eq:rad1} can be turned into a system of two equations and two unknowns (where $\tau_{2803} = 1/2\tau_{2796}$ and F$_{\rm 2803, int} = 1/2 F_{\rm 2796, int}$ from \autoref{eq:optical_depth} and \autoref{eq:rat}). This system of equations can be solved, assuming that both the continuum and emission lines have the same optical depth, to give three real solutions. The only solution that leads to a positive, non-zero $\tau_{\rm 2803, cont}$ is 
\begin{equation}
\tau_{\rm 2803, cont} = \ln{\left(\frac{F_{2803}
+\sqrt{F_{2803}^2 -F_{2796}F_{\rm cont}}}{F_{2796}}\right)} .\label{eq:tau_wcont}
\end{equation}
If $F_{\rm cont} = 0$, this reduces to the previous $\tau_{2803} = - \ln(R/2)$ relation, given in \autoref{eq:r_tau}.  

\begin{figure}
\includegraphics[width = 0.5\textwidth]{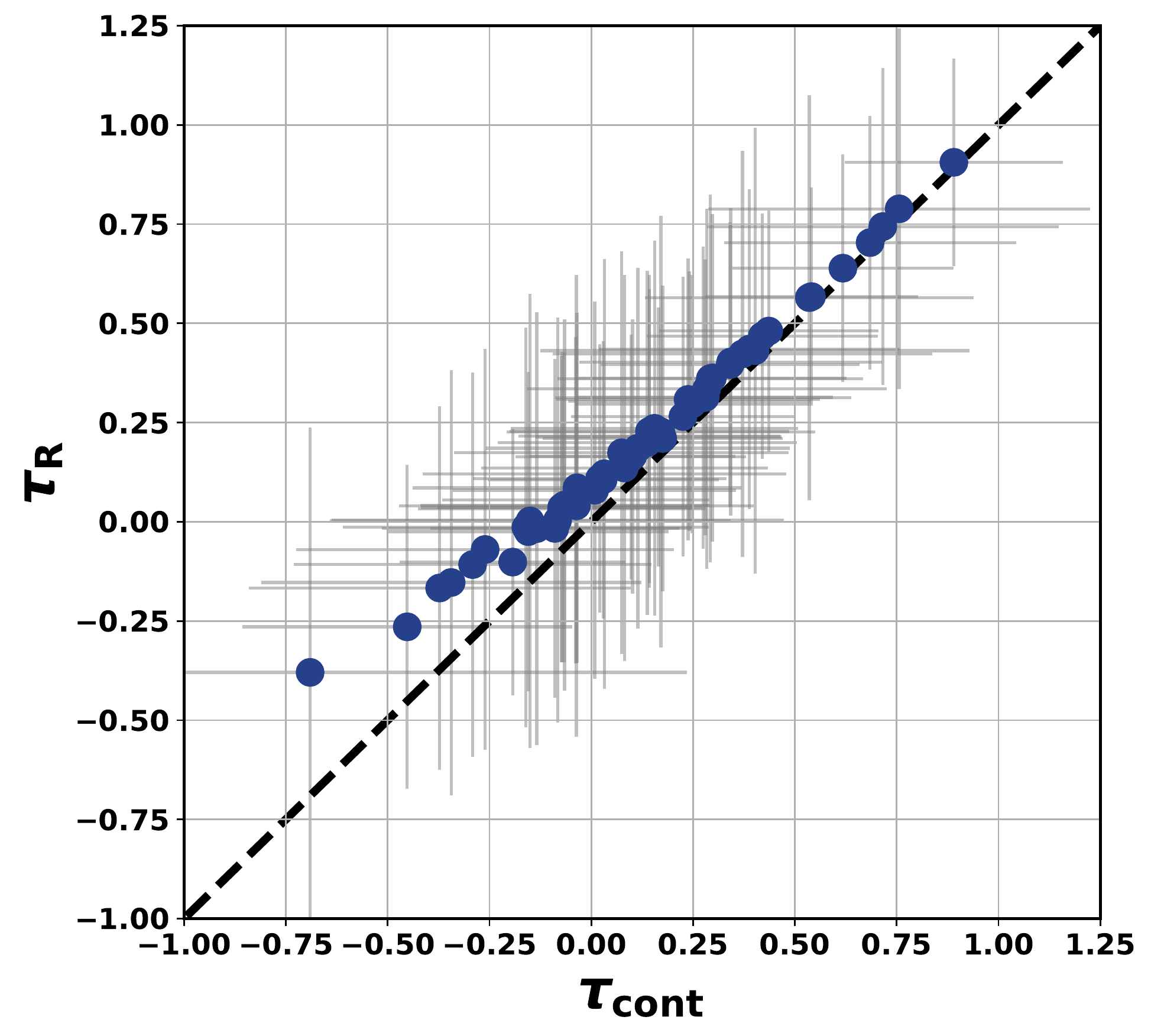}
\caption{Plot of the optical depth calculated assuming that the \mgii\ emission is much stronger than the continuum ($\tau_{R}$) versus the optical depth calculated solving the full radiative transfer equation (\autoref{eq:tau_wcont}). The values follow the one-to-one line over nearly the entire $\tau$ range. The values slightly deviate at low optical depth, but remain consistent within the errors. Thus, our simplification throughout the paper that $F_{\rm int} \gg F_{\rm cont}$ is valid.   }
\label{fig:tau_test}
\end{figure}
\autoref{fig:tau_test} compares the \mgii\ optical depth calculated assuming that the \mgii\ is much brighter than the continuum ($\tau_{\rm R}$, the value used in the paper) and explicitly accounting for the continuum attenuation ($\tau_{\rm cont}$; \autoref{eq:tau_wcont}). The optical depths are nearly identical for most $\tau$, and marginally deviate at the lowest $\tau$. However, the two $\tau$ values always remain statistically consistent with each other. The behavior at low optical depth is expected from \autoref{eq:tau_wcont}, where $F_{2803}$ must be less than $F_{2796}$ and $F_{2803}^2 << F_{2796} F_{\rm cont}$ to produce a low $\tau_{\rm 2803, cont}$. This indicates that $F_{\rm cont}$ impacts $\tau$ the most when the $\tau$ is low. The assumption that $F_{\rm int} \gg F_{\rm cont}$, and the subsequent results relating $R$ and \nhi, is valid for J1503.
\end{document}